\documentclass[twocolumn,english,showpacs,preprintnumbers,amsmath,amssymb,floatfix,nofootinbib]{revtex4-1}

\usepackage[T1]{fontenc}
\usepackage[latin9]{inputenc}
\usepackage{color}
\usepackage{array}
\usepackage{amstext}
\usepackage{subfig}
\usepackage{graphicx}
\usepackage{esint}
\usepackage{rotating}
\usepackage{slashed}
\usepackage{siunitx}
\usepackage[normalem]{ulem}
\usepackage[font=small,labelfont=bf]{caption}
\usepackage{lipsum}
\usepackage{xcolor}
\usepackage[colorlinks = true,
linkcolor = magenta,
urlcolor  = blue,
citecolor = red,
anchorcolor = blue]{hyperref}

\newcommand{\pom} {I\!\!P}
\newcommand{\reg} {I\!\!R}
\newcommand{\xpom}{x_{\xpom}}

\makeatletter

\providecommand{\tabularnewline}{\\}

\@ifundefined{textcolor}{}
{%
 \definecolor{BLACK}{gray}{0}
 \definecolor{WHITE}{gray}{1}
 \definecolor{RED}{rgb}{1,0,0}
 \definecolor{GREEN}{rgb}{0,1,0}
 \definecolor{BLUE}{rgb}{0,0,1}
 \definecolor{CYAN}{cmyk}{1,0,0,0}
\definecolor{MAGENTA}{cmyk}{0,1,0,0}
 \definecolor{YELLOW}{cmyk}{0,0,1,0}
 }

\@ifundefined{definecolor}
{\usepackage{color}}{}
\@ifundefined{definecolor}
{\usepackage{color}}{}
\makeatother
\usepackage{babel}

\def\Re{{\cal R \mskip-4mu \lower.1ex \hbox{\it e}\,}}
\def\Im{{\cal I \mskip-5mu \lower.1ex \hbox{\it m}\,}}

\def\tev{\,{\ifmmode\mathrm {TeV}\else TeV\fi}}
\def\gev{\,{\ifmmode\mathrm {GeV}\else GeV\fi}}
\def\mev{\,{\ifmmode\mathrm {MeV}\else MeV\fi}}
\def\to{\rightarrow}

\begin{document}

%
%
\title{Determination of diffractive PDFs from global QCD analysis of inclusive diffractive DIS and dijet cross-section measurements at HERA} 
%
%

\author{Maral~Salajegheh$^{1}$} 
\email{Maral@hiskp.uni-bonn.de }

\author{Hamzeh~Khanpour$^{2,3,4,5}$}
\email{Hamzeh.Khanpour@cern.ch}

\author{Ulf-G.~Mei{\ss}ner$^{1,6,7}$}
\email{meissner@hiskp.uni-bonn.de }

\author{Hadi~Hashamipour$^{5}$}
\email{ H\_Hashamipour@ipm.ir}

\author{Maryam~Soleymaninia$^{5}$}
\email{Maryam\_Soleymaninia@ipm.ir}

\affiliation {
$^{1}$Helmholtz-Institut f\"ur Strahlen-und Kernphysik and Bethe Center for Theoretical Physics, Universit\"at Bonn, D-53115 Bonn, Germany.  \\
$^{2}$Dipartimento Politecnico di Ingegneria ed Architettura, University of Udine, Via della Scienze 206, 33100 Udine, Italy. \\
$^{3}$International Centre for Theoretical Physics (ICTP), Strada Costiera 11, 34151 Trieste, Italy. \\
$^{4}$Department of Physics, University of Science and Technology of Mazandaran, P.O.Box 48518-78195, Behshahr, Iran.  \\
$^{5}$School of Particles and Accelerators, Institute for Research in Fundamental Sciences (IPM), P.O.Box 19395-5531, Tehran, Iran.   \\
$^{6}$Institute for Advanced Simulation and Institut f\"ur Kernphysik, Forschungszentrum J\"ulich, D-52425 J\"ulich, Germany. \\
$^{7}$Tbilisi State University, 0186 Tbilisi, Georgia.
}

\date{\today}

%
\begin{abstract}

We  present an updated set of {\tt SKMHS} 
diffractive parton distribution functions (PDFs).
In addition to the diffractive deep-inelastic 
scattering (diffractive DIS) data sets, the recent 
diffractive dijet cross sections measurements by the H1
experiment from the HERA collider are added to the data sample.
The new set of diffractive PDFs, entitled  {\tt SKMHS23} and  {\tt SKMHS23-dijet},
are presented at  next-to-leading order (NLO) and 
next-to-next-to-leading order (NNLO) accuracy in perturbative QCD. 
Since the gluons directly contribute to jet production through 
the boson-gluon fusion process, the data on 
diffractive dijet production in inclusive DIS 
help to constrain the gluon density, allowing for the determination of 
both the quark and gluon densities with better accuracy.
The NLO and NNLO theory predictions calculated using both {\tt SKMHS23} and  
{\tt SKMHS23-dijet} are 
compared to the analyzed data showing excellent agreements.
The effect arising from the inclusion of diffractive dijet 
data and higher order QCD corrections on the extracted 
diffractive PDFs and data/theory agreements are clearly examined 
and discussed.

\end{abstract}
%

\maketitle
\tableofcontents{}

%
\section{Introduction}\label{sec:introduction}
%

In deep inelastic scattering (DIS) process, 
the diffractive reactions of the type $ep \to eXY$, where 
$X$ indicates a high-mass hadronic final state, 
represent about 8-10\% of 
the events at HERA. Such type of process provides rich
experimental input to test  quantum chromodynamics (QCD) 
in the diffractive regime~\cite{ZEUS:2008xhs,Frankfurt:2022jns,H1:2006zxb,H1:2006zyl,ZEUS:2009uxs}.

According to the QCD factorization theorem~\cite{Collins:1989gx,Collins:1997sr}, 
calculations of the diffractive
cross sections with high enough Q$^2$ factorizes into two main different parts: a set of 
process-independent diffractive parton distribution functions (PDFs) and  
a process-dependent hard scattering coefficient function.

The diffractive PDFs need to be determined  from a QCD fit to the 
measured inclusive diffractive cross sections by 
applying the standard DGLAP evolution equations~\cite{Gribov:1972ri,Lipatov:1974qm,Dokshitzer:1977sg,Altarelli:1977zs}, 
while the hard scattering 
coefficient functions are calculable in perturbative QCD.
The QCD factorization is proven to hold both for the inclusive 
and the dijet diffractive processes~\cite{Collins:1989gx,Collins:1997sr}.
However, in the case of low photon virtuality, some 
non-perturbative quantities such as 
higher twist (HT) effects need to be taken into account. 

From an phenomenological point of view, the diffractive PDFs are 
determined by assuming an additional factorization that depends 
on the structure of a colorless exchange objects.
This assumption is known as proton vertex factorization~\cite{H1:2006zxb}.
In a diffractive DIS process, 
the Pomeron and Reggeon flux-factors
in the proton are introduced, and for the case of diffractively exchanged objects the 
universal Parton densities are assumed.
Several measurements on the diffraction in DIS suggest the 
validity of the proton vertex factorization assumption in diffractive DIS~\cite{H1:2006zxb}. 
Diffractive PDFs are universal quantities for all diffractive DIS reactions, with
the hardness of the DIS process being ensured by 
the virtuality of the exchanged photon, Q$^2$~\cite{H1:2006zxb}.

Nearly all recent progress in the extraction of diffractive PDFs
stems from the widely used H1 and ZEUS diffractive 
DIS cross section measurements. 
Over the past few years, some groups have reported their   
sets of diffractive PDFs with uncertainties, 
such as {\tt H1-2006-DPDF}~\cite{H1:2006zyl}, {\tt ZEUS-2010-DPDF}~\cite{ZEUS:2009uxs}, 
{\tt GKG18}~\cite{Goharipour:2018yov}, 
{\tt HK19}~\cite{Khanpour:2019pzq}, {\tt MMKG19}~\cite{Maktoubian:2019ppi}, 
and the most recent analysis by {\tt SKMHS22}~\cite{Salajegheh:2022vyv}.
Among these diffractive PDFs determinations,  the {\tt HK19} and {\tt SKMHS22} are 
performed up to NNLO accuracy in perturbative QCD, while the former are limited to the NLO.
The {\tt GKG18} and {\tt SKMHS22} are performed in the framework of {\tt xFitter}~\cite{Alekhin:2014irh} in 
order to achieve a more reliable estimate of the diffractive PDFs uncertainties.
In addition, {\tt GKG18}, {\tt HK19} and {\tt SKMHS22} also analyzed the 
most recent H1/ZEUS combined diffractive DIS cross section measurements.

Up to now, predictions for diffractive DIS, and in
particular the diffractive dijet production, were performed
at NLO and NNLO order in QCD.
ZEUS-2010-DPDF analyzed also the diffractive dijet production 
data at NLO~\cite{ZEUS:2009uxs}, and most recently the 
predictions for the dijet production is provided at NNLO  in Ref.~\cite{Britzger:2018zvv}.

In this paper, we present {\tt SKMHS23-dijet},  
a new determination of diffractive PDFs using the 
previous analyzed inclusive diffractive DIS 
measurements by the H1 and ZEUS Collaborations, including for the first time  
the dijet production cross-section measurements in   
diffractive $ep$ scattering data collected in the 
years 2005-2007 with the H1 detector at HERA.
The  {\tt SKMHS23-dijet}  is extracted from QCD analysis at NLO and NNLO 
accuracy in perturbative QCD.
In order to analyze the dijet production data, the well-established {\tt Alpos} 
framework~\cite{Britzger:2019lvc,Alpos1,Alpos2} 
supplemented with  {\tt APFEL}~\cite{Bertone:2013vaa},  {\tt NNLOJET} and  
{\tt fastNLO}~\cite{Britzger:2012bs,Wobisch:2011ij} is used which is an
object-oriented data to theory comparison fitting tool. 
The statistical analysis of the theory predictions for both diffractive DIS  and dijet production are 
also performed using this program.
The diffractive dijet production data which are included in {\tt SKMHS23-dijet} help to 
constrain the gluon density, allowing for a good accuracy determination 
of both the quark and gluon densities.
In order to examine the effect of dijet data on the extracted densities, 
we also present the {\tt SKMHS23} 
analysis in which the dijet data are excluded from the data sample. 
Finally, the NLO and NNLO theory predictions are compared to the analyzed data.
The effect arising from the inclusion of diffractive dijet data and 
higher order QCD corrections on the extracted diffractive 
PDFs and data/theory agreement are also examined and discussed.

The rest of the paper is organized as follows. 
The theoretical framework considered in {\tt SKMHS23} is introduced
in Sec.~\ref{Theoretical-Framework}. This section also discusses the
QCD prediction for the diffractive dijet production in an 
electron-proton $(ep)$ scattering process, and the corresponding factorization theorem. 
The details of the {\tt SKMHS23} are presented in Sec.~\ref{global-QCD-analysis} which
includes the experimental input, the {\tt SKMHS23} parameterizations, the heavy
quark contributors to the diffractive DIS process, and 
finally the fitting framework and minimization strategy. 
The {\tt SKMHS23} fit results and main findings of this work are scrutinized and
discussed in Sec.~\ref{sec:results}.  
Finally, Sec.~\ref{Conclusion} summarizes the 
findings, and outlines possible future developments as well.

%
\section{Theoretical Framework}\label{Theoretical-Framework}
%

In this section, we describe in detail the standard 
theoretical framework for diffractive DIS processes in which 
the perturbative QCD framework is applied for the event 
with a large rapidity gap (LRG) in the rapidity distribution of the outgoing hadrons.
We thoroughly discuss the calculation of diffractive dijet cross sections in 
inclusive DIS processes as well and
the relevant factorization theorem. 
We also provide the details of the factorization of the proton diffractive 
PDFs.

%
\subsection{QCD prediction for diffractive dijet production in $(ep)$ scattering}\label{sec:Jet-production-cross-section}
%

The diffraction, $\gamma ^*+ p \rightarrow X + p$, in the 
single diffractive process such as inclusive diffractive DIS, $e(k) + p(P) 
\rightarrow e(k^\prime) + p(P^\prime)+X$,  is observed when the 
virtual photon $\gamma^*$ dissociates into 
the hadronic system $X$ whereas the proton remains intact. 
The diffractive reaction in DIS is described by the DIS 
kinematic invariants which are given by
\begin{align}
\label{eq:DISkin}
Q^2&=-q^2=(k-{k^{\prime}}^2),\nonumber\\
x&=\frac{-q^2}{2P\cdot q},\nonumber\\
y&=\frac{P\cdot q}{P\cdot k},
\end{align}
where $Q^2$ is the virtuality of photon, $x$ is the longitudinal 
fraction of the proton momentum carried by the struck quark 
(same as the Bjorken scaling variable), and $y$ indicates  the inelasticity. 
These quantities are related via $Q^2 = xys$, where 
the electron-proton center-of-mass energy squared is denoted by $s$.

In addition, the new quantities for diffractive kinematics are 
defined in  relation to the scattered protons. 
One of them is the longitudinal momentum fraction of 
the exchanged Pomeron:
\begin{align}
\label{eq:x_ip}
x_{\pom} = \frac{q\cdot (P-P^\prime)}{q\cdot P}\,.
\end{align}
The second variable is the squared 
four-momentum transfer at the proton vertex:
\begin{align}
\label{eq:t}
t=(P^\prime -P)^2\,.
\end{align}
Finally, the last one is the fractional momentum of the 
diffractive exchange carried by the parton inside the Pomeron:
\begin{align}
\label{eq:beta}
\beta=\frac{x}{x_{\pom}}=\frac{Q^2}{2q\cdot (P-P^{\prime})}\,.
\end{align}

\begin{figure}[t]
	\vspace{0.20cm}
	\resizebox{0.40\textwidth}{!}{\includegraphics{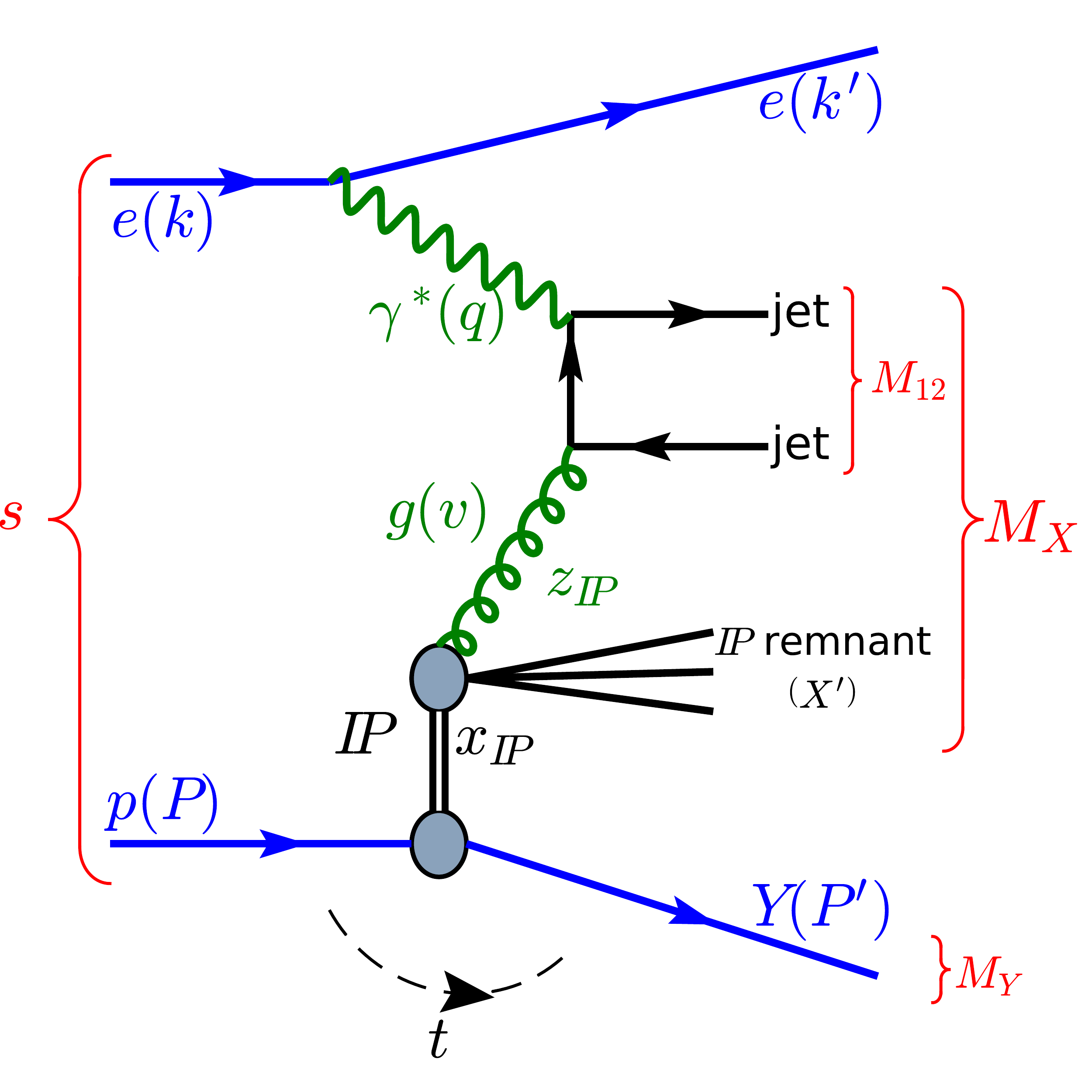}}
	\begin{center}
		\caption{{\small
The Feynman diagram describing the diffractive dijet 
production in an electron-proton collision at HERA.
			} 
			\label{fig:Feynman}}
	\end{center}
\end{figure}

The cross section of diffractive dijet ($jj$) production, $e + p \rightarrow e + p + jj + X^\prime$,  
is an important obsevable which can affect the behavior of diffractive PDFs. 
The inclusion of these data
in the analysis is one of the main objective of this study.
The Feynman diagram describing the diffractive dijet production 
in an electron-proton collision at HERA is
shown in Fig.~\ref{fig:Feynman}.
%
%
%
%
For the diffractive dijet production, an additional variable needs to be introduced. 
According to the Feynman diagram presented in Fig.~\ref{fig:Feynman}, 
in the hard subprocess, $v$  is the four-momentum of 
the gluon emitted from the Pomeron.

The longitudinal momentum fraction of  
gluon is denoted by $z_{\pom}$, the new invariant:
\begin{align}
\label{eq:Dijetvariables}
z_{\pom}=\frac{q \cdot v}{q \cdot (P - P^{\prime})}.
\end{align}
It should be noted here that for the dijet production process 
the variable $x$ is not the momentum fraction of the parton entering to the hard subprocess. 
This fraction is denoted as $\tilde{x}$ and it is 
momentum fraction of the interacting parton with respect to the proton. 
Further, $x_{\pom}$ is the momentum fraction of the Pomeron with respect to the proton. 
The momentum fraction of the parton with respect to 
the Pomeron is denoted by the $z_{\pom}$.
It can be shown that for the dijet production one can write:
\begin{align}
\label{eq:Dijet2}
z_{\pom}=\frac{\tilde{x}}{x_{\pom}}.
\end{align}

At leading order (LO), the center-of-mass energy of hard 
subprocess is equal to the invariant mass of the 
dijet system $M_{12}$
\begin{align}
\label{eq:COM}
M_{12}^2=(q+v)^2.
\end{align}

In the next section, the factorization theorem will be presented and discussed.

%
\subsection{Factorization theorem in diffractive dijet production}\label{Factorization-theorem }
%

The factorization theorem of QCD can be employed for the 
diffractive processes so that the cross section of dijet production
is given by  the convolution of diffractive PDFs for proton $f^D_{i/p}$ with
the partonic  cross sections $d\hat{\sigma}$~\cite{Collins:1989gx,Collins:1997sr}, 
\begin{align}
\label{eq:Dijetcross}
&d\sigma (e+p\rightarrow e+p+jj+X^{\prime})=\nonumber \\
&\sum _i \int dt \int dx_{\pom} \int dz_{\pom}\nonumber \\
&\times f^D_{i/p}(z_{\pom},\mu _F^2,x_{\pom},t) 
\otimes d \hat{\sigma}_{ei\rightarrow jj}
(\hat{s},\mu^2_{R},\mu ^2_F).
\end{align}
Here, the hadronic system $X^\prime$ is  
what remains of the hadronic system $X$ after 
removing the two jets. 
In addition, the integrals are performed over the accepted phase space and 
the sum runs over all the partons contributing to the cross section.
The first argument of the diffractive PDFs $f^D_{i/p}$ is the momentum of the 
parton with respect to the Pomeron.
$\mu _F$ and $\mu _R$ represent the factorization and 
renormalization scales, respectively. 
The invariant energy squared in the subprocess is defined as 
\begin{align}
\label{eq:energy}
\bar{s}\sim x_{\pom}z_{\pom}ys-Q^2.
\end{align}
In the DIS region in which $Q^2 \gg \Lambda ^2$, the only relevant contribution 
to the dijet production cross section is the 
direct process defined in Eq.~\eqref{eq:Dijetcross}.

According to the proton vertex factorization theorem, 
the diffractive PDFs can be 
factorized into the product of two distinct terms.
The first term depends on the $x_{\pom}$ and $t$,   
while the second term depends only to the $z_{\pom}$ and $\mu _F$.
Hence, the diffractive PDFs $f_{i/p}^D(z_{\pom}, \mu _F^2; x_{\pom}, {t})$ is given by,  
\begin{align}
\label{eq:DPDFs}
f_{i/p}^D(z_{\pom},\mu _F^2; x_{\pom}, {t})= & 
f_{\pom/p}(x_{\pom}, {t})
f_{i/\pom}(z_{\pom},\mu _F^2) \nonumber\\
+&n_{\reg}f_{\reg/p}(x_{\pom}, {t}) 
f_{i/\reg}(z_{\pom},\mu _F^2)\,,
\end{align}
where the Pomeron and 
Reggeon flux-factors are denoted by $f_{\pom/p}(x_{\pom}, {t})$ and 
$f_{\reg/p}(x_{\pom}, {t})$, respectively. 
The flux-factors describe the emission of the 
Pomeron and Reggeon from the proton target. 
The Reggeon contributes significantly at low $z_{\pom}$ and large $x_{\pom}$. 
The global normalization of the Reggeon contribution is $n_{\reg}$, which 
taken as free parameter in the fit. 
The Pomeron and Reggeon partonic distribution functions are indicated 
by $f_{{i}/\pom}(z_{\pom},\mu _F^2)$ and $f_{i/\reg}(z_{\pom},\mu _F^2)$, respectively. 

The parametrization and determination of these distribution functions will be 
discussed in detail in section~\ref{subsec:param}.

Many properties of  diffractive PDFs are similar to the 
non-diffractive PDFs. 
Despite the fact that the presence of the leading proton in the final state 
leads to an additional constraint for the calculation of 
diffractive PDFs, they still obey 
the standard DGLAP evolution equation like the ordinary 
PDFs~\cite{Altarelli:1977zs,Dokshitzer:1977sg,Gribov:1972ri}. 
In the analysis of diffractive PDFs the cross section for diffractive processes has a $t$-dependence,
which one usually integrates out. Consequently, the $t$-dependence of diffractive PDFs is
restricted to $|t|<1.0~\mathrm{GeV}^2$ here.

As mentioned before, our aim in this paper is to include the dijet production 
cross section in the diffractive DIS analysis up to NNLO. 
The calculations of partonic cross sections of diffractive dijet production at NNLO accuracy  
and the one from dijet production in DIS are the same. 
Recently, the latter calculations have been used to describe the inclusive dijet cross 
section in DIS~\cite{Currie:2016ytq,Currie:2017tpe}.

According to Eq.~\eqref{eq:Dijetcross}, 
to calculate the diffractive dijet cross section one 
needs to convolute the partonic cross section $d \hat{\sigma}_{ei\rightarrow jj}$ 
with the diffractive PDFs $f^D_{i/p}(z_{\pom}, \mu _F^2, x_{\pom}, t)$. 
In our work, the hard (partonic) cross section is calculated using the 
\texttt{NLOJet++} package~\cite{Nagy:2001xb}, however, to account for 
the additional dependence of the cross section on $x_{\pom}$ and $t$ some 
adjustments are required as specified in Ref.~\cite{Britzger:2018zvv}.
This calculation can be very time-consuming using conventional methods 
such as Monte Carlo integration, specially if one requests a high precision.
Nevertheless, in a QCD analysis one should repeatedly 
evaluate this convolution for different values of diffractive PDF parameters. 
To overcome this difficulty, an interface to the \texttt{fastNLO} package is 
implemented in the \texttt{Alpos} framework~\cite{Britzger:2019lvc,Alpos1,Alpos2}.
By using the methodology of the \texttt{fastNLO}, the calculation of matrix elements is 
done only once and the convolution integral turns into a 
summation over a grid of integration variables~\cite{Britzger:2012bs}. 
Additional information and details of the \texttt{fastNLO} formalism can 
be found in e.g.~\cite{Britzger:2013kia}.

%
\section{Details of the {\tt SKMHS23} QCD analysis}\label{global-QCD-analysis} 

In this section, we present the details of the {\tt SKMHS23} QCD analysis, including 
the experimental data sets analyzed in this work, the diffractive PDFs parametrization,
the minimization procedures, and the diffractive PDF uncertainty method.

%
\subsection{Experimental data sets}\label{subsec:data}

This section deals with the experimental data sets 
used in the {\tt SKMHS23} global QCD analysis, focusing on the 
diffractive dijet cross-section measurements at HERA.
The details of the inclusive diffractive DIS experimental 
input are discussed in detail in our 
previous studies~\cite{Salajegheh:2022vyv,Goharipour:2018yov} and we will present a short review here.

The determination of diffractive PDFs relies mainly on the inclusive diffractive DIS 
cross-section measurements  by the H1 and ZEUS collaborations.
The inclusive diffractive DIS and dijet data sets which are listed 
in Tables~\ref{tab:DDISdata} and \ref{tab:DIJETdata}
include the following:
\begin{itemize}

\item 
The H1 measurements of the inclusive diffractive DIS cross section, {\tt  H1-LRG-11}, at 
the $\sqrt{s} = 225, 252, 319$~GeV$^2$ which covers the phase space of
$4.0~ \text{GeV}^2 < \text{Q}^2 < 44.0 ~ \text{GeV}^2$  and  
$5.0 \times 10^{-4} < x_{\pom} < 3.0 \times 10^{-3}$~\cite{H1:2011jpo}.

\item 
The inclusive measurement of diffractive DIS by the H1 Collaboration, called {\tt H1-LRG-12}~\cite{H1:2012pbl}.
This measurement covers the phase space $3.0~ \text{GeV}^2 < \text{Q}^2 < 1600 ~ \text{GeV}^2$ 
of the photon virtuality,
and the squared four-momentum transfer of $|t| < 1.0 ~ \text{GeV}^2$.

\item 
The most recent published data on the diffractive DIS 
cross-section come from the H1 and ZEUS combined measurement which 
is useful to determine precise diffractive PDFs with reliable uncertainty.
The kinematic range of
these measurements is $2.5~ \text{GeV}^2 < \text{Q}^2 < 200 ~ \text{GeV}^2$
for the photon virtuality, 
$3.5 \times 10^{-4} < x_{\pom} < 9.0 \times 10^{-2}$ for the proton fractional momentum loss, 
$1.8 \times 10^{-3} < \beta < 0.816$ in scaled fractional 
momentum variable and finally
$0.09 ~ \text{GeV}^2 < |t| < 0.55 ~ \text{GeV}^2$ in the squared 
four-momentum transfer at the proton vertex~\cite{H1:2012xlc}.

As discussed in detail in Refs.~\cite{Khanpour:2019pzq,Goharipour:2018yov}, 
the H1 and ZEUS combined data 
are subject to two different corrections which are the proton
dissociation background and the global normalization factor 
for the extrapolation from $0.09 ~ \text{GeV}^2 < |t| < 0.55 ~ \text{GeV}^2$ to 
$|t| < 1.0 ~ \text{GeV}^2$.

\item
The single differential dijet cross sections measurements 
in diffractive DIS  published by the H1 Collaboration
at HERA which correspond to an integrated luminosity of 290~pb$^{-1}$~\cite{H1:2014pjf}. 
The phase space of these measurements is spanned by the photon virtuality 
of $4.0~ \text{GeV}^2 < \text{Q}^2 < 100 ~ \text{GeV}^2$, and by the
fractional proton longitudinal momentum loss $x_{\pom} < 3.0 \times 10^{-2}$. 
As will be discussed in detail below, the diffractive DIS dijet data
is used in the {\tt SKMHS23} QCD analysis for the first time.
The effect arising from the inclusion of this data on the extracted 
diffractive PDFs and 
data/theory agreements will also be discussed.

\end{itemize}

\begin{table*}
	\caption{\small List of all diffractive DIS data points with their properties 
		used in the {\tt SKMHS23}  global QCD analysis. 
		For each data set we  provide the 
		kinematical coverage of ${\beta}$, ${{x}_{\pom}}$, and ${Q}^{2}$. The number of data points is 
		displayed as well. 
		The details of the kinematical cuts applied on these data sets are explained in the text. } 
	\label{tab:DDISdata}
	\begin{tabular}{l | c  c  c  c  c  c }
		\hline\hline
		\tt{\text{Experiment}} ~&~ \tt{\text{Observable}} ~&~ [$\beta^{{\text{min}}}, {\beta}^{{\text{max}}}$] ~&~ [${x_{\pom}}^{\rm{min}}, {x_{\pom}}^{\rm{max}}$]  ~&~ ${Q}^{2}\,[{\text{GeV}}^2]$  ~&~ \tt{\# of points} ~&~ \tt{\text{Reference}}
		\tabularnewline
		\hline\hline
		\tt{\text{H1-LRG-11}} $\sqrt{s}={225}$ GeV & ${\sigma}_{r}^{D(3)}$ & [$0.033$--$0.88$]    & [$5{\times} 10^{-4}$ -- $3{\times} 10^{-3}$] & 4--44 & \textbf{22}  & \cite{H1:2011jpo} 
		\\
		\tt{\text{H1-LRG-11}} $\sqrt{s}={252}$ GeV & ${\sigma}_{r}^{D(3)}$ & [$0.033$--$0.88$]    & [$5{\times} 10^{-4}$ -- $3{\times} 10^{-3}$] & 4--44 & \textbf{21}  &  \cite{H1:2011jpo} 
		\\
		\tt{\text{H1-LRG-11}} $\sqrt{s}={319}$ GeV & ${\sigma}_{r}^{D(3)}$ & [$0.089$--$0.88$]    & [$5{\times} 10^{-4}$ -- $3{\times} 10^{-3}$] & 11.5--44 & \textbf{14} &  \cite{H1:2011jpo}   
		\\	
		\tt{\text{H1-LRG-12}} & ${\sigma}_{r}^{D(3)}$ & [$0.0017$--$0.80$]   & [$3{\times} 10^{-4}$ -- $3{\times} 10^{-2}$] & 3.5--1600 & \textbf{277}   &  \cite{H1:2012pbl} 
		\\	
		\tt{\text{H1/ZEUS combined}} & ${\sigma}_{r}^{D(3)}$  &   [$0.0018$--$0.816$]   & [$3{\times} 10^{-4}$ -- $9 {\times} 10^{-2}$] & 2.5--200 & \textbf{192}  & \cite{H1:2012xlc}   
		\\			
		\hline \hline
		\multicolumn{1}{c}{\tt\textbf{Total data}} ~~&~~ &~~ &~~& ~~&~~\textbf{526}  \\  \hline  \hline
	\end{tabular}
\end{table*}

\begin{table*}
	\caption{\small Dijet data set 
		used in the {\tt SKMHS23}  global QCD analysis.} 
	\label{tab:DIJETdata}
	\begin{tabular}{l | c  c  c  c  c  c }
		\hline\hline
		\tt{\text{Experiment}} ~&~ \tt{\text{Observable}} ~&~ \tt{DIS range} ~&~ \tt{dijet range}  ~&~ \tt{Diffractive range}  ~&~ \tt{\# of points} ~&~ \tt{\text{Reference}}
		\tabularnewline
		\hline\hline
		\tt{\text{H1-LRG (HERA II)}} & $d^2\sigma/dp_T^{*jet1}dQ^2$ & $0.1<y<0.7$    & $p_T^{*jet1}>5.5$ GeV & $x_{\pom}<0.03$  &\textbf{15} &\cite{H1:2014pjf} 
		\\
		 &  & $4<Q^2<100$ GeV$^2$    & $p_T^{*jet2}> 4.0 $ GeV& $|t|<1$ GeV$^2$  &  
		\\
		&  &    & $-1<\eta^{jet}<2$ & $M_Y<1.6GeV$ &  &    
		  
		\\			
		\hline \hline
		\multicolumn{1}{c}{\tt\textbf{Total data}} ~~&~~ &~~ &~~& ~~&~~\textbf{15}  \\  \hline  \hline
	\end{tabular}
\end{table*}

Finally, in order to avoid the contributions from  
higher twist (HT) and some other 
nonperturbative effects, one needs to implement 
some kinematical cuts to all 
diffractive DID data sets mentioned above. 
To this end, we follow the formalism presented in 
Refs.~\cite{Goharipour:2018yov,Salajegheh:2022vyv,ZEUS:2009uxs} and 
consider some cuts on the data samples.
We consider $M_X \geqslant 2~\mathrm{GeV}$, the data sets for $\beta \geqslant 0.81$ are excluded.
A $\chi^2$ scan is performed in Ref.~\cite{Goharipour:2018yov}, 
to find an optimum value for the $Q^2$.
In this work, the region with $Q^2 = Q^2_{\text{min}}  \geqslant 9~\mathrm{GeV}^2$
only is included in the fit,  which 
shows the best data/theory description.

%
\subsection{{\tt SKMHS23} diffractive PDFs parametrization}\label{subsec:param}
%

Like the standard PDFs, the diffractive PDFs are nonperturbative quantities 
and should be determined by  a QCD global analysis. 
As it has been mentioned before, diffractive PDFs are the sum of Pomeron and secondary 
Reggeon contributions neglecting the possible interference terms. 
We consider the parametrization form for diffractive PDFs with unknown parameters 
at a starting scale $\mu_0^2 = 1.69$~GeV$^2$, which is  
less than the squared charm mass ($m_c^2$) threshold.

Due to the lack of experimental data  
for the case of diffractive processes, a 
somewhat less flexible parametrization form for the diffractive PDFs 
is employed in our work. 
For the same reason the Pomeron PDFs at the initial scale $f_{i/\pom}(z_{\pom},\mu _0^2) $ should 
be the same for all light partons $i = u = d = s = \bar{u} = \bar{d} = \bar{s}$, 
while the gluon distribution is considered separately. 
The contribution of the Reggeon PDFs becomes 
important at large values of $x_{\pom}$ and it is equal to the pion PDF.
For the leading Pomeron pole at starting scale $Q_0^2$, we parametrize the input
gluon and quark-singlet diffractive PDFs as follows:
\begin{align}
zf_{g}(z, {Q}_0^2) = & \alpha_gz^{{\beta}_{g}}(1 - {z})^
{{\gamma}_g}({1} + {\eta}_g\sqrt{z}), 
\label{eq:paramformq} \\
zf_{q}(z, {Q}_{0}^{2}) = & 
{\alpha}_qz^{{\beta}_q}({1} - {z})^
{{\gamma}_q}({1} + {\eta}_q\sqrt{z}).
\label{eq:parameformg} 
\end{align}
The longitudinal momentum fraction, $z$, at the lowest order of the hard 
process is equal to $\beta$ ($z = \beta$), then by including the higher 
orders we have $0<\beta<z$. 
To ensure that the diffractive PDFs vanish at $z=1$, the above equations are 
multiplied by factor $e^{\frac{-0.01}{1-z}}$, 
which is required for the DGLAP equations to be solvable~\cite{H1:2006zyl,ZEUS:2009uxs}.
The $x_{\pom}$-dependence of the diffractive PDFs is then
determined by Pomeron and Reggeon flux factors 
which are paremetrized such that, 
\begin{align}
\label{eq:flux}
f_{\pom/p,\reg/p}(x_{\pom}, {t})=A_{\pom , \reg}\frac{e^{B_{\pom , \reg}t}}
{x^{2\alpha _{\pom , \reg}(t)-1}_{\pom}}.
\end{align}
We assume  linear trajectories of the form
\begin{align}
\label{eq:traj}
\alpha _{\pom , \reg}(t)=\alpha _{\pom , \reg}(0)+\alpha ^{\prime} _{\pom , \reg}t.
\end{align}
Furtehr, $A_{\pom , \reg}$ are the normalizations of the Pomeron and Reggeon terms,
respectively, and are treated in the same way as in Ref.~\cite{H1:2006zyl}. 

After assessing the fits using Eqs.~\eqref{eq:paramformq} and \eqref{eq:parameformg}, 
we found that the parameters $\eta_q$ and $\eta_g$ can not be 
well constraint by the diffractive data, therefore, we set them to zero.
In Eqs.~\eqref{eq:flux} and \eqref{eq:traj}, we set the Reggeon flux parameters 
to the same value as in~\cite{H1:2006zyl,H1:2007oqt}. 
For the Pomeron flux parameters $\alpha^\prime_{\pom}$ and $B_{\pom}$
we use the latest value from Ref.~\cite{Aaron:2010aa} and leave  $\alpha_{\pom}(0)$ to 
be free. It needs to be determined from the QCD fit. 
Therefore, in total we have $8$ free parameters $\alpha_q$, $\beta_q$, $\gamma_q$, 
$\alpha_g$, $\beta_g$, $\gamma_g$, $n_{\reg}$ and $\alpha_{\pom}(0)$.
These will be determined from the fit to the experimental data.
For the initial inputs, we adopt the world average 
value for $\alpha_s^{n_f=5}(M_Z^2) = 0.1185$~\cite{ParticleDataGroup:2018ovx}, 
the charm and bottom masses are set to $m_c=1.40$~GeV and $m_b = 4.50$~GeV 
for both NLO and NNLO accuracy.

The heavy flavors 
will be generated through the evolution equations at ${\text{Q}}^2 > m_{c, b}^2$.
For the contribution of the heavy flavors in diffractive DIS, we employ the 
FONLL scheme implemented in the {\tt APFEL} package~\cite{Bertone:2013vaa}. 
The FONLL is a general-mass variable flavor number scheme (GM-VFNS) and 
the abbreviation stands for
``Fixed-Order plus Next-to-Leading Log orders''. 
This approach introduced first in Ref.~\cite{Cacciari:1998it} to investigate 
the production of heavy quarks in hadro-production,
then extended to DIS~\cite{Forte:2010ta} and also to Higgs boson production~\cite{Forte:2015hba}. 
This method is used to combine a fixed order calculation which corresponds to the
massive $\mathcal{O}(\alpha^3)$ cross section with a NLL resumed computation of
cross section in the massless limit. For more details, we refer the reader to
Ref.~\cite{Cacciari:1998it} and references therein.
In the {\tt SKMHS23} QCD analysis, we choose the {\tt FONLL-A} scheme at NLO accuracy, 
while for the case of NNLO, the {\tt FONLL-C} is considered. 
More details of these schemes can be found in Ref.~\cite{VBertone-thesis}.

%
%
\subsection{Minimization and diffractive PDF uncertainty method}\label{subsec:uncertainies}

As already discussed, the {\tt SKMHS23} QCD analysis of the
diffractive PDFs is presented at NLO and NNLO in perturbative QCD 
with as much data as possible. 
As a phenomenological study of diffractive PDFs, the  {\tt SKMHS23} analysis should 
answer three questions adequately:
1) how to adjust the free fit parameters of the model  
2) how to predict observables precisely, and  3) how precise are our distributions and 
observable predictions. 
As mentioned, a QCD analysis should have a sound approach to 
finding the best fit parameters and evaluate their uncertainty. 
In order to find the best values of the free parameters, a $\chi^2$ function defined as follows, 
is minimized~\cite{H1:2014cbm};
\begin{eqnarray}
\chi^2 = \vec{p}^\mathrm{T}\mathbf{C}^{-1}\vec{p}+\sum_{k}^{N_{\mathrm{sys}}} 
\varepsilon_k^2,
\end{eqnarray}	
where $\mathbf{C}$ denotes the covariance matrix of the relative uncertainties, 
and $i$th element of $\vec{p}$ is defined as the logarithm of the ratio 
of a measured observable to its theoretical prediction,
\begin{eqnarray}
p_i = \log\left[\frac{\mathcal{E}_i}{\mathcal{T}_i}\right]-\sum_{k}^{N_{\mathrm{sys}}}E_{i,k},
\end{eqnarray}
which means that the experimental data are distributed 
according to the log normal distribution and where $E_{i,k}$ is defined as,
\begin{eqnarray}
E_{i,k} = \sqrt{f_k^C}\left(\frac{\delta_{\mathcal{E}_i}^{k,+}-\delta_{\mathcal{E}_i}^{k,-}}{2}\varepsilon_k
+\frac{\delta_{\mathcal{E}_i}^{k,+}+\delta_{\mathcal{E}_i}^{k,-}}{2}\varepsilon_k^2\right).
\end{eqnarray}
The parameter $f_k^C$  denotes the 
fraction of the systematic errors from the source $k$ which are considered as 
correlated uncertainty and the parameters $\delta_{\mathcal{E}_i}^{k,-}$ 
and $\delta_{\mathcal{E}_i}^{k,+}$ are the relative uncertainty of the $\mathcal{E}_i$ measurement. 
The nuisance parameters $\varepsilon_k$ will be treated as free parameters, and will be determined by the
$\chi^2$ minimization.

Concerning the question of calculating the observables and evolution of the diffractive PDFs, 
we have used the \texttt{Alpos} package~\cite{Alpos1,Alpos2}, 
which also provides an interface to the CERN \texttt{MINUIT} package~\cite{James:1975dr} 
that is responsible for the $\chi^2$ minimization. 
For the uncertainty of diffractive PDF distributions and theoretical predictions, 
we have used the well stablished optimized Hessian method as 
described in \cite{Nadolsky:2008zw} and implemented in the \texttt{Alpos} package. 
In the next section we will discuss  in detail the $\chi^2$ values extracted from {\tt SKMHS23}
QCD fits,  and the resulting diffractive PDFs in terms of their  predictions and  uncertainties.

%
\section{{\tt SKMHS23} fit results}\label{sec:results}

This section focuses on the main results of the {\tt SKMHS23} QCD analysis and on the new
features and improvements that are introduced in this work.
We first present the {\tt SKMHS23}  diffractive PDFs and the fitted parameters.
Then, we focus on the improvements arising from the inclusion of the higher order QCD correction.
We also stress and discuss the effect and impact of the diffractive dijet production data on 
the extracted diffractive PDFs.
Finally, we present and discuss the quality of the {\tt SKMHS23} QCD fit in terms of both
individual and total data sets. Data-theory comparisons will be presented as well.

In Tab.~\ref{tab:NLO-NLO} we present the {\tt SKMHS23}  best fit parameters 
and their errors extracted from the QCD analysis 
at NLO and NNLO accuracy using the inclusive diffractive DIS data. 
The  best fit parameters extracted from the global QCD analysis 
at NLO and NNLO accuracy using both inclusive diffractive DIS 
and diffractive dijet data sets, entitled {\tt SKMHS23-dijet}, are presented in
Tab.~\ref{tab:NLO-NLO-dijet}. 

%
%
\begin{table*}[ht]
	\begin{center}
		\caption{\small The {\tt SKMHS23}  best fit parameters and their 
			errors extracted from the QCD analysis 
			at NLO and NNLO accuracy using the inclusive diffractive DIS data. 
			Values marked with (*) are fixed in the QCD fit since the analyzed data sets 
			do not constrain these parameters well enough. The input values for
                        $\alpha_s$, $m_c$ and $m_b$ are also given.}
		\begin{tabular}{ c | c | c    }
			\hline \hline
			Parameters	    & {\tt SKMHS23 (NLO)}    &    {\tt SKMHS23 (NNLO)}  \\  \hline \hline
			$\alpha_g$     & $0.355 \pm 0.084$  & $0.497 \pm 0.108$  \\ 
			$\beta_g$       & $0.201 \pm 0.101$  & $0.291 \pm 0.087$ \\ 
			$\gamma_g$       & $0.018 \pm 0.206$   & $0.353 \pm 0.233$\\ 
			$\eta_g$         & $0.0^*$            & $0.0^*$\\   
			$\alpha_q$     & $0.728 \pm 0.055$  & $0.979 \pm 0.091$ \\ 
			$\beta_q$       & $1.525 \pm 0.071$  & $1.705 \pm 0.096$\\ 
			$\gamma_q$      & $0.437 \pm 0.036$  & $0.558 \pm 0.044$\\ 
			$\eta_q$            & $0.0^*$            & $0.0^*$ \\     
			$\alpha_{\pom}(0)$    & $1.099 \pm 0.0039$ & $1.01 \pm 0.0040$  \\
			$n_{\reg}$     & $0.00055 \pm 0.000004$   & $0.00055 \pm 0.000004$  \\ \hline 
			$\alpha_s(M_Z^2)$             & $0.1185^*$          &  $0.1185^*$  \\
			$m_c$ [GeV]                  & $1.40^*$            &  $1.40^*$  \\
			$m_b$ [GeV]                   & $4.5^*$            &  $4.5^*$  \\ 	\hline \hline
		\end{tabular}
		\label{tab:NLO-NLO}
	\end{center}
\end{table*}
%
%
%
%
\begin{table*}[ht]
	\begin{center}
		\caption{\small The {\tt SKMHS23-dijet} best fit parameters and their 
			errors extracted from global the  QCD analysis 
			at NLO and NNLO accuracy using both inclusive  diffractive DIS 
			and diffractive dijet data sets. 
			Values marked with (*) are fixed in the QCD fit since the analyzed datasets 
			do not constrain these parameters well enough. The input values for
                        $\alpha_s$, $m_c$ and $m_b$ are also given.}
		\begin{tabular}{ c | c | c    }
			\hline \hline
			Parameters	    & {\tt SKMHS23-dijet (NLO)}    &    {\tt SKMHS23-dijet (NNLO)}  \\  \hline \hline
			$\alpha_g$     & $0.323 \pm 0.069$  & $0.477 \pm 0.094$  \\ 
			$\beta_g$       & $0.169 \pm 0.094$  & $0.278 \pm 0.083$ \\ 
			$\gamma_g$       & $-0.099 \pm 0.163$   & $0.303 \pm 0.189$\\ 
			$\eta_g$         & $0.0^*$            & $0.0^*$\\   
			$\alpha_q$     & $0.747 \pm 0.055$  & $0.986 \pm 0.085$ \\ 
			$\beta_q$       & $1.560 \pm 0.068$  & $1.719 \pm 0.085$\\ 
			$\gamma_q$      & $0.442 \pm 0.036$  & $0.560 \pm 0.043$\\ 
			$\eta_q$            & $0.0^*$            & $0.0^*$ \\     
			$\alpha_{\pom}(0)$    & $1.100 \pm 0.0029$ & $1.101 \pm 0.0038$  \\
			$n_{\reg}$     & $0.00075 \pm 0.000004$   & $0.00073 \pm 0.000004$  \\ \hline 
			$\alpha_s(M_Z^2)$             & $0.1185^*$          &  $0.1185^*$  \\
			$m_c$ [GeV]                   & $1.40^*$            &  $1.40^*$  \\
			$m_b$ [GeV]                   & $4.5^*$            &  $4.5^*$  \\ 	\hline \hline
		\end{tabular}
		\label{tab:NLO-NLO-dijet}
	\end{center}
\end{table*}
%
%

In total we have 10 parameters that need to be extracted from the QCD fit, 
which include 4 for both the 
gluon and total singlet densities and two for the Reggeon flux.
For both gluon and singlet PDFs, the parameter $\eta_g$ and $\eta_g$ are set to zero 
during the QCD fits since the analyzed data sets 
do not constrain these parameters well enough.
As one can see from  Tab.~\ref{tab:NLO-NLO} and Tab.~\ref{tab:NLO-NLO-dijet}, 
all the shape parameters
are will determined, except for the case of 
$\gamma_g$  which comes with large errors. 
This again reflects the lack of data to constrain all shape parameters. 
We prefer to keep $\gamma_g$ free in the fit to give the gluon density enough flexibility. 
The extracted value for the $n_{\reg}$ is rather small as expected~\cite{Goharipour:2018yov}.
The consistency of the parameters extracted from different QCD fits presented in 
Tables~\ref{tab:NLO-NLO} and \ref{tab:NLO-NLO-dijet}
are acceptable, however, the $\gamma_g$ is mostly affected by 
the higher order QCD corrections and the dijet data as well.

Now we are in a position to present and discuss the  {\tt SKMHS23} and  {\tt SKMHS23-dijet}  
diffractive PDFs and their uncertainties, focusing on the perturbative
convergence upon inclusion of the higher-order QCD corrections and the effect arising 
form the inclusion of the diffractive dijet production to the data sample. 

\begin{figure*}[htb]
	\vspace{0.50cm}
	\centering
	\subfloat{\includegraphics[width=0.33\textwidth]{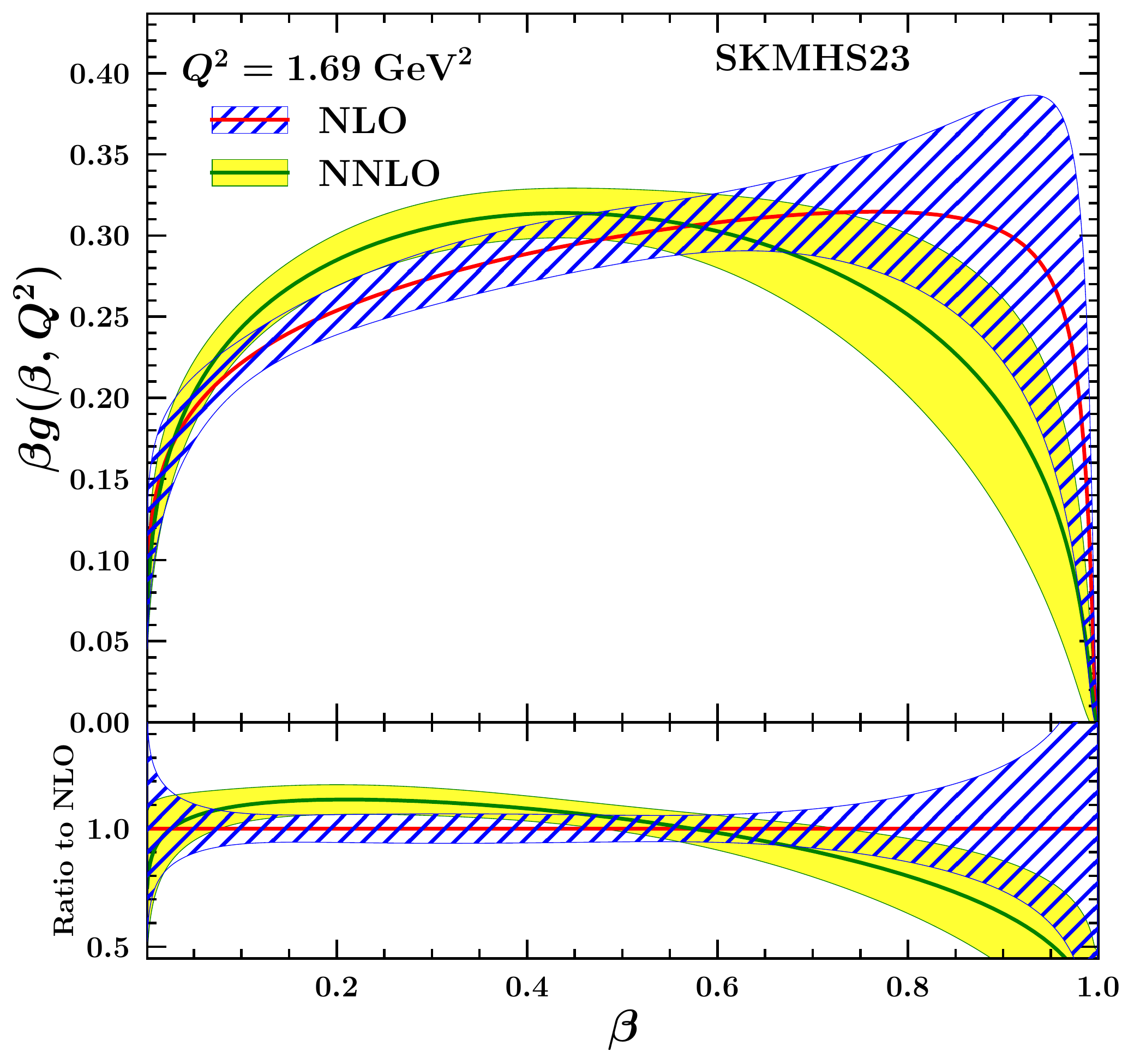}} 	
	\subfloat{\includegraphics[width=0.33\textwidth,height=0.235\textheight]{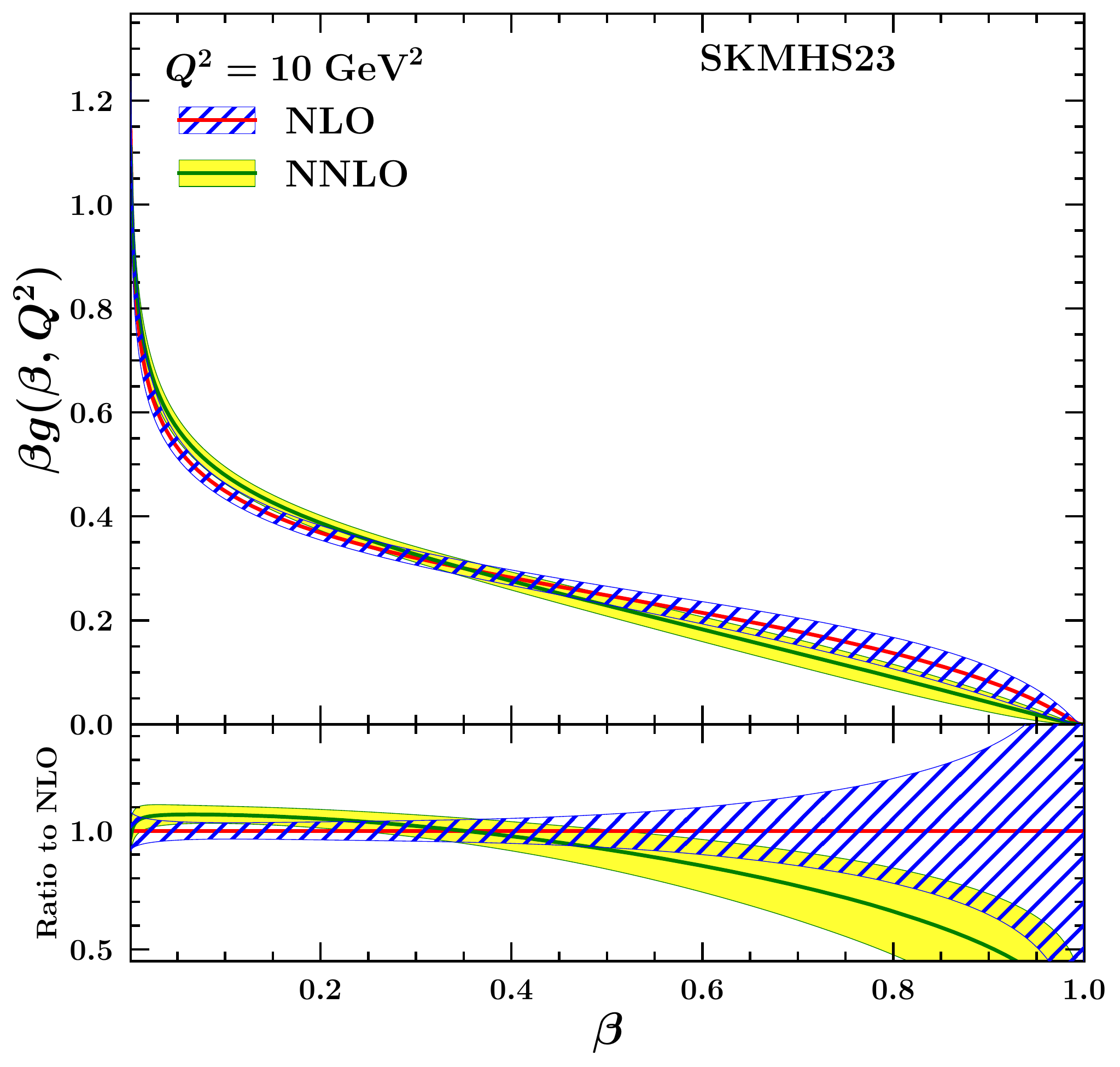}}
	\subfloat{\includegraphics[width=0.33\textwidth]{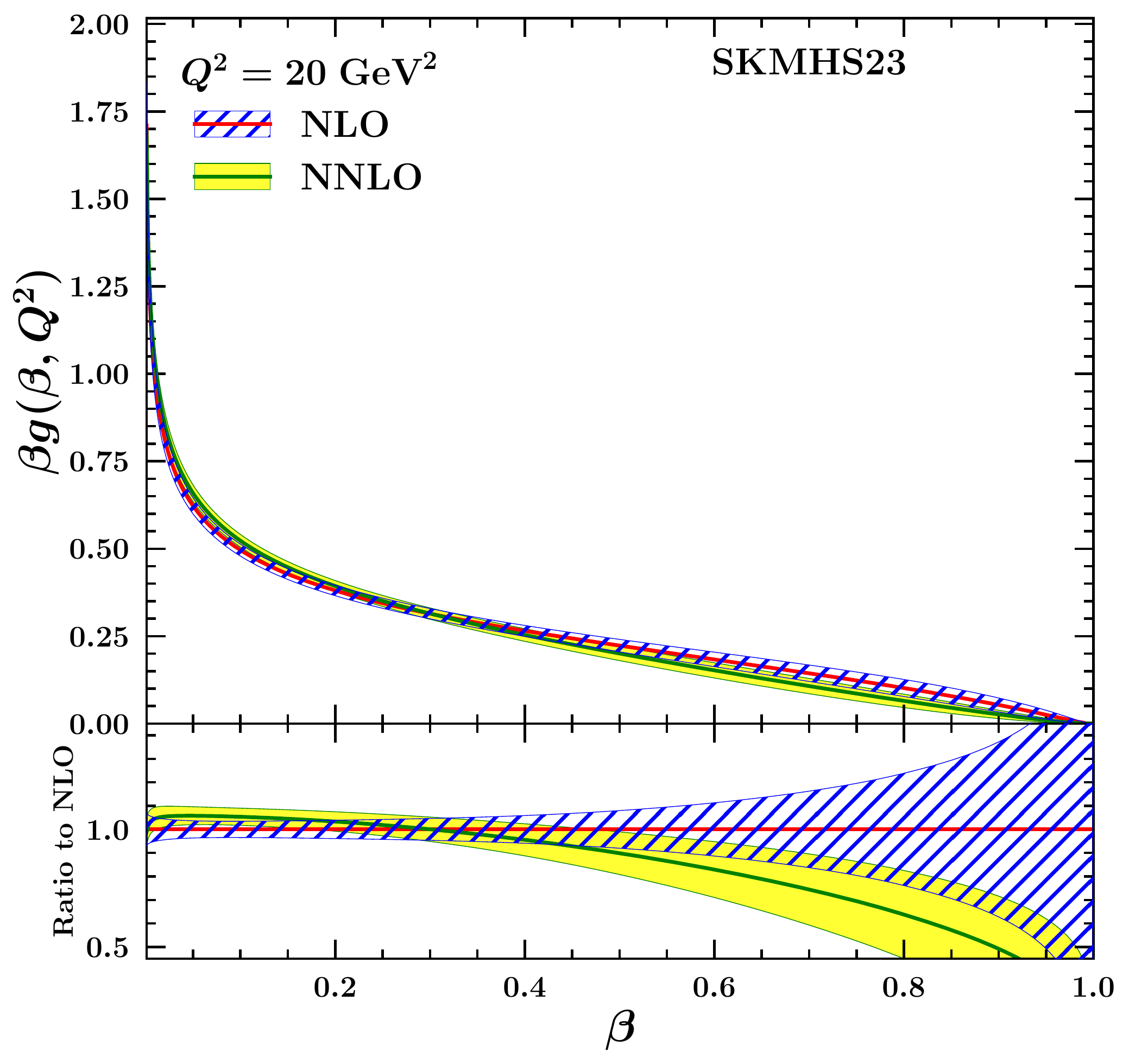}}\\	
	\subfloat{\includegraphics[width=0.33\textwidth,height=0.235\textheight]{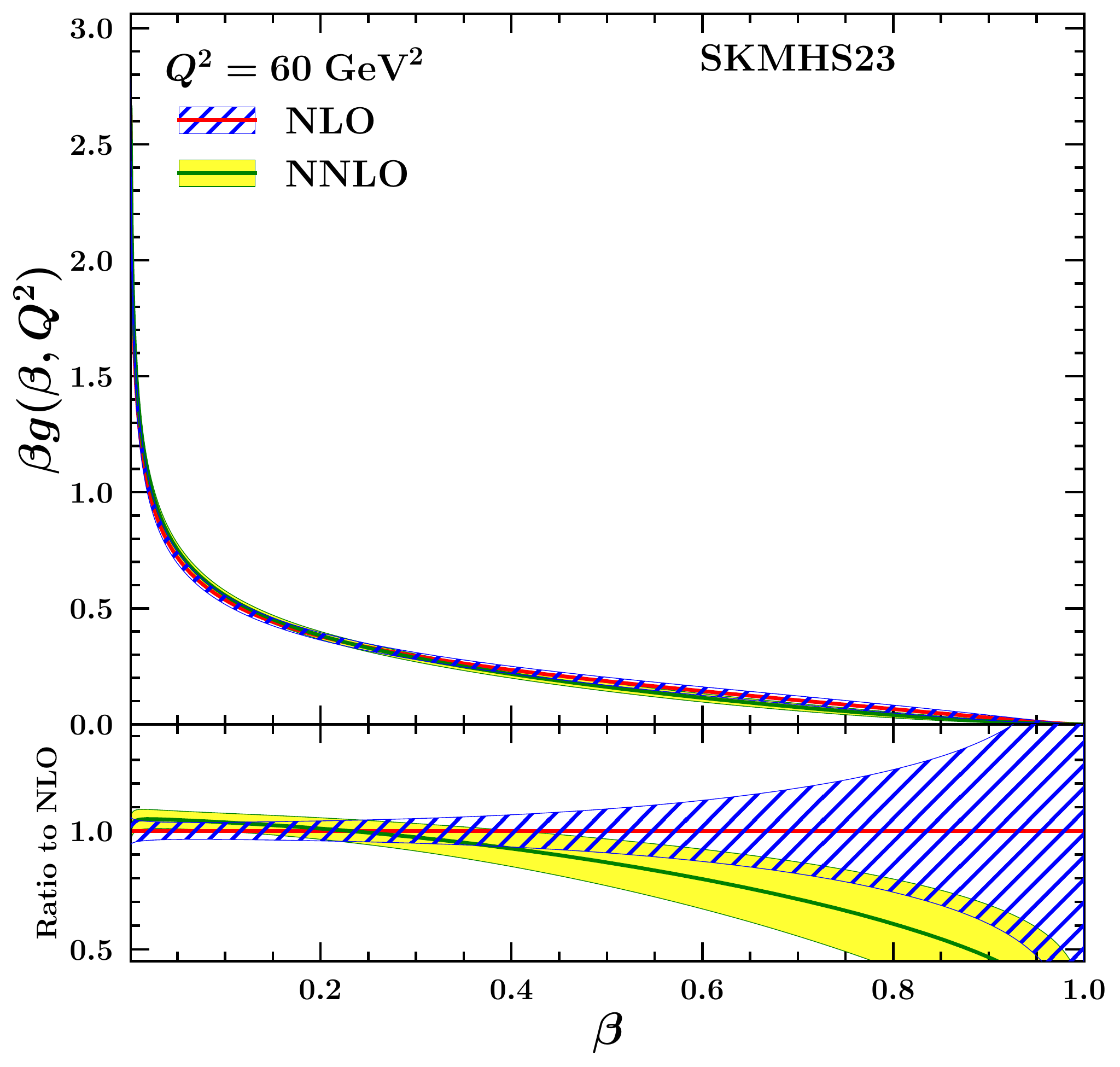}} 
	\subfloat{\includegraphics[width=0.33\textwidth]{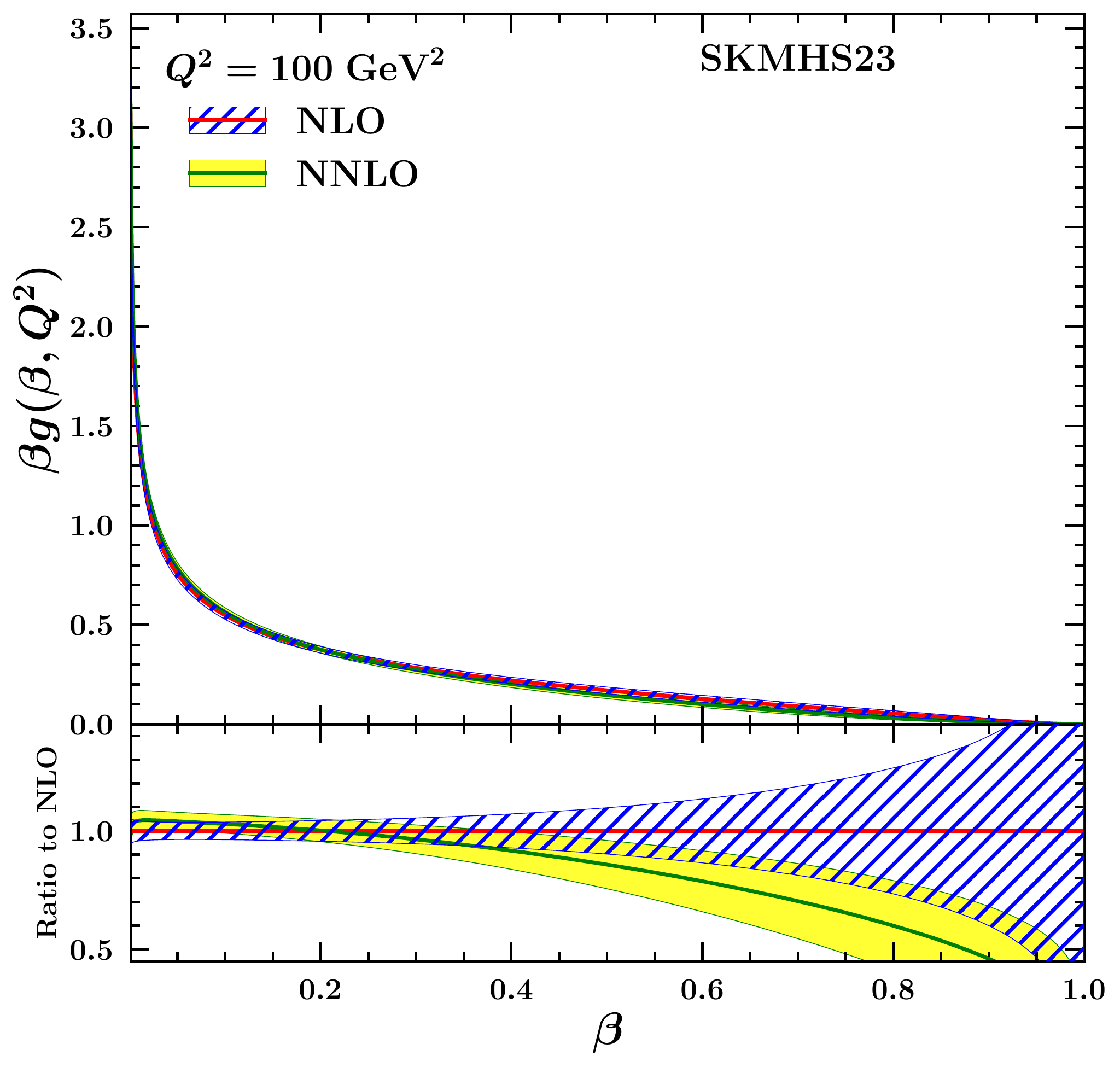}}	
	\subfloat{\includegraphics[width=0.33\textwidth]{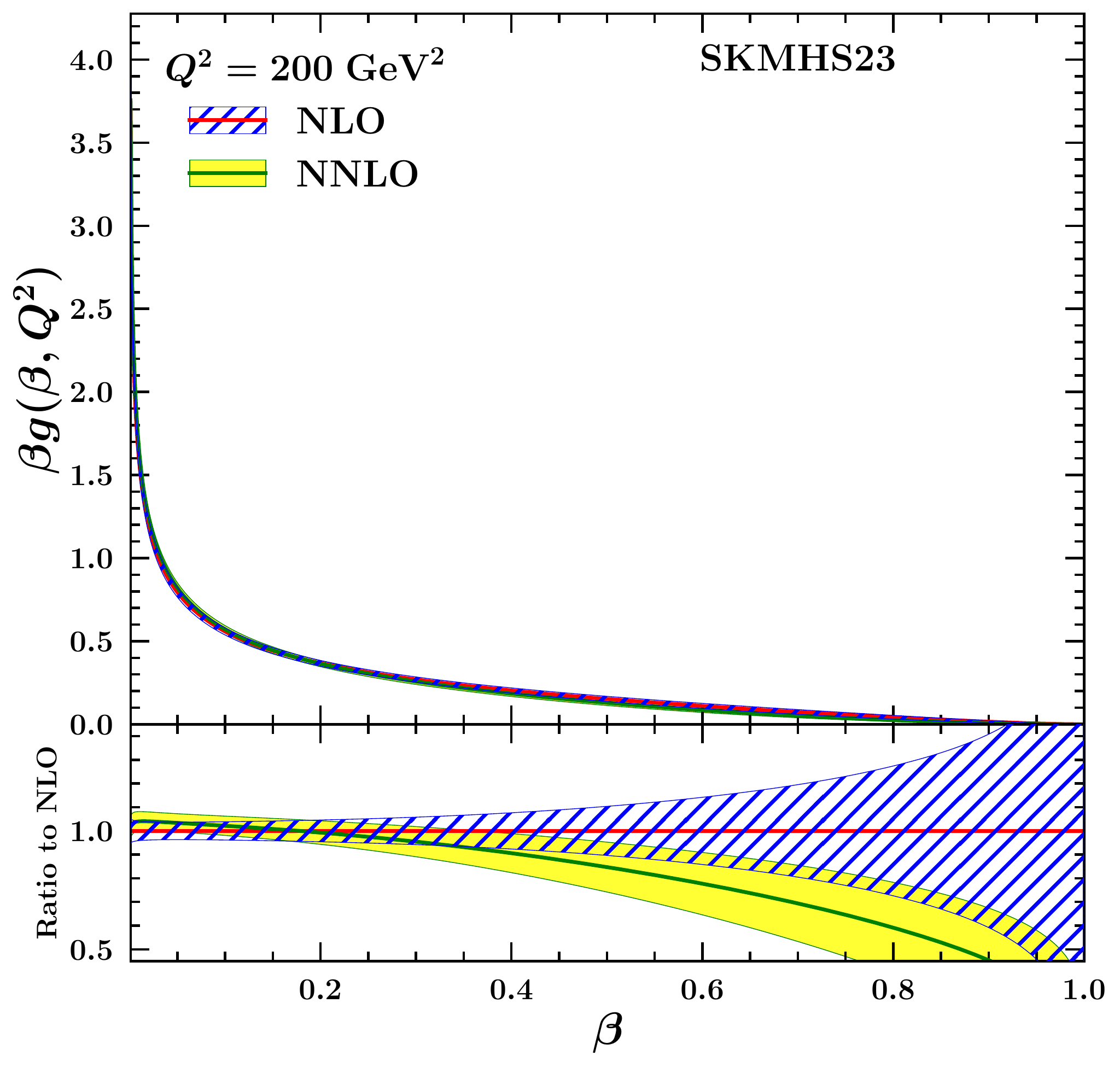}}	
	\begin{center}
		\caption{ \small 
The {\tt SKMHS23} gluon distribution at the input scale $Q_0^2 = 1.69$ GeV$^2$, and at higher 
energy value of 10, 20, 60, 100 and 200~GeV$^2$.
The extracted uncertainties determined using the Hessian method also are shown as well. 
We show both the absolute distributions and ratios to the NLO results.  }
		\label{fig:DPDF-g-Q0_withoutdijet}
	\end{center}
\end{figure*}

\begin{figure*}[htb]
	\vspace{0.50cm}
	\centering
	\subfloat{\includegraphics[width=0.33\textwidth]{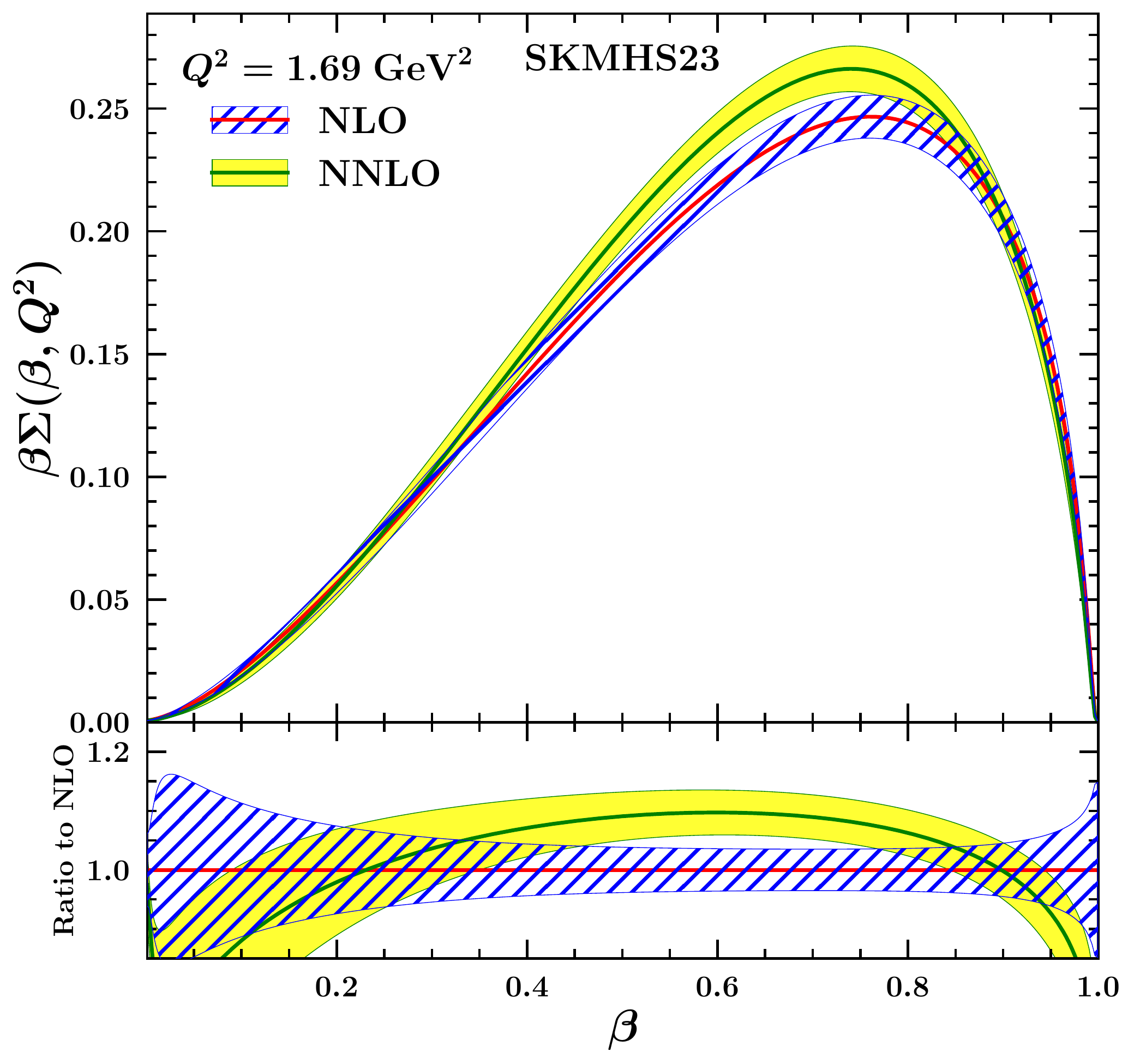}} 	
	\subfloat{\includegraphics[width=0.33\textwidth,height=0.235\textheight]{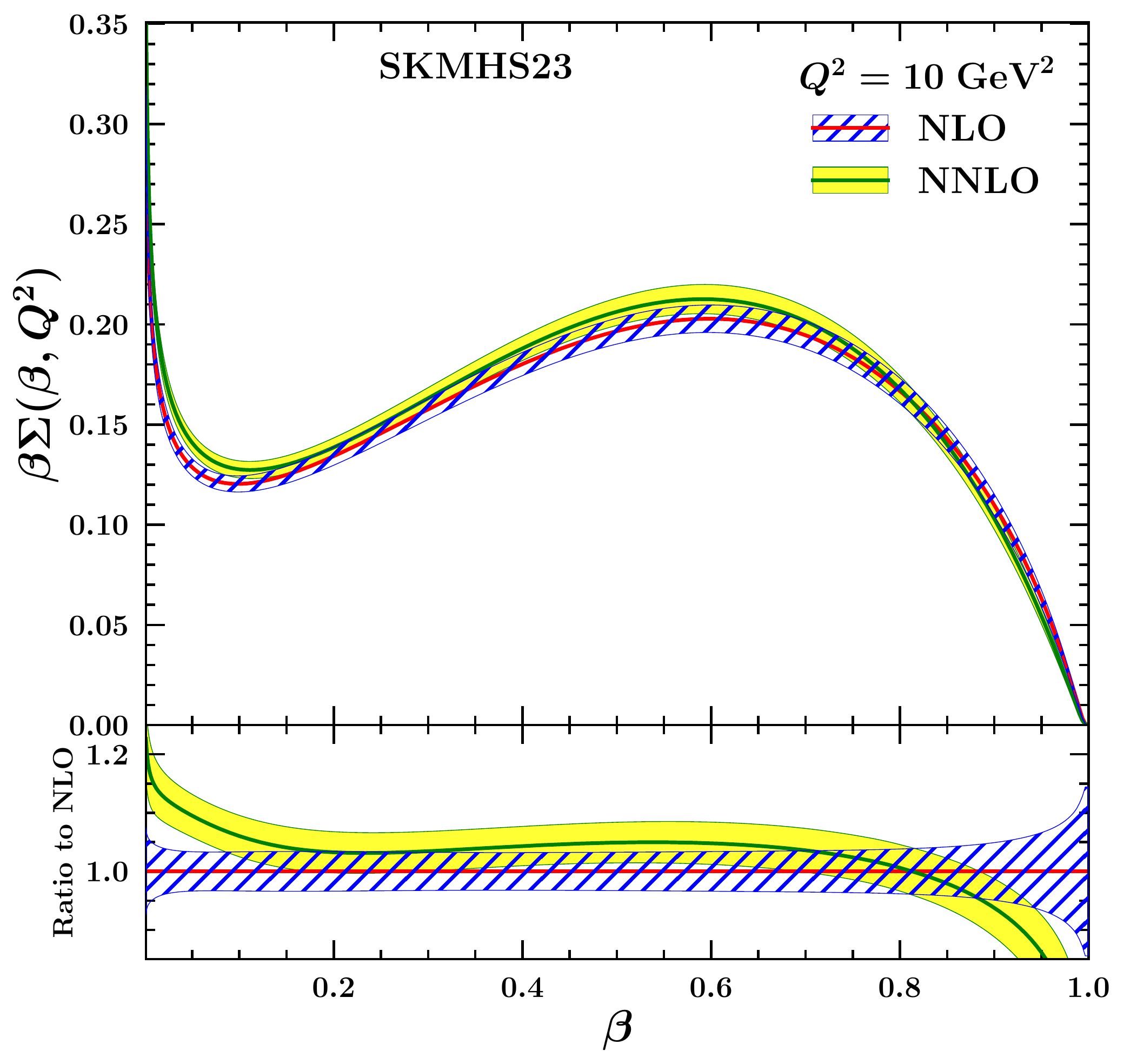}}
	\subfloat{\includegraphics[width=0.33\textwidth]{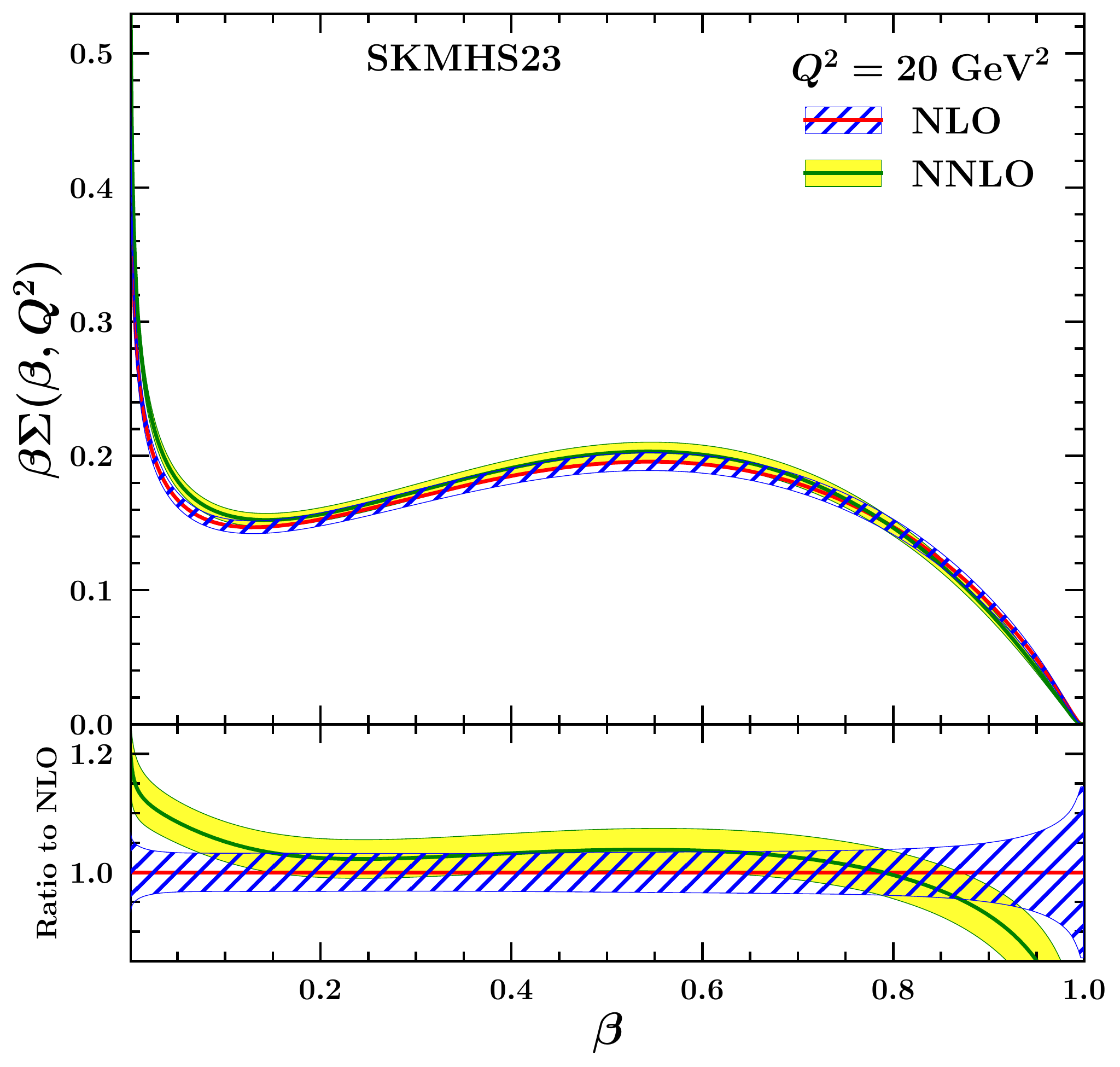}}\\	
	\subfloat{\includegraphics[width=0.33\textwidth,height=0.235\textheight]{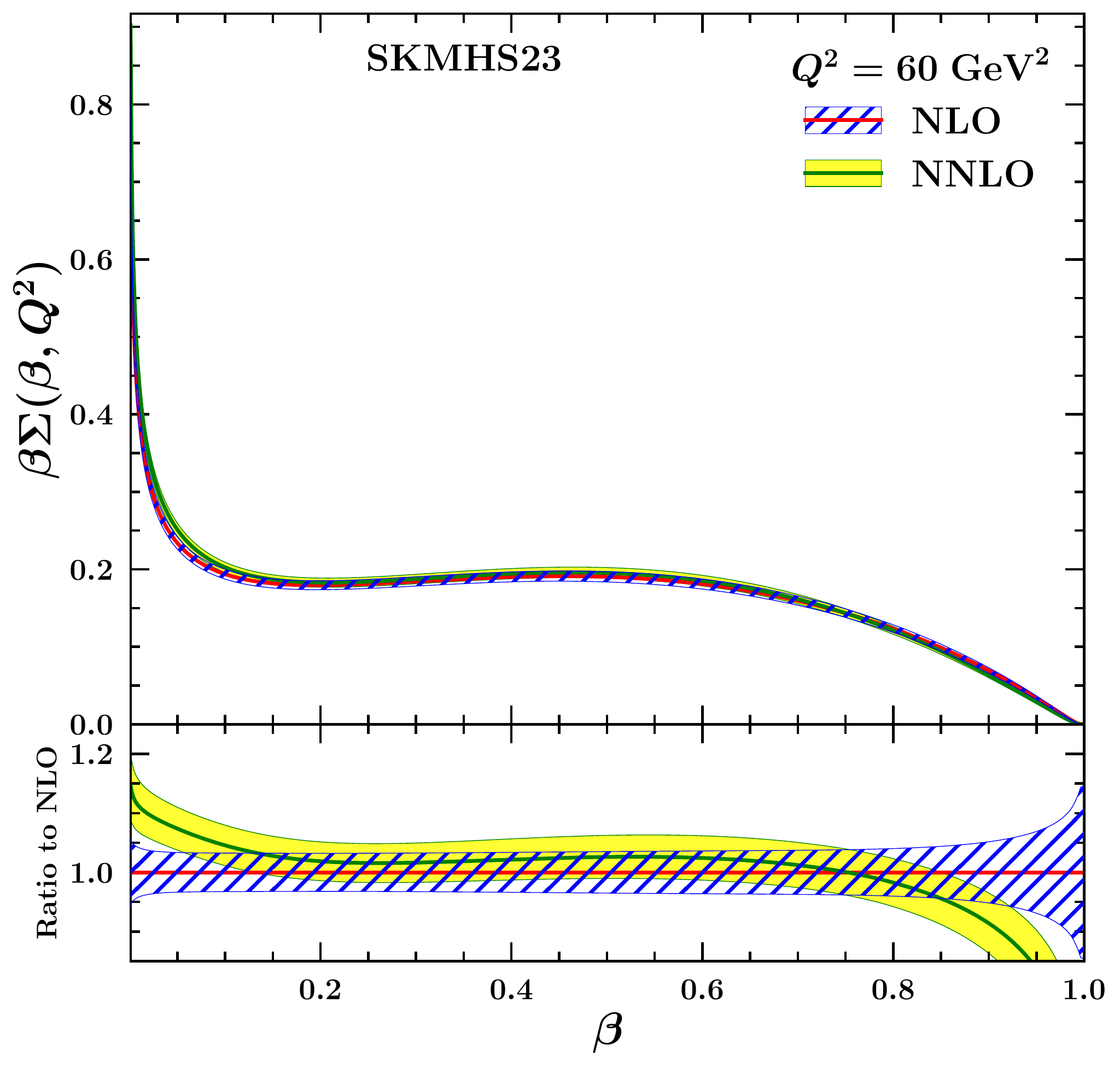}} 
	\subfloat{\includegraphics[width=0.33\textwidth]{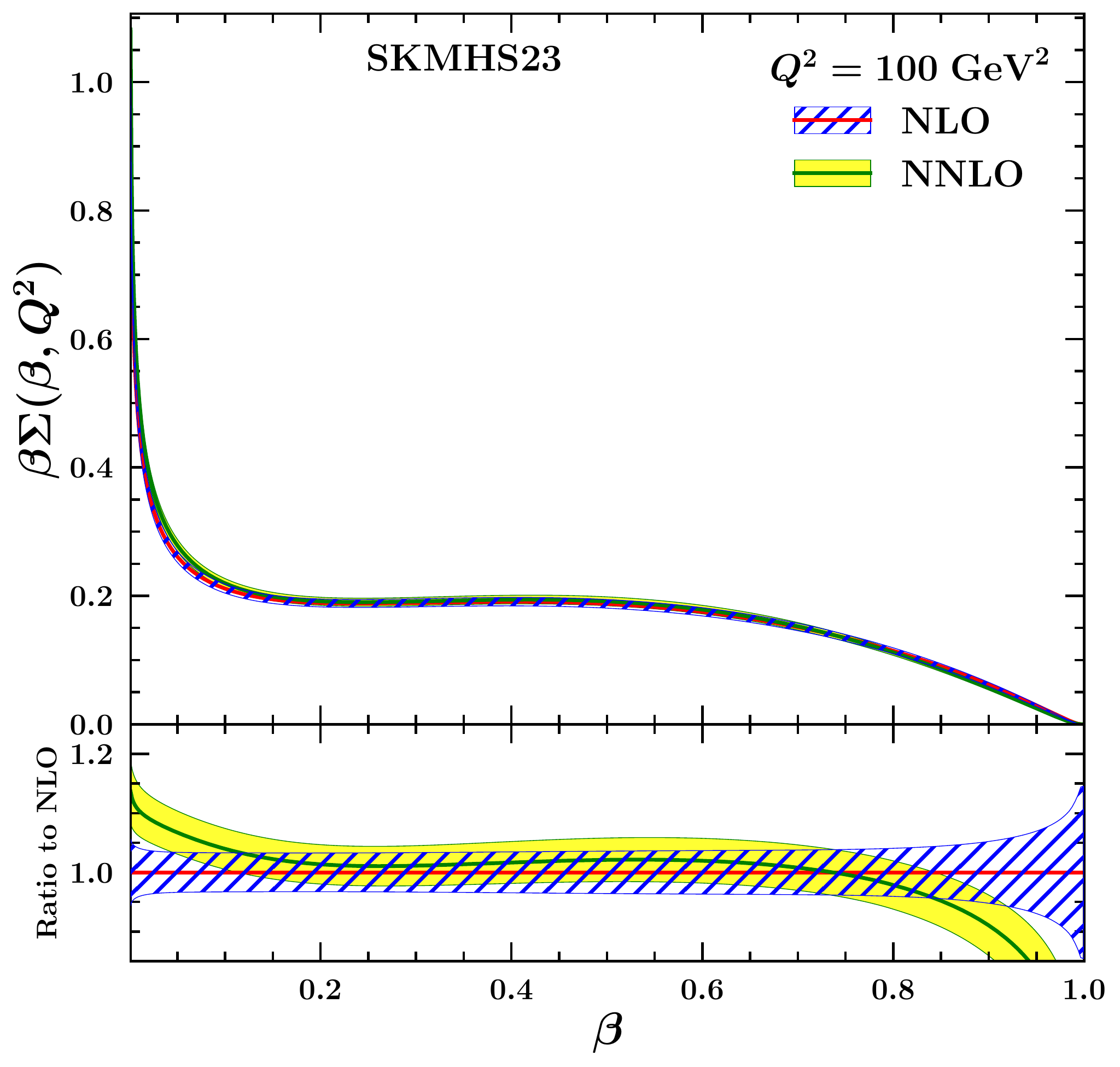}}	
	\subfloat{\includegraphics[width=0.33\textwidth]{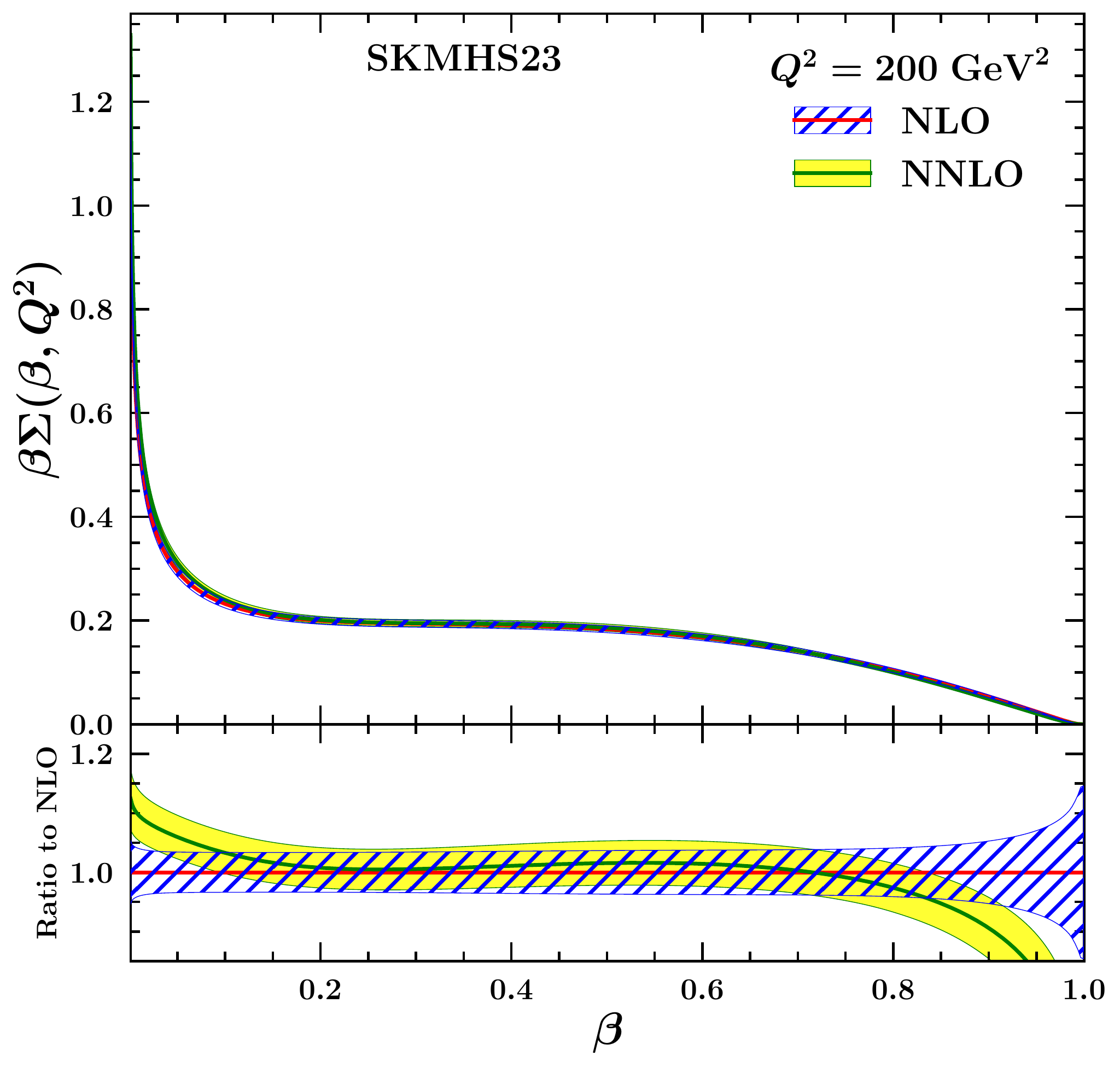}}	
	\begin{center}
		\caption{ \small 
Same as Fig.~\ref{fig:DPDF-g-Q0_withoutdijet} but this time for the {\tt SKMHS23} singlet 
distribution with their including uncertainties.}
		\label{fig:DPDF-S-Q0_withoutdijet}
	\end{center}
\end{figure*}

In Fig.~\ref{fig:DPDF-g-Q0_withoutdijet}, we present the NLO and NNLO {\tt SKMHS23} 
gluon distributions at the input scale $Q_0^2 = 1.69$~GeV$^2$.
The results for the higher energy value of 10, 20, 60, 100 and 200~GeV$^2$ are also shown as well. 
The extracted uncertainties determined using the Hessian method also are shown. 
We show both the absolute distributions and ratios to the NLO results.
The NLO and NNLO {\tt SKMHS23} singlet distribution with their including uncertainties 
are shown in Fig.~\ref{fig:DPDF-S-Q0_withoutdijet}.

Considering the results presented in Fig.~\ref{fig:DPDF-g-Q0_withoutdijet} and \ref{fig:DPDF-S-Q0_withoutdijet}, 
a few remarks are in order. 
A remarkable feature of the distributions shown in these plots is their perturbative convergence. 
As one can see, a difference can be seen between the NLO and NNLO results for both the gluon 
and the singlet densities for  medium to large value of $\beta$.
For the gluon density, the NLO results are larger than the NNLO ones for a high value of
$\beta$ and smaller for the small region of $\beta$.
As can bee seen, a significant reduction for the uncertainty bands are 
achieved after including the higher order QCD corrections
showing the effect of the inclusion the NNLO accuracy in the diffractive PDFs determination.
The differences between the NLO and NNLO diffractive  PDFs
are rather small when going to the higher values of Q$^2$.

\begin{figure*}[htb]
\vspace{0.50cm}
	\centering
	\subfloat{\includegraphics[width=0.33\textwidth]{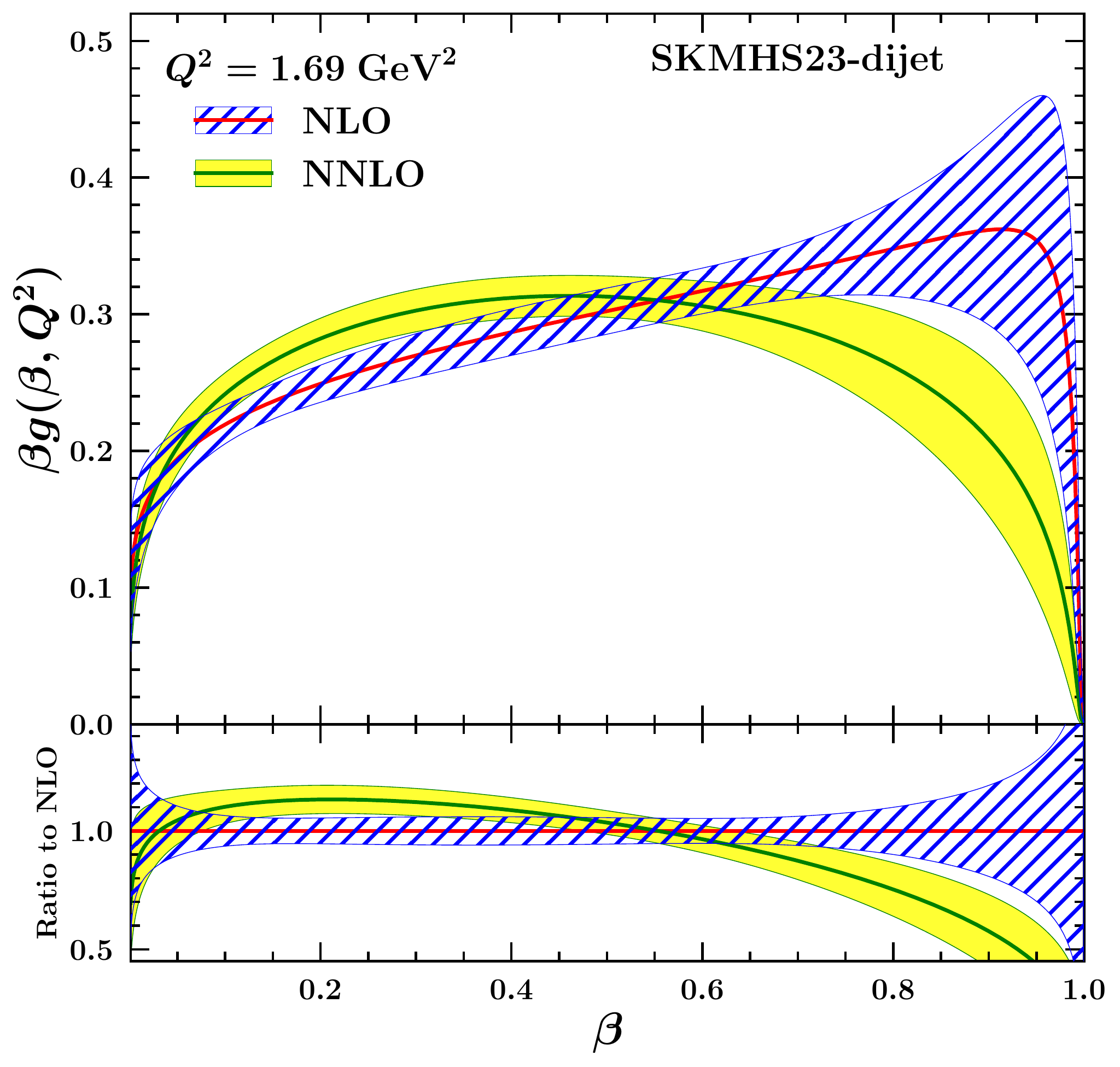}} 	
	\subfloat{\includegraphics[width=0.33\textwidth,height=0.235\textheight]{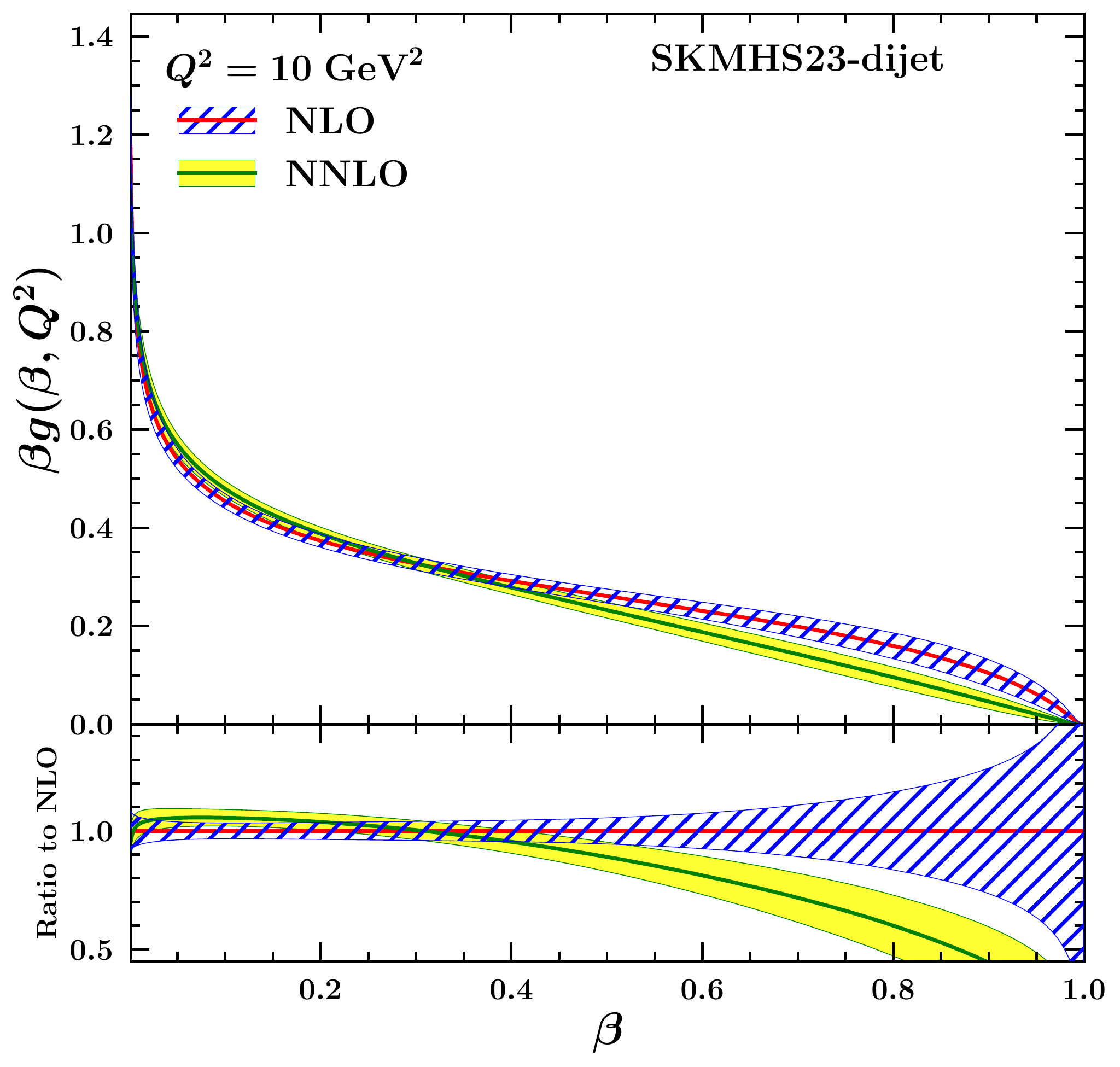}}
	\subfloat{\includegraphics[width=0.33\textwidth]{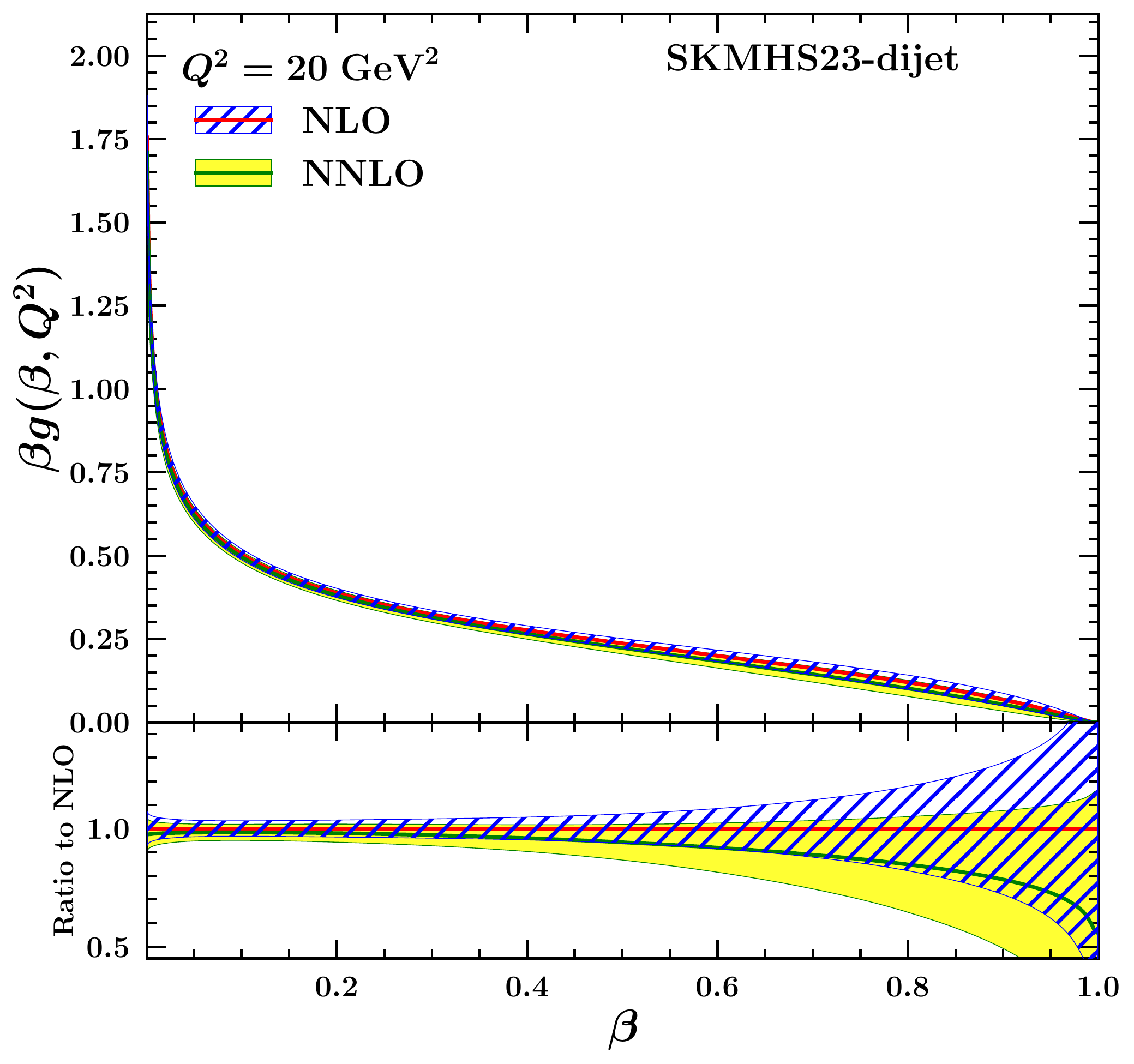}}\\	
	\subfloat{\includegraphics[width=0.33\textwidth,height=0.235\textheight]{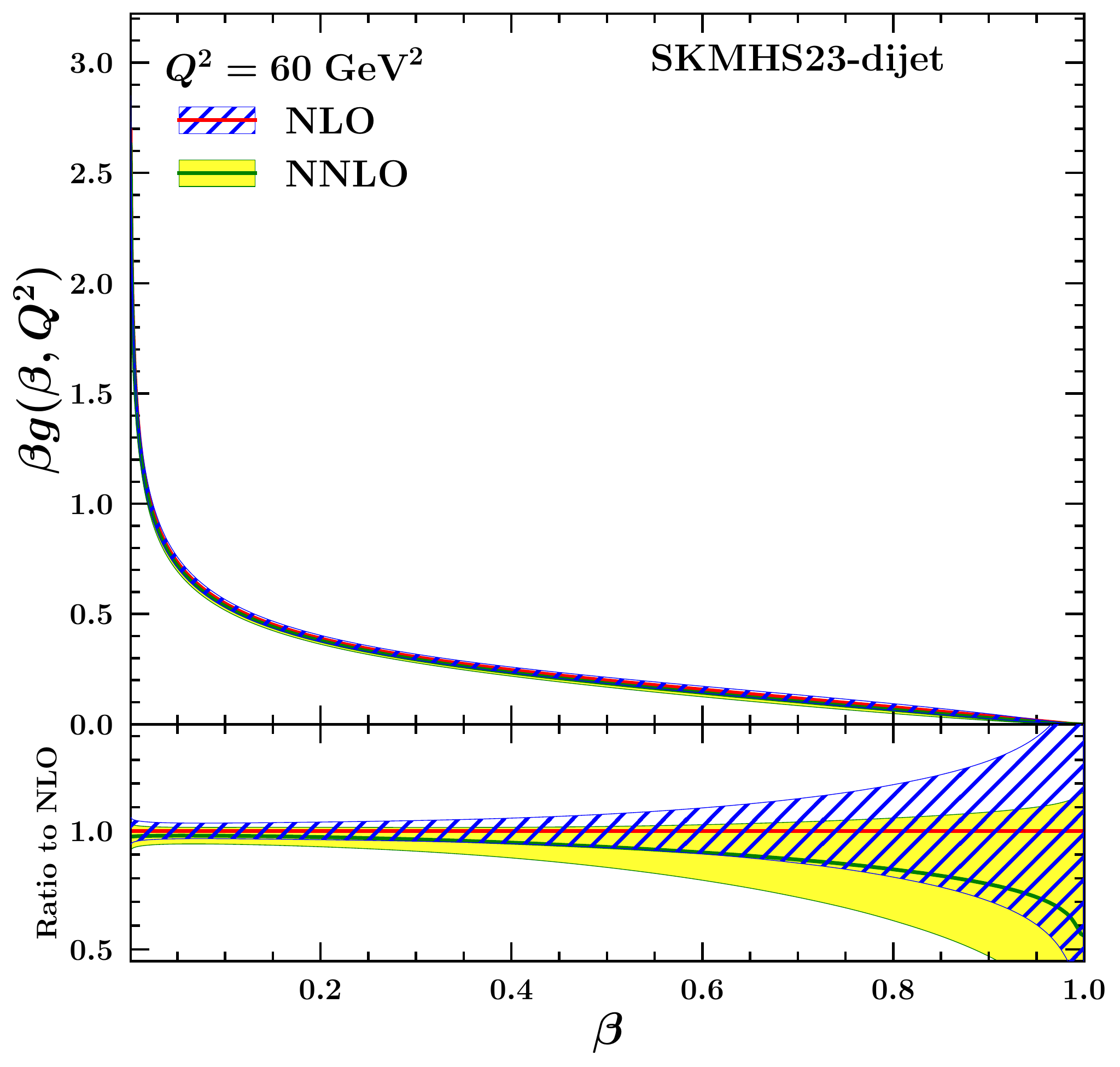}} 
	\subfloat{\includegraphics[width=0.33\textwidth]{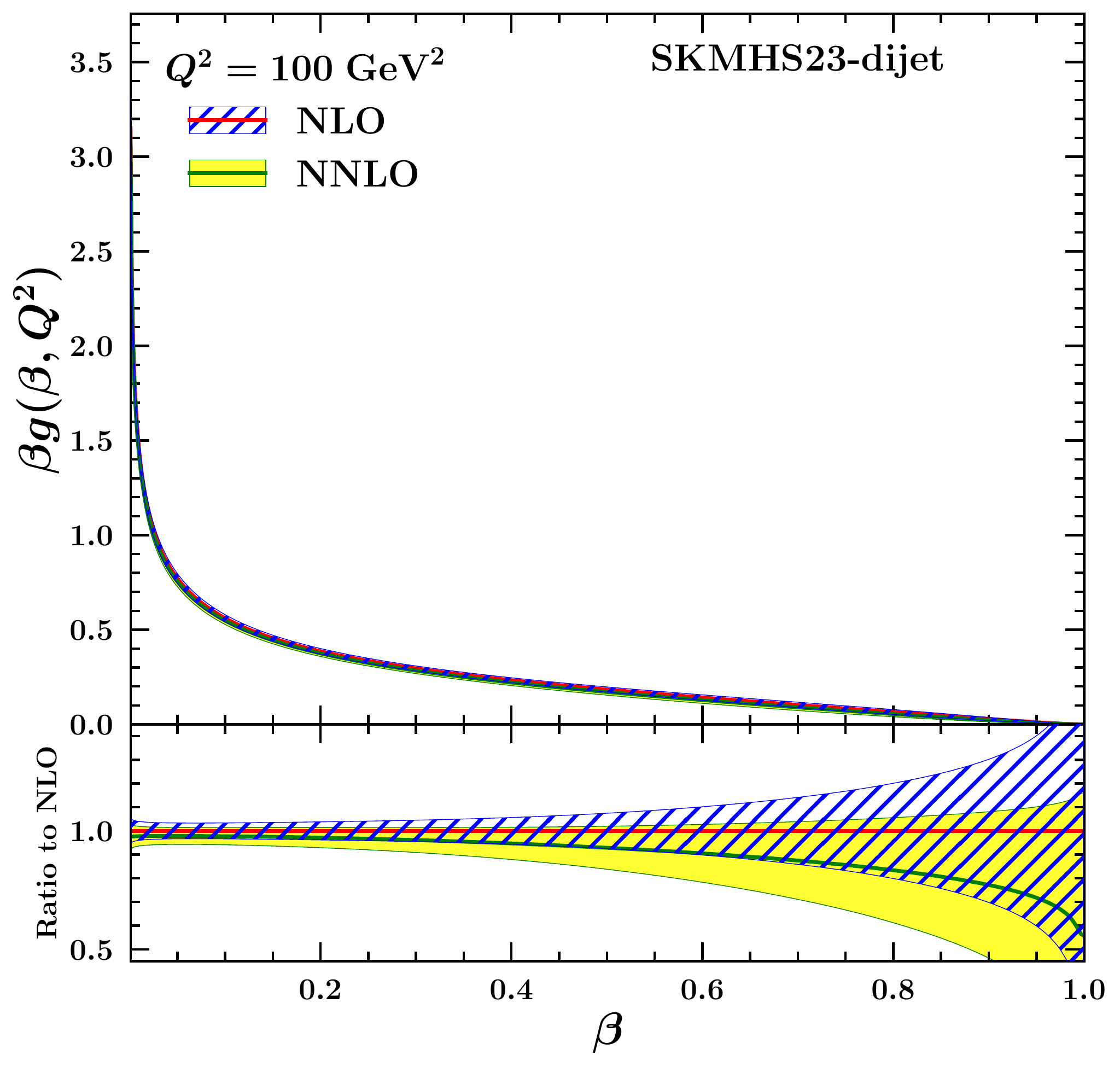}}	
	\subfloat{\includegraphics[width=0.33\textwidth]{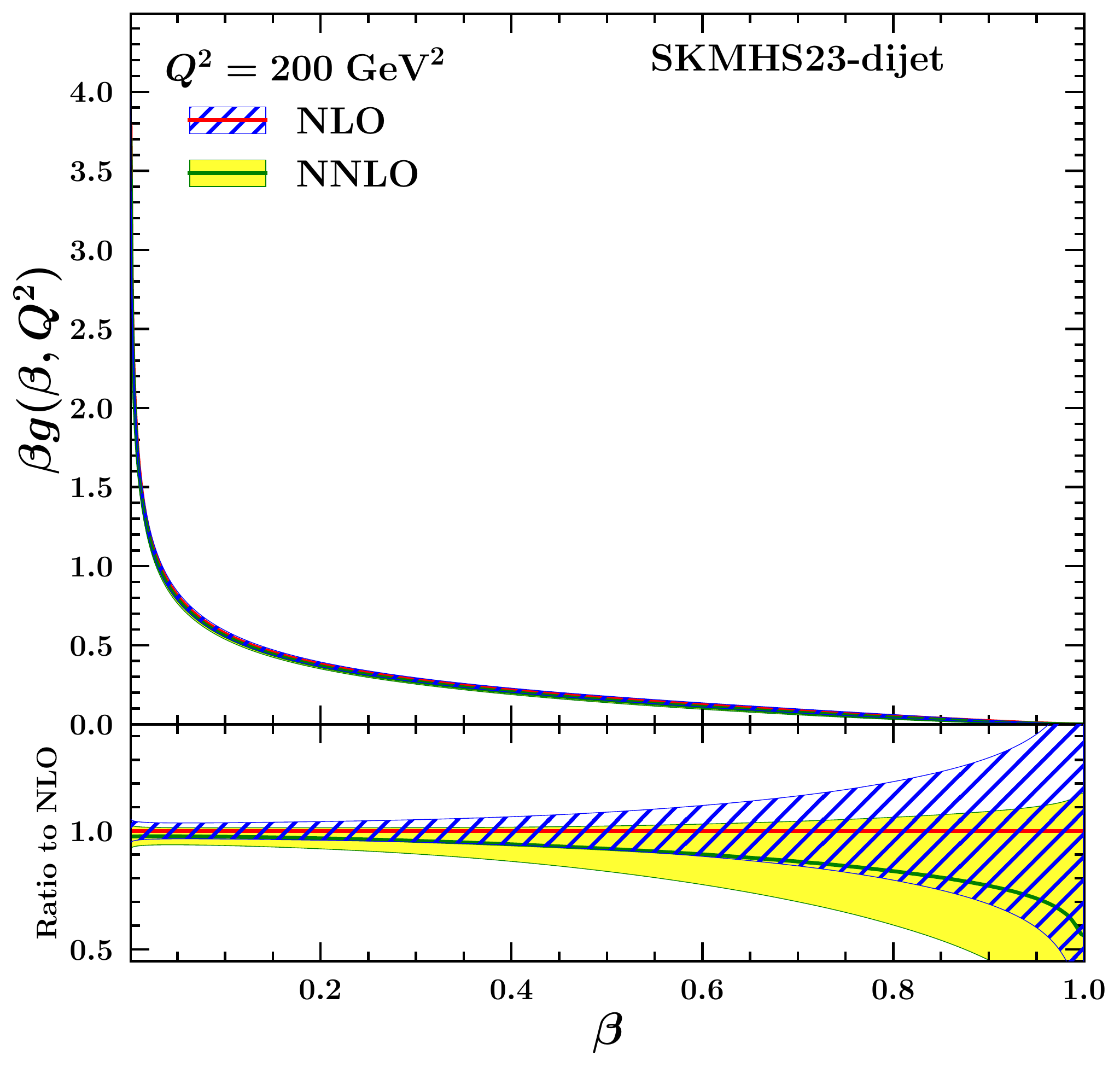}}	
\begin{center}
\caption{ \small 
The {\tt SKMHS23-dijet}  gluon distribution at the input scale $Q_0^2 = 1.69$ GeV$^2$, 
and at higher energy value of 10, 20, 60, 100 and 200~GeV$^2$.
The extracted uncertainties determined using the Hessian method also are shown as well. 
We show both the absolute distributions and ratios to the NLO results.}
\label{fig:DPDF-g-Q0_withdijet}
\end{center}
\end{figure*}

\begin{figure*}[htb]
	\vspace{0.50cm}
	\centering
	\subfloat{\includegraphics[width=0.33\textwidth]{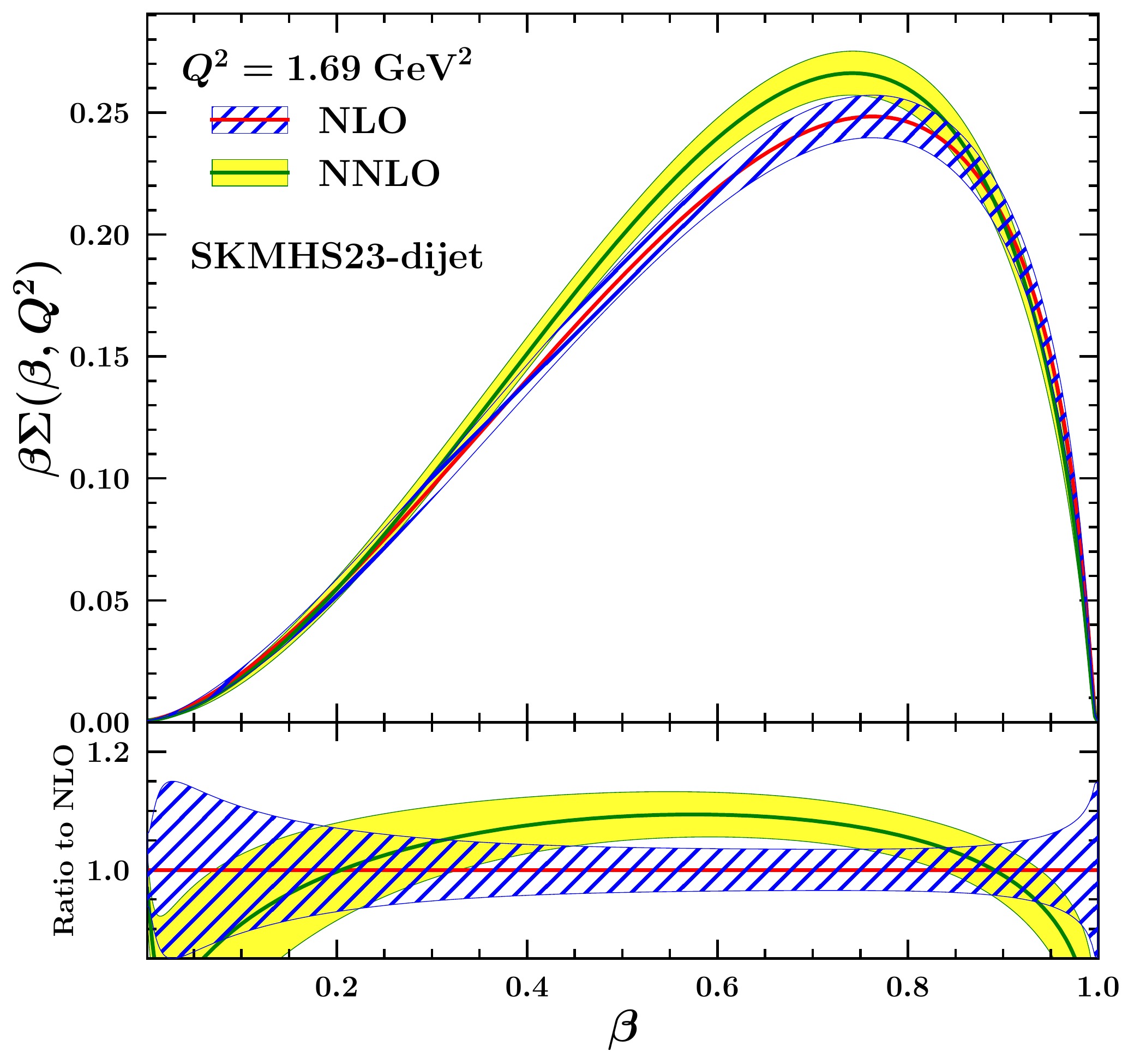}} 	
	\subfloat{\includegraphics[width=0.33\textwidth,height=0.235\textheight]{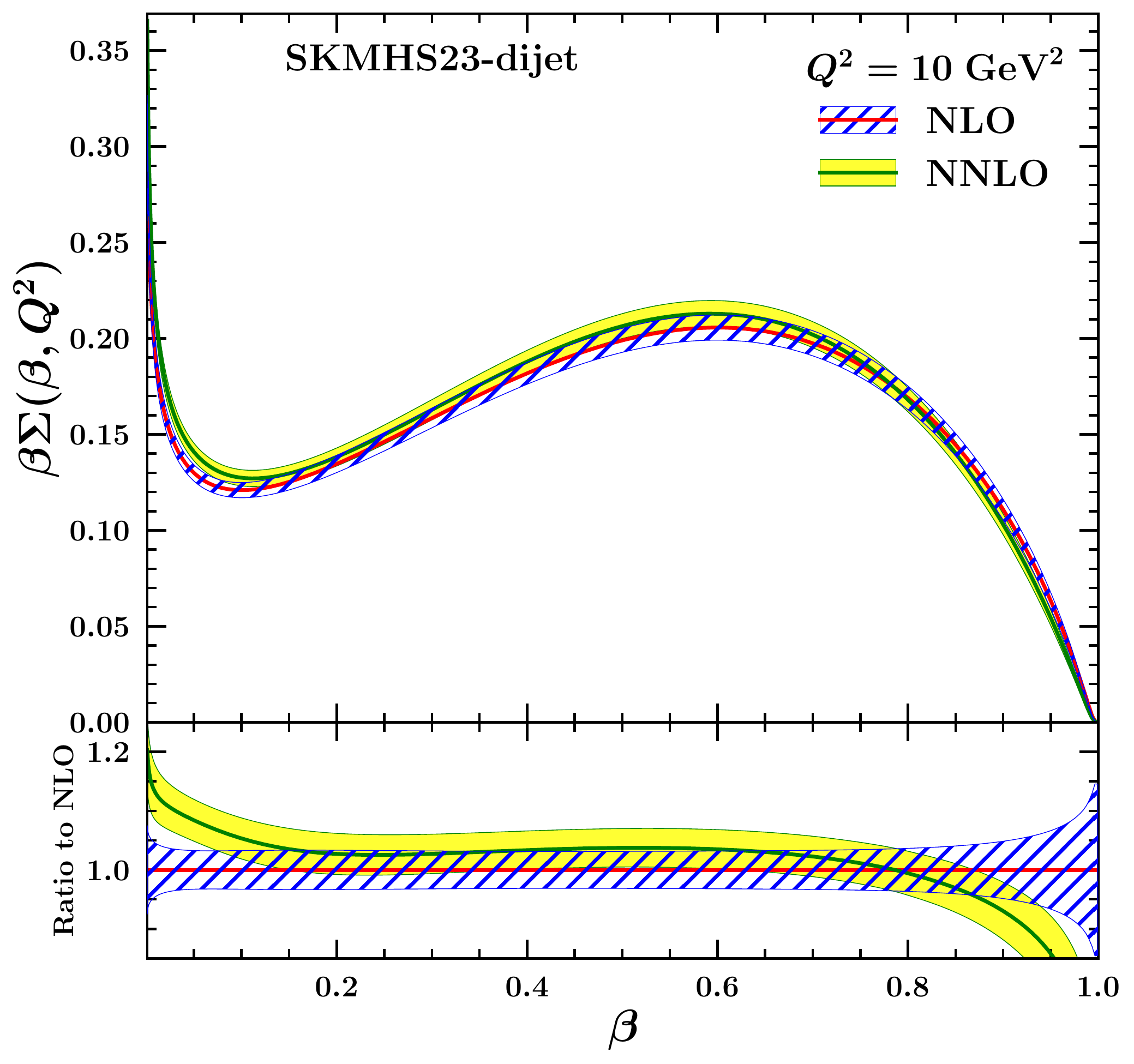}}
	\subfloat{\includegraphics[width=0.33\textwidth]{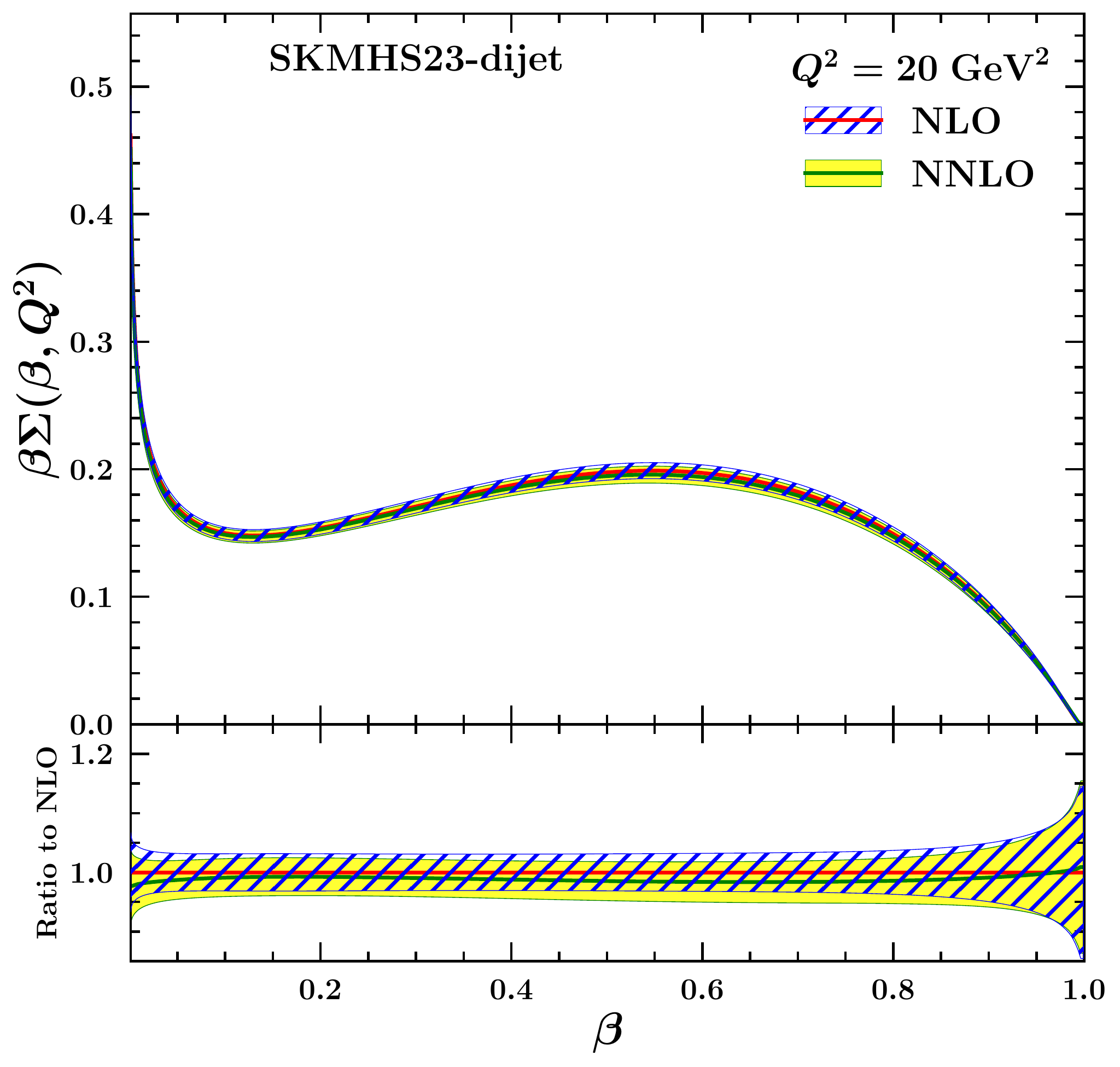}}\\	
	\subfloat{\includegraphics[width=0.33\textwidth,height=0.235\textheight]{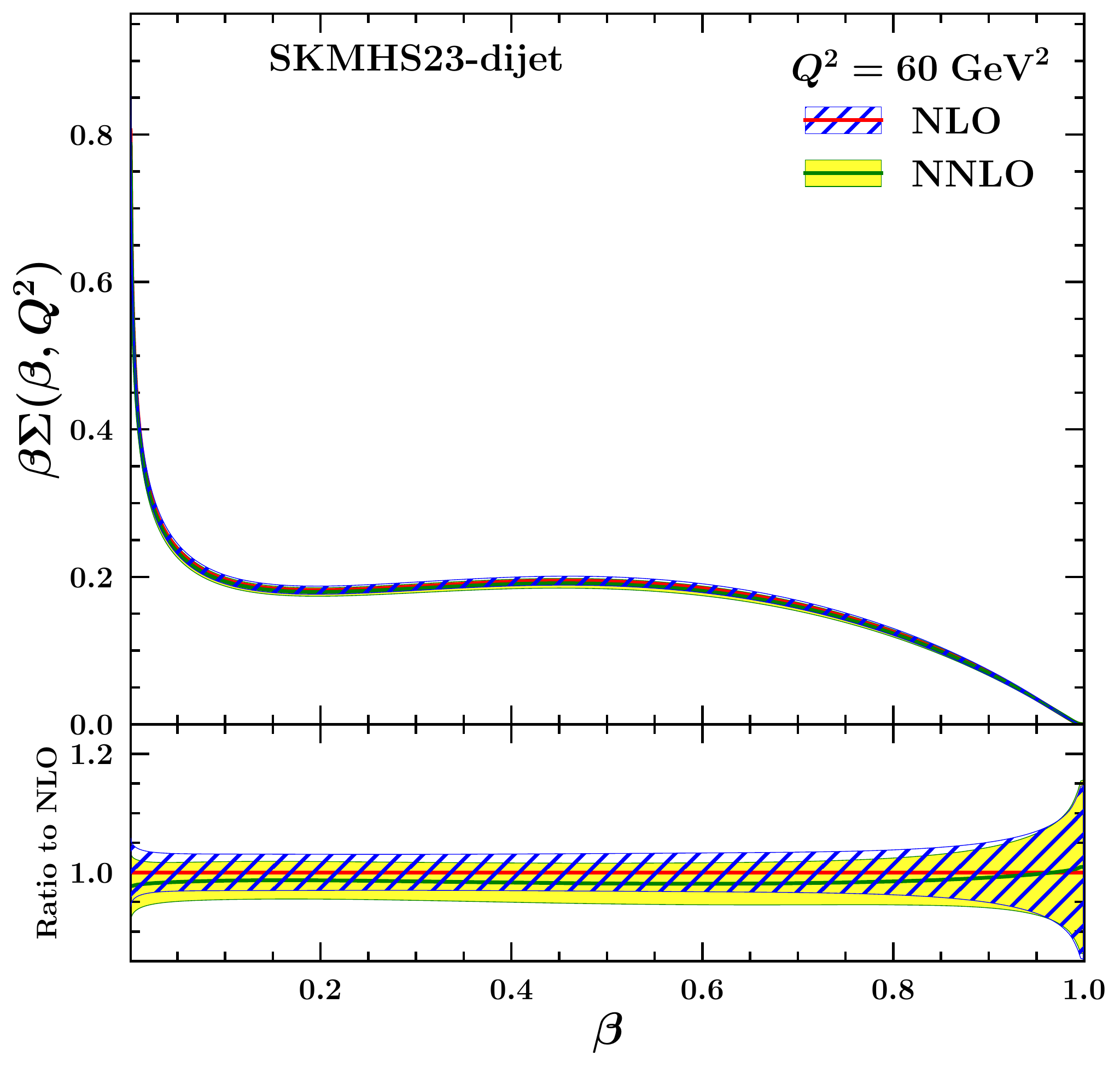}} 
	\subfloat{\includegraphics[width=0.33\textwidth]{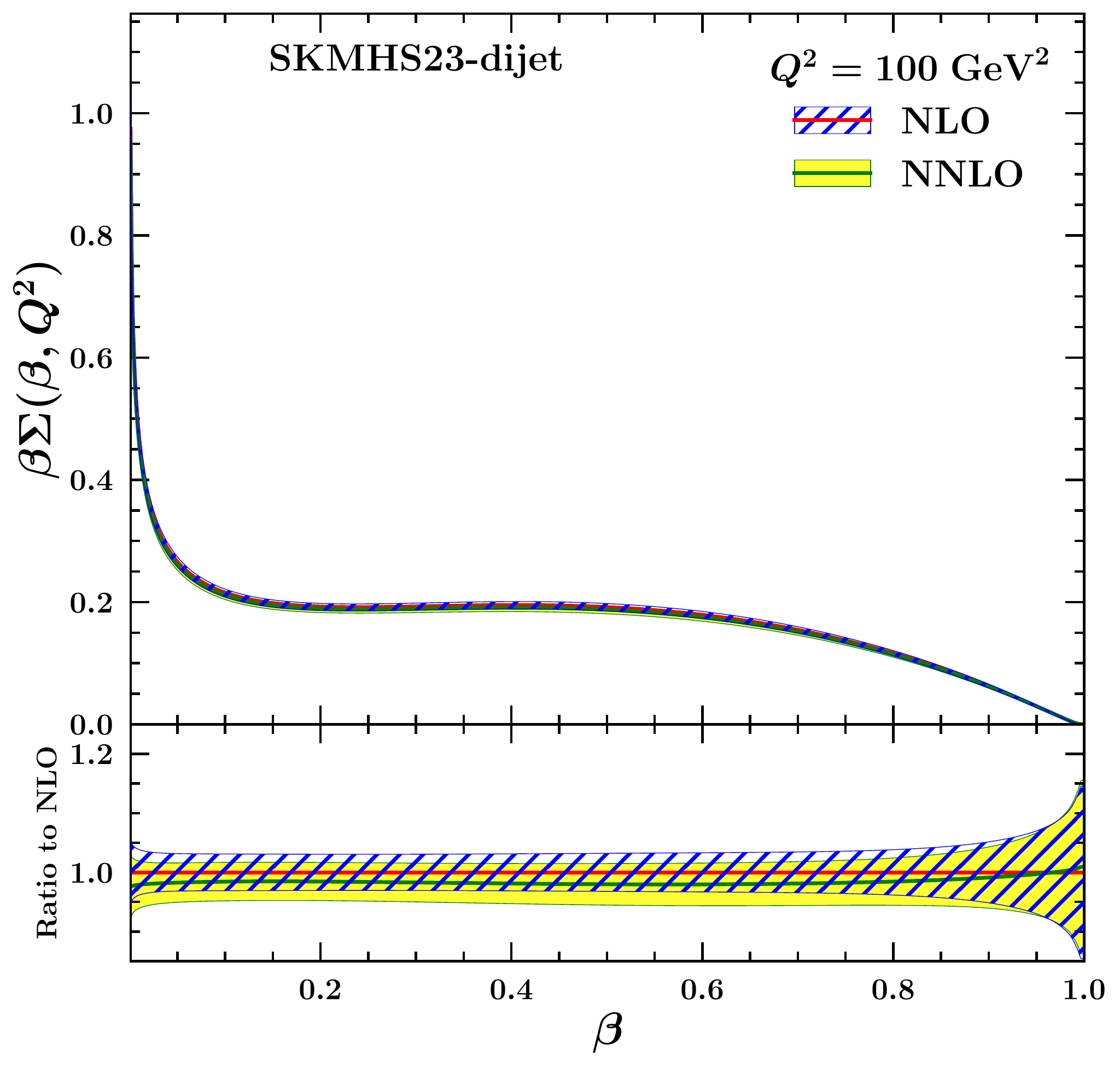}}	
	\subfloat{\includegraphics[width=0.33\textwidth]{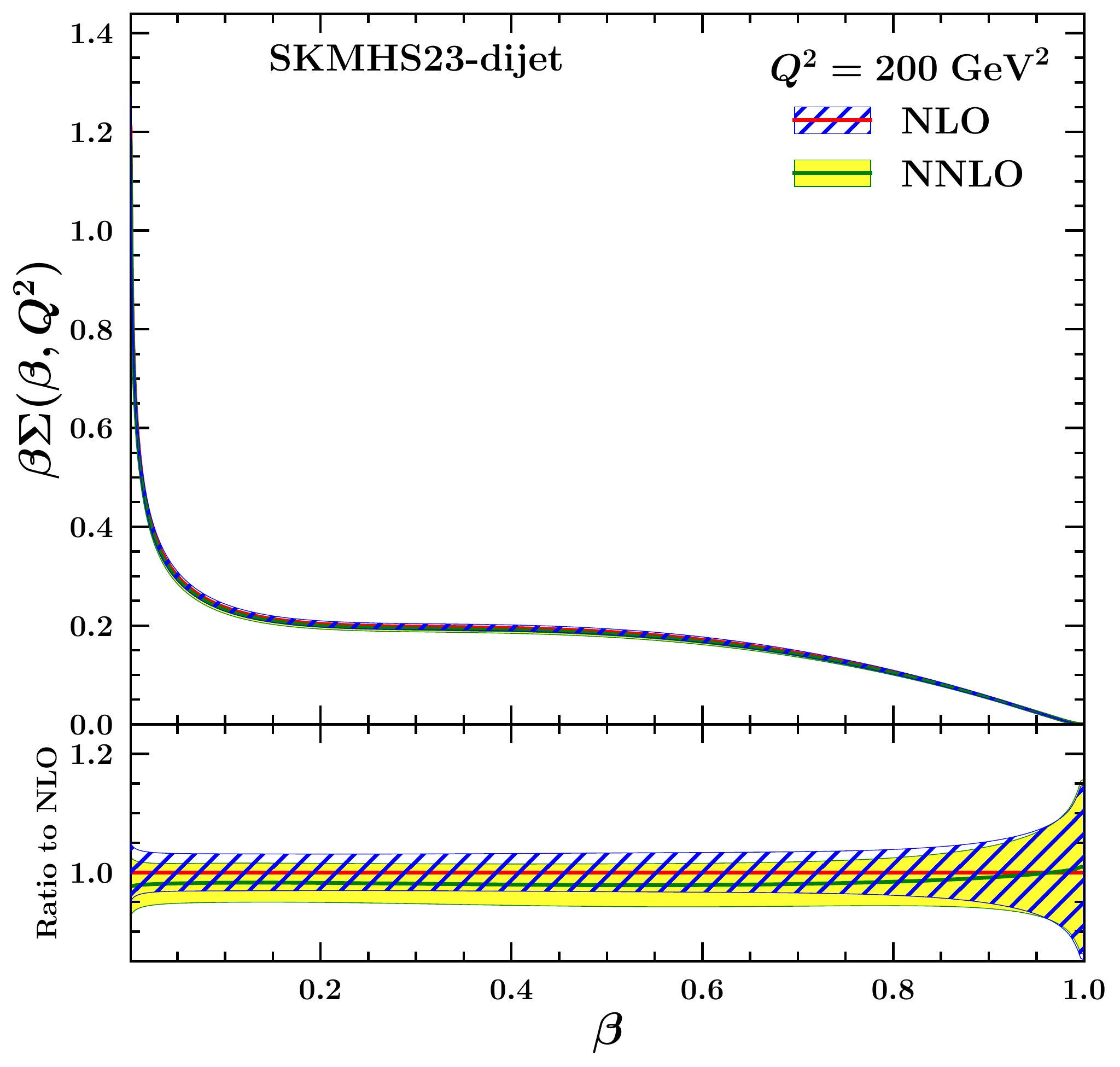}}	
	\begin{center}
		\caption{ \small 
Same as Fig.~\ref{fig:DPDF-g-Q0_withdijet} but this time for the {\tt SKMHS23-dijet} 
singlet distribution with their including uncertainties.}
		\label{fig:DPDF-S-Q0_withdijet}
	\end{center}
\end{figure*}

In Figs.~\ref{fig:DPDF-g-Q0_withdijet} and \ref{fig:DPDF-S-Q0_withdijet}, we show the 
the NLO and NNLO  {\tt SKMHS23-dijet} gluon and singlet distributions with their 
uncertainties determined using the Hessian method  
at the input scale $Q_0^2 = 1.69$~GeV$^2$.
The results for the higher energy value of 10, 20, 60, 100 and 200~GeV$^2$ are also shown.
We show both the absolute distributions and ratios to the NLO results.
The same findings as in the case of the {\tt SKMHS23} also hold for the {\tt SKMHS23-dijet}.
A significant reduction for the uncertainty bands can bee seen at the NNLO accuracy, mostly for  
large values of $\beta$.

\begin{figure*}[htb]
	\vspace{0.50cm}
	\centering
	\subfloat{\includegraphics[width=0.4\textwidth]{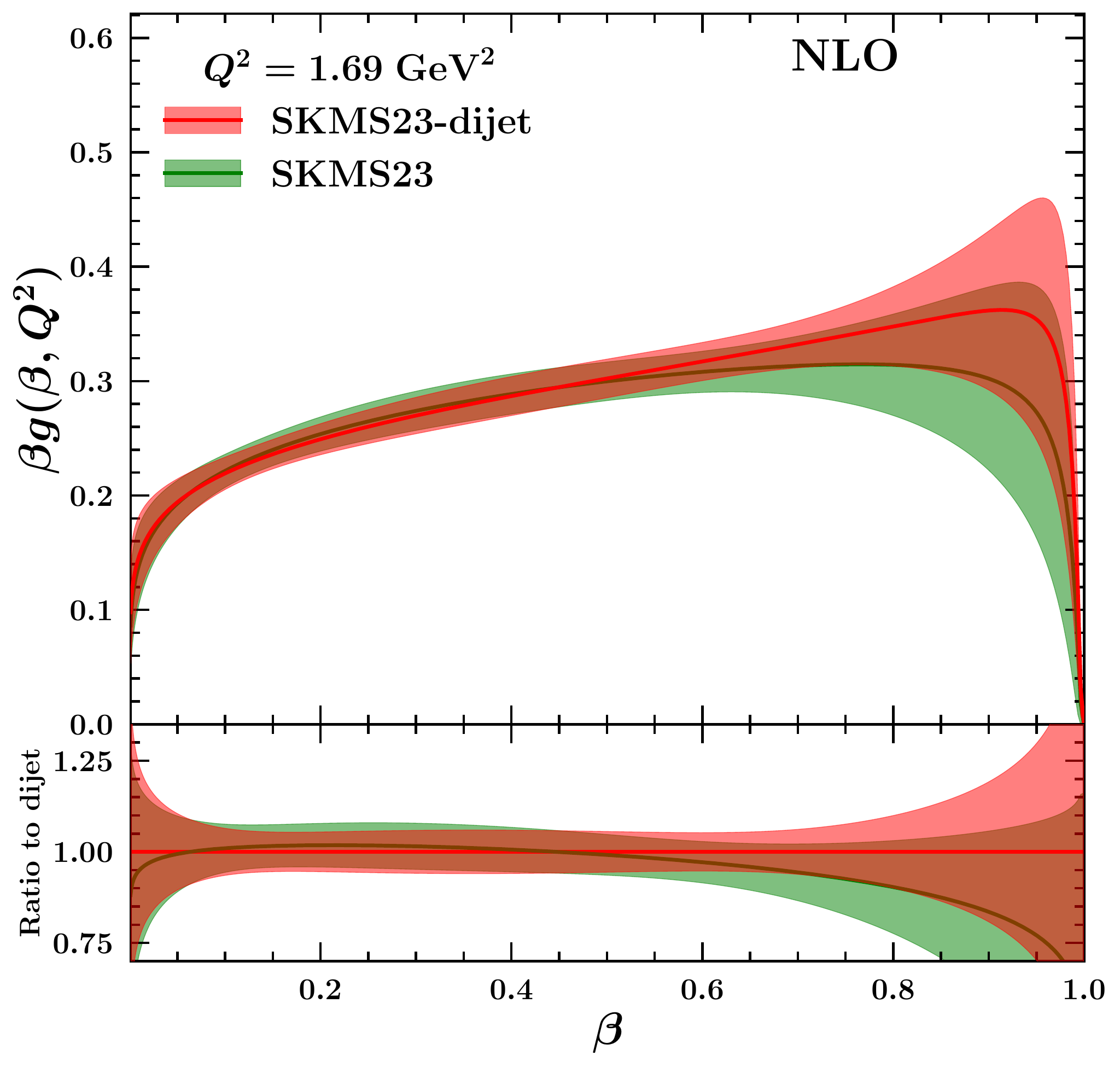}} 	
	\subfloat{\includegraphics[width=0.4\textwidth]{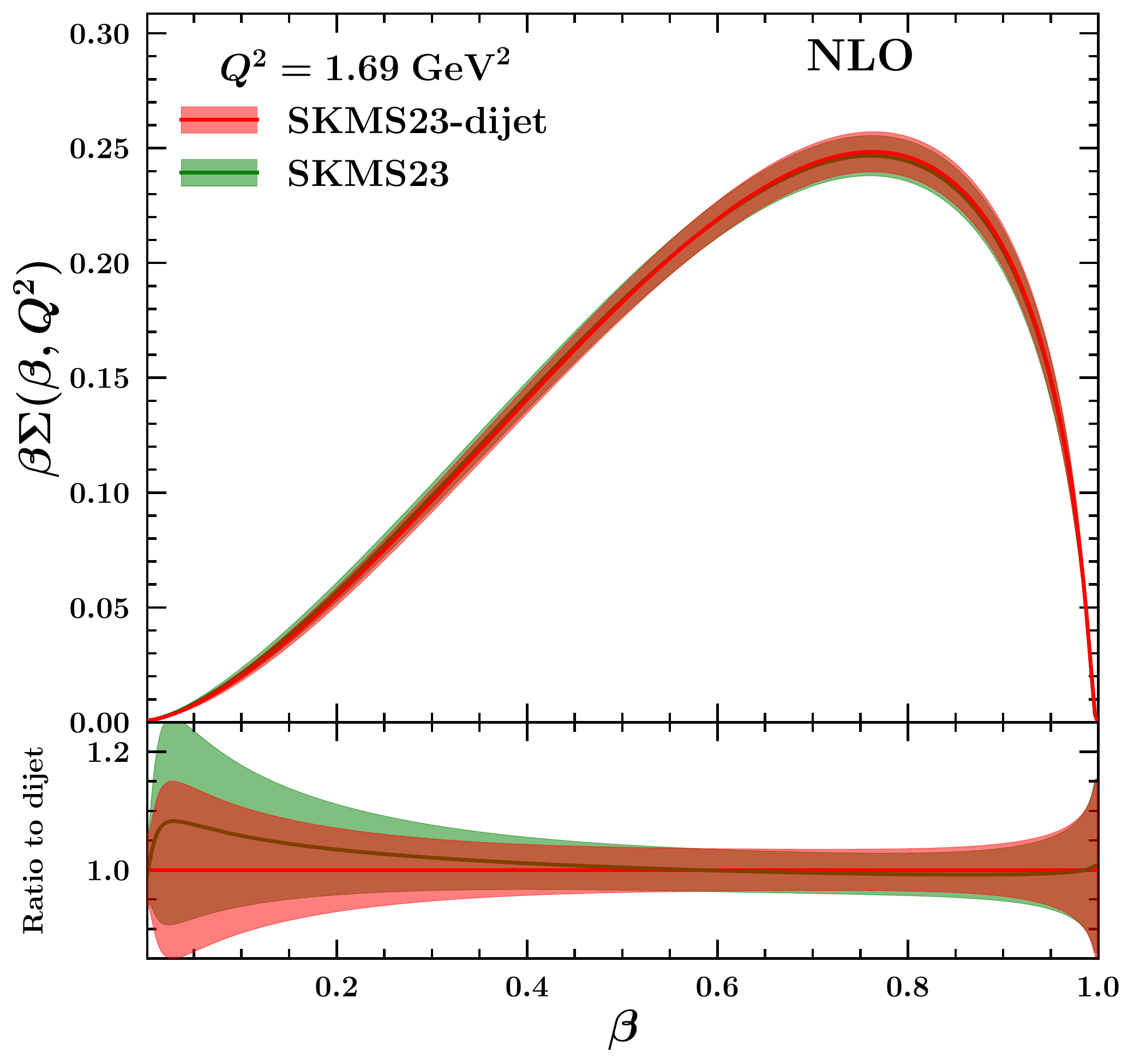}}	\\	
    \subfloat{\includegraphics[width=0.4\textwidth]{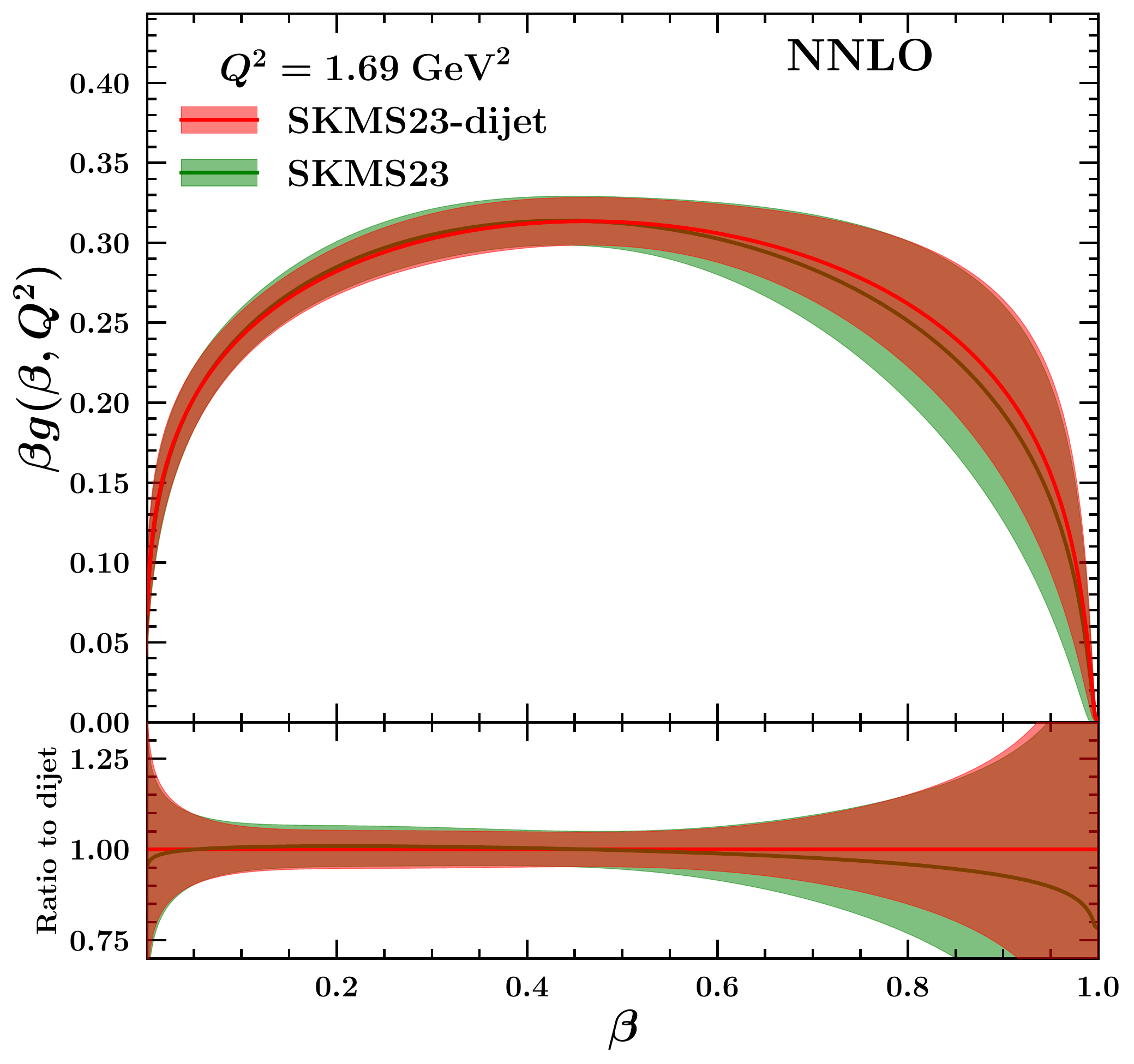}}	
	\subfloat{\includegraphics[width=0.4\textwidth]{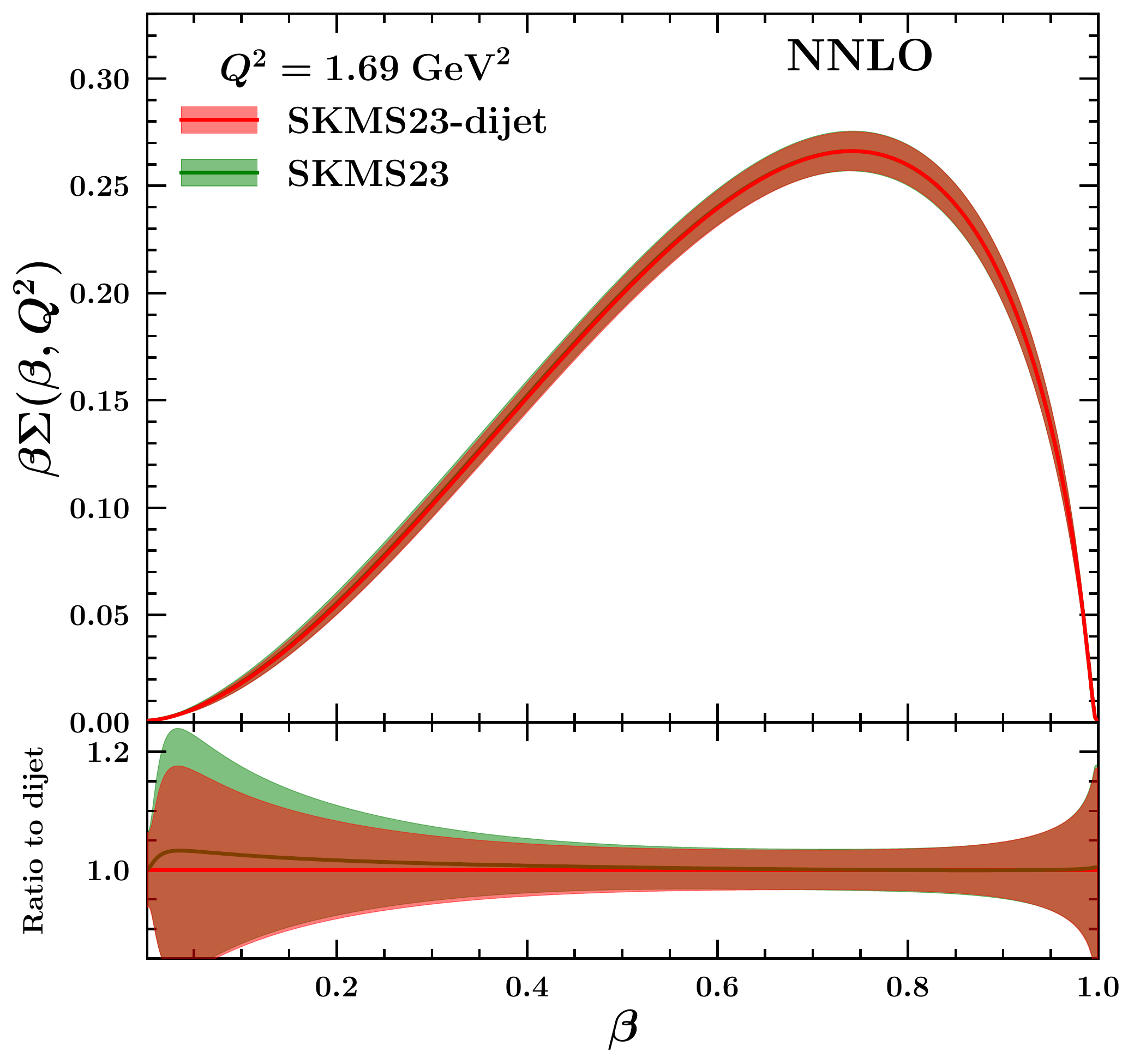}} 	
	\begin{center}
		\caption{ \small 
Comparison of the NLO and NNLO results for the {\tt SKMHS23} and {\tt SKMHS23-dijet} 
at the $Q_0^2 = 1.69$~GeV$^2$ for the gluon and singlet distributions. 
The lower panels display the ratio to the {\tt SKMHS23-dijet}. }
		\label{fig:DPDF-compare}
	\end{center}
\end{figure*}

In order to further scrutinize  the results presented in this 
work and to examine the effect
arising form the inclusion of the inclusive DIS dijet production data 
on the extracted diffractive PDFs, we
present a comparison of the NLO and NNLO results for the {\tt SKMHS23} and 
{\tt SKMHS23-dijet} in Fig.~\ref{fig:DPDF-compare} 
 at  $Q_0^2 = 1.69$ GeV$^2$ for the gluon and singlet distributions.
The upper panel of each plot displays the absolute distributions,
while the lower panel displays the {\tt SKMHS23}/{\tt SKMHS23-dijet} ratios. 
As one can see, the inclusion of the dijet data mostly affects the shape of the gluon distribution
for large value of $\beta$.
It also affects the uncertainty bands of the extracted diffractive PDFs. 
It causes a reduction of the error bands for the gluon density 
at large values of $\beta$, and for small values of $\beta$ for the case of the total singlet density.

In Tables~\ref{tab:chi2-all-SKMHS22} and \ref{tab:chi2-all-SKMHS22dihet} we 
present the values of the $\chi^2$ per data point
for both the individual and the total inclusive diffractive data sets 
included in the our analysis. 
The values for the {\tt SKMHS23} QCD fit are presented in Tab.~\ref{tab:chi2-all-SKMHS22}, and the 
values for our {\tt SKMHS23-dijet} global QCD fit  which includes 
the inclusive diffracvie dijet production 
are presented in Tab.~\ref{tab:chi2-all-SKMHS22dihet}.
The values are shown at NLO and NNLO for all the QCD analyses.

Concerning the fit quality of the total data set,
the most noticeable feature of the {\tt SKMHS23} and {\tt SKMHS23-dijet} analyses 
is the slight improvement upon the inclusion of the higher-order corrections.
Such kind of improvement also can be achieved after including the inclusive 
diffractive dijet production data in the QCD fit. 
As one can see, the inclusion of the dijet data improves the total ${\chi}^{2}/{\text{dof}}$ from 1.11 to  
1.09 for our NLO analysis, and from 1.10 to 1.07 for the NNLO case.
The improvement of the total $\chi^2$  is particularly pronounced when the dijet data
are added in the  NNLO fit. 
These findings demonstrate that both the inclusion of the NNLO corrections and considering the  
dijet data improve the description of the data. 
These findings are also consistent with the perturbative convergence and the 
uncertainty estimation discussed above
after considering the NNLO accuracy. 
Concerning the fit quality of the individual experiments, 
the general trend of the $\chi^2$ per data point is the same as that of the
total one for all QCD analyses, with the two main exceptions.
The $\chi^2$ per data point for the \text{H1-LRG-12}, despite remaining good, but 
increases slightly as higher-order QCD corrections are included in {\tt SKMHS23} fit.
For the case of \text{H1-LRG-11} $\sqrt{s} = {252}$~GeV, this value remains unchanged 
after inclusion of the NNLO correction in our {\tt SKMHS23} fit. 

For both  {\tt SKMHS23} and {\tt SKMHS23-dijet} analyses, 
the $\chi^2$ per data point for the case of 
the \text{H1/ZEUS combined}~\cite{H1:2012xlc} data set, are still 
large for the NLO and NNLO analysis.
This treatment is discussed in detail in Ref.~\cite{Goharipour:2018yov}.
To decrease the $\chi^2$  for this specific data set, one needs to impose 
a minimum cut on the Q$^2$ value at around 16~GeV$^2$.
In this work, we prefer to consider Q$^2 \ge$ Q$^2_{\rm min}$ with Q$^2_{\rm min} = 9$~GeV$^2$.

%
\begin{table*}[htb]
	\caption{ \small The values of $\chi^2/N_{\text{pts}}$ for both the individual and the total data sets
		included in the {\tt SKMHS23} QCD fit.} \label{tab:chi2-all-SKMHS22}
	\begin{tabular}{l c c c  c }
		\hline \hline
		&             & ~~{\tt SKMHS23 (NLO)}~~  &  ~~{\tt SKMHS23 (NNLO)}~~  \\ \hline
		\text{Experiment} & Process  & ${\chi}^{2}/{N}_{\text{pts}}$  & ${\chi}^2/{N}_{\text{pts}}$ 
		\tabularnewline
		\hline
		\tt{\text{H1-LRG-11}} $\sqrt{s} = {225}$~GeV~\cite{H1:2011jpo} & inclusive DDIS &  10/13  & 9/13    \\	
		\tt{\text{H1-LRG-11}} $\sqrt{s} = {252}$~GeV~\cite{H1:2011jpo} & inclusive DDIS &  19/12  &  19/12\\		
		\tt{\text{H1-LRG-12}}~\cite{H1:2012pbl}        & inclusive DDIS &  134/165  & 136/165 \\	
		\tt{\text{H1/ZEUS combined}}~\cite{H1:2012xlc} & inclusive DDIS &  141/96   & 140/96 \\	 \hline \hline
		\multicolumn{1}{c}{~\textbf{${\chi}^{2}/{\text{dof}}$}~}  &   &    $~308/278=1.11~$ &   $~306/278=1.10~$ \\  \hline
	\end{tabular}
\end{table*}
%
%

%
\begin{table*}[htb]
	\caption{ \small The values of $\chi^2/N_{\text{pts}}$ for both 
		the individual and the total data sets
		included in the {\tt SKMHS23-dijet} global QCD fit.} \label{tab:chi2-all-SKMHS22dihet}
	\begin{tabular}{l c c c  c }
		\hline \hline
		&             & ~~{\tt SKMHS23-dijet (NLO)}~~  &  ~~{\tt SKMHS23-dijet (NNLO)}~~  \\ \hline
		\text{Experiment} & Process  & ${\chi}^{2}/{N}_{\text{pts}}$  & ${\chi}^2/{N}_{\text{pts}}$ 
		\tabularnewline
		\hline
		\tt{\text{H1-LRG-11}} $\sqrt{s} = {225}$~GeV~\cite{H1:2011jpo} & inclusive DDIS &  11/13  & 10/13    \\	
		\tt{\text{H1-LRG-11}} $\sqrt{s} = {252}$~GeV~\cite{H1:2011jpo} & inclusive DDIS &  19/12  &  18/12\\		
		\tt{\text{H1-LRG-12}}~\cite{H1:2012pbl}        & inclusive DDIS &  135/165  & 135/165 \\	
		\tt{\text{H1/ZEUS combined}}~\cite{H1:2012xlc} & inclusive DDIS &  141/96   & 139/96 \\	 
		\tt{\text{H1-LRG (HERA II)}}~\cite{H1:2014pjf} & inclusive dijet production &  12/15   & 10/15 \\	 	\hline \hline
		\multicolumn{1}{c}{~\textbf{${\chi}^{2}/{\text{dof}}$}~}  &   &    $~320/293=1.09~$ &   $~314/293=1.07~$ \\  \hline
	\end{tabular}
\end{table*}
%
%

We now in a position to compare our diffractive PDFs to the most recent determinations
available in the literature, namely the {\tt GKG18}~\cite{Goharipour:2018yov} and 
our previous work {\tt SKMHS22-tw2-tw4-RC}~\cite{Salajegheh:2022vyv}. 

In the analysis by {\tt GKG18}, they presented the first QCD analysis of 
diffractive PDFs in the framework of {\tt xFitter}~\cite{Alekhin:2014irh}, and 
analyzed for the first time the H1/ZEUS combined data sets~\cite{H1:2012xlc}. 
In our most recent work, {\tt SKMHS22}, we presented a new set of diffractive PDFs 
and their uncertainties at NLO and NNLO accuracy in perturbative QCD within the {\tt xFitter} framework. 
The  diffractive PDFs has been we extracted considering the standard twist-2 
contribution, the twist-4 correction, 
and the contribution of subleading Reggeon exchange.

\begin{figure*}[htb]
	\vspace{0.50cm}
	\centering
	\subfloat{\includegraphics[width=0.5\textwidth]{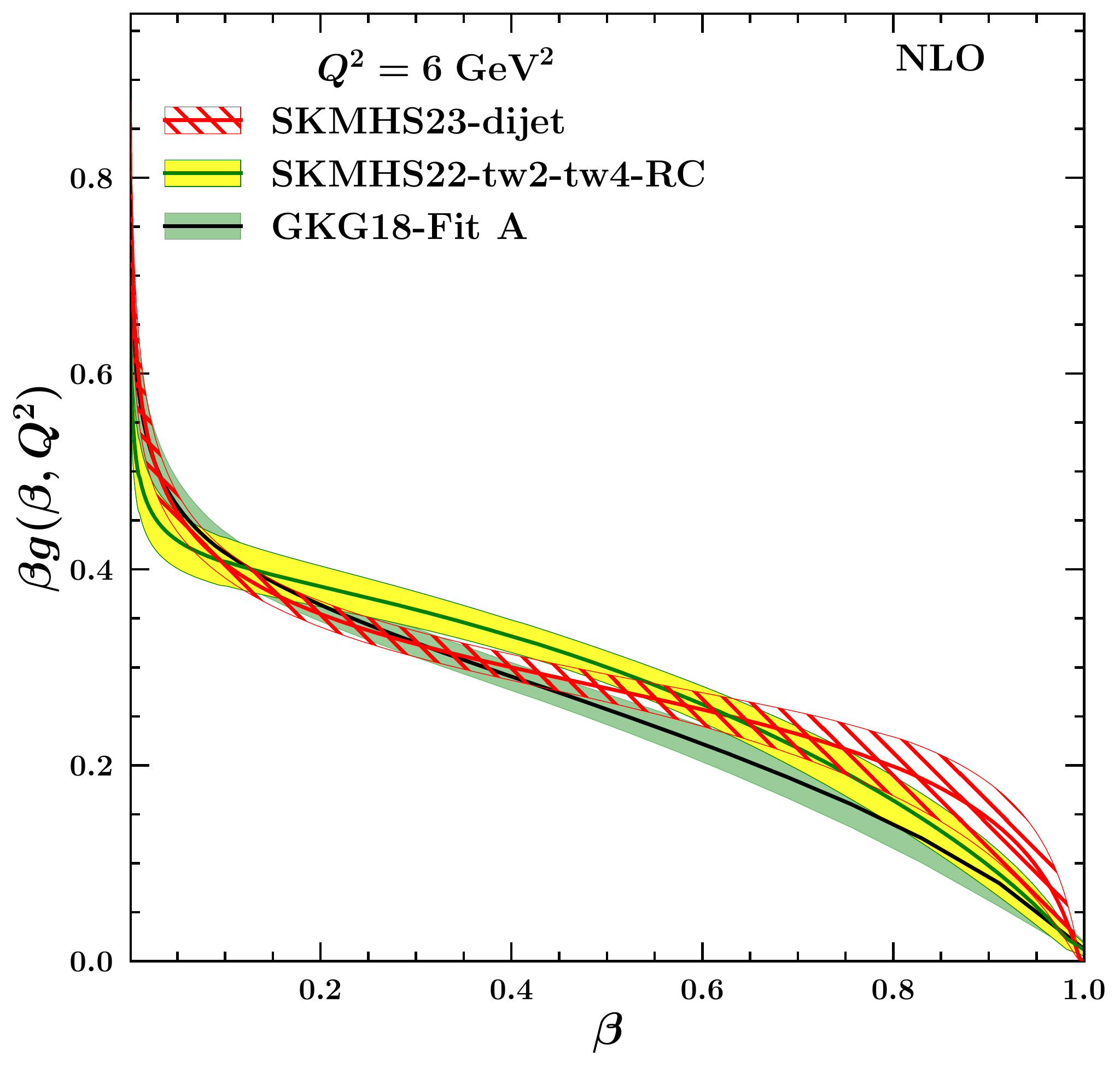}} 	
	\subfloat{\includegraphics[width=0.5\textwidth,height=0.355\textheight]{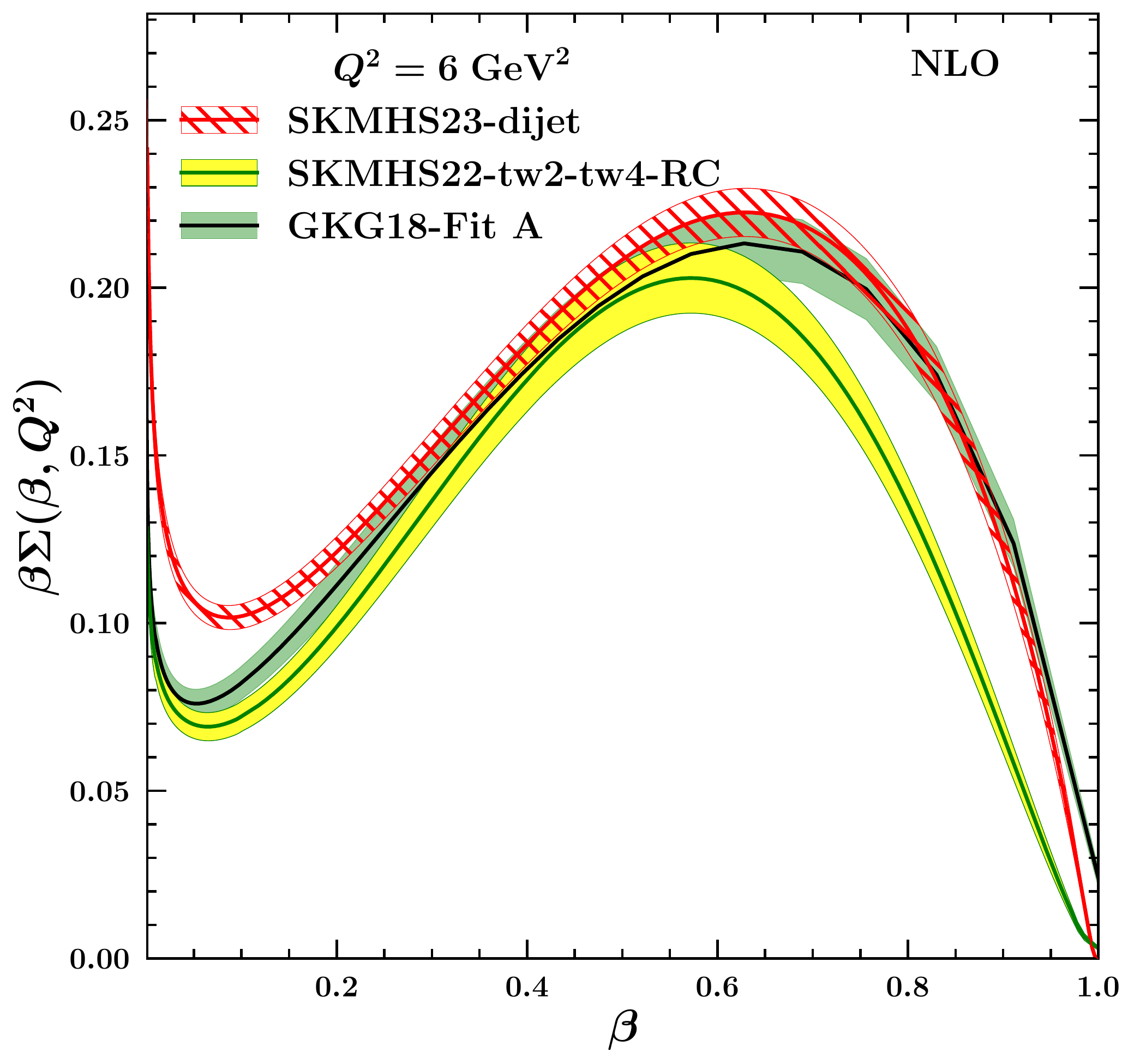}}
	\begin{center}
		\caption{ \small 
			Comparison between {\tt SKMHS23-dijet}, {\tt GKG18}~\cite{Goharipour:2018yov} 
			and {\tt SKMHS23}~\cite{Salajegheh:2022vyv} at Q$^2$ = 6~GeV$^2$ as a function 
			of $\beta$, for  gluon (left) and total singlet distributions (right). }
		\label{fig:DPDF-Comparison}
	\end{center}
\end{figure*}

Since the {\tt GKG18} analysis was performed only at NLO accuracy, we limit the comparison
to this order.  
Such a comparison is shown in Fig.~\ref{fig:DPDF-Comparison} at Q$^2$ = 6~GeV$^2$ as 
a function of $\beta$, for both gluon and total singlet distributions. 
Concerning the shapes of the diffractive PDFs and their error bands, a number of interesting
differences and similarities between these three sets can be seen from the comparisons
in Fig.~\ref{fig:DPDF-Comparison}.
For the case of the gluon density, overall good agreements between these three sets 
can be seen. However, the new analysis mostly affects the gluon 
density function over the large value of momentum fraction $\beta$.
The differences in shape among the three diffractive PDFs sets are more marked 
in the case of the total singlet.
The {\tt SKMHS23-dijet} analysis is in fairly good agreement with 
{\tt GKG18} analysis over the medium to large value of $\beta$. 
Both {\tt GKG18} and {\tt SKMHS22-tw2-tw4-RC} are more suppressed at small 
value of $\beta$ with respect to the {\tt SKMHS23-dijet}.

Concerning the diffractive PDFs uncertainties, we observe that for both the
gluon and total singlet distributions the three sets are in good agreement
in the region covered by the high $\beta$ data, roughly $\beta>0.4$. 
Conversely, over the small region of $\beta$, differences are more significant.
Typically, the uncertainties of the {\tt SKMHS23-dijet} are
smaller than those of both  {\tt GKG18} and {\tt SKMHS22-tw2-tw4-RC}.

\begin{figure*}[htb]
\vspace{0.50cm}
\centering
\includegraphics[width=0.8\textwidth]{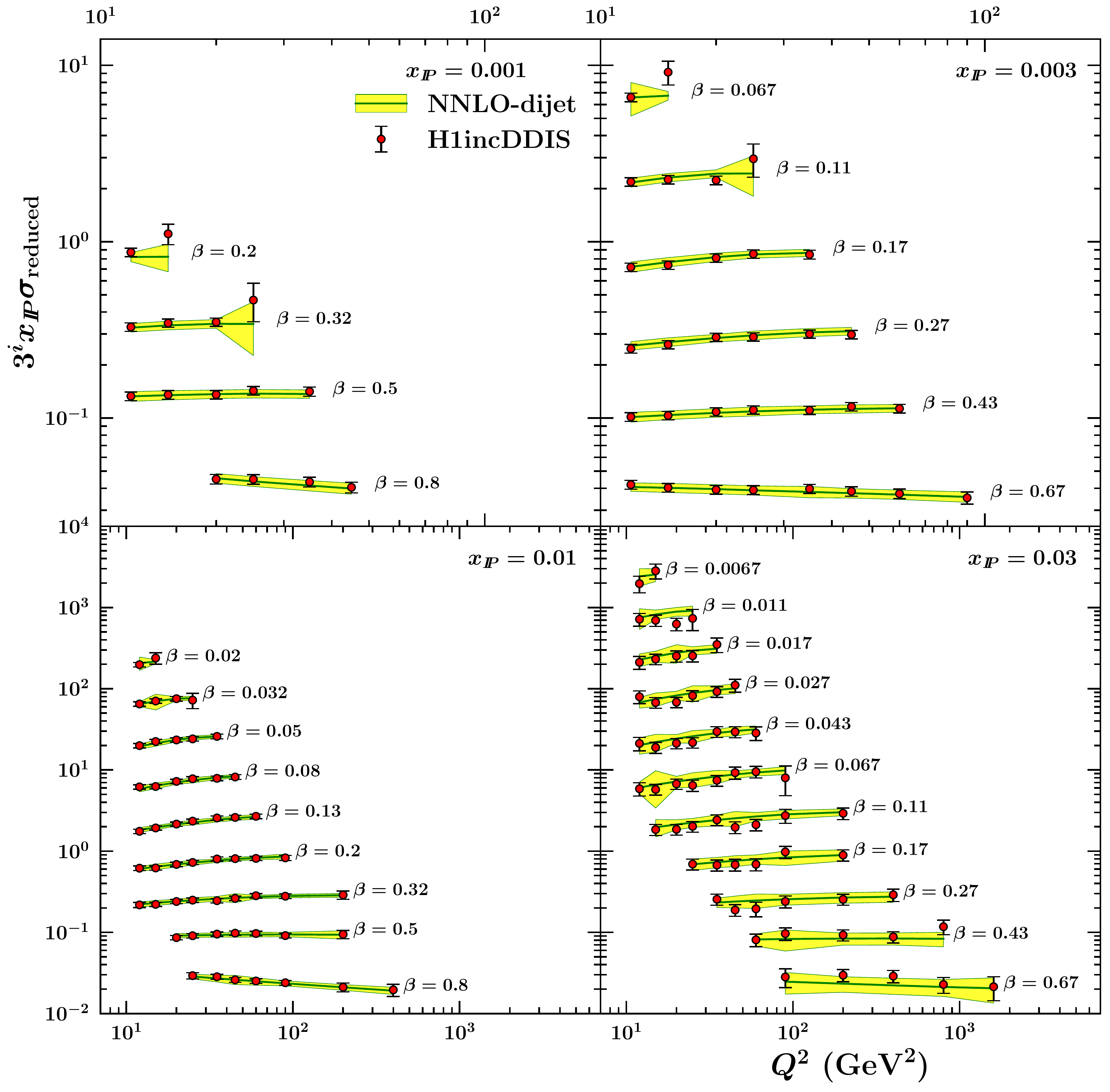}
\begin{center}
\caption{ \small 
The NNLO theory prediction obtained using the {\tt SKMHS23-dijet} global QCD fit  
in comparison with the inclusive diffractive DIS data sets as a 
function of Q$^2$ and for two different selected bins of $x_{\pom}$ = 0.001, 0.003, 0.01 and 0.03. 
The shaded area indicates to the experimental uncertainty.  }
\label{fig:H1incDDIS}
\end{center}
\end{figure*}

Furthermore, we now
present a comparison of the data sets used in our analysis to the
corresponding NNLO theoretical predictions obtained using
the NNLO {\tt SKMHS23-dijet} fit. 
In Fig.~\ref{fig:H1incDDIS} such a comparison is displayed for the 
NNLO theory prediction calculated using the {\tt SKMHS23-dijet} global QCD fit 
with the inclusive diffractive DIS data sets.
The comparisons are presented as a function of Q$^2$ and for four
different selected bins of $x_{\pom}$ = 0.001, 0.003, 0.01 and 0.03, and 
several values of $\beta$. The shaded area indicates to the experimental uncertainty.
As can be seen, in general an overall very good agreement between the data
and the NNLO theoretical predictions is achieved for all diffractive experiments, 
which is consistent with the $\chi^2$ values per 
data points reported in Tab.~\ref{tab:chi2-all-SKMHS22dihet}.
Remarkably, the {\tt SKMHS23-dijet} NNLO theoretical predictions and the 
inclusive diffractive data are in 
good agreement over the whole kinematical region.

\begin{figure}[htb]
\vspace{0.50cm}
\centering
 \subfloat{\includegraphics[width=0.215\textwidth,height=0.153\textheight]{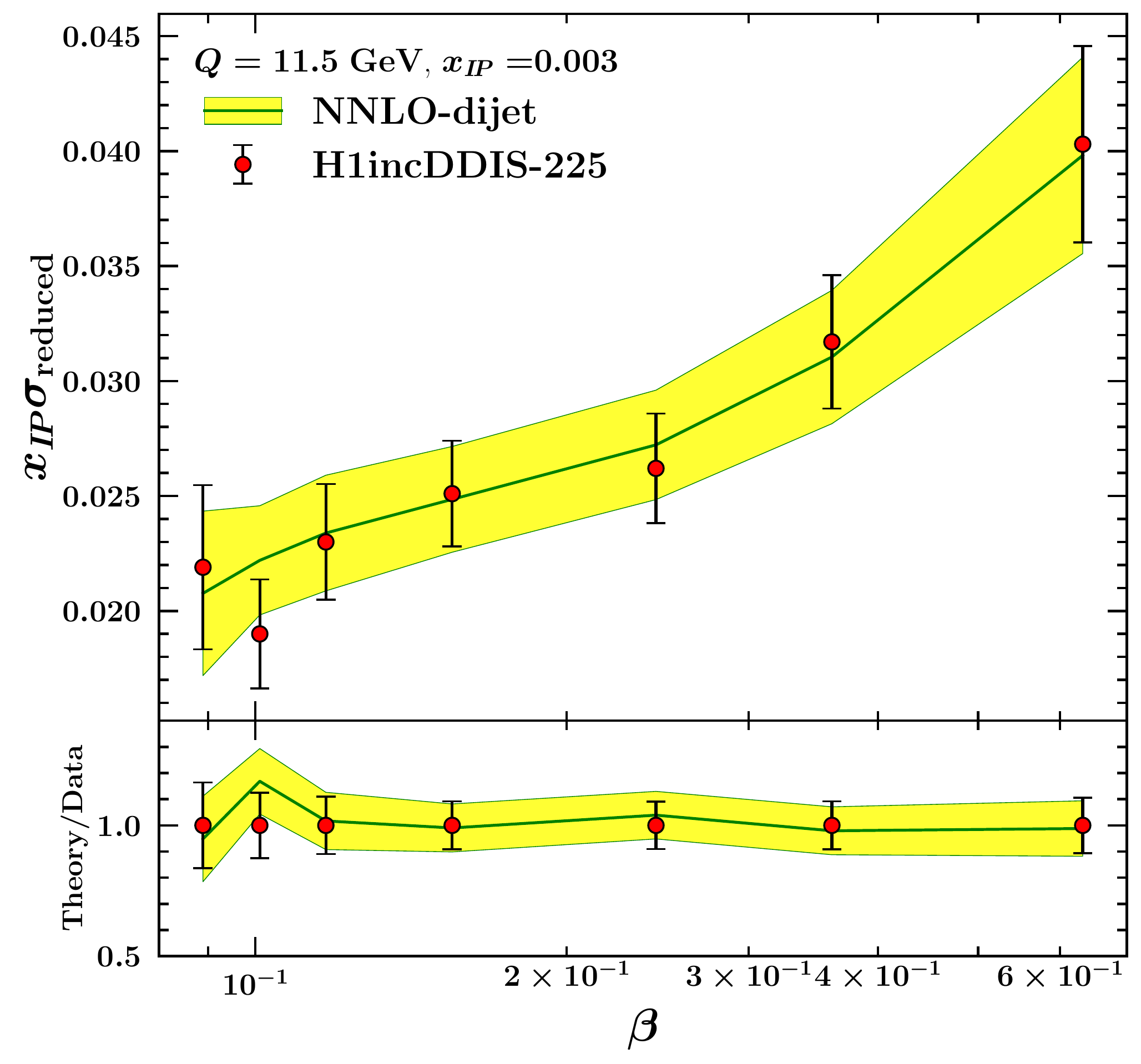}} 	
 \subfloat{\includegraphics[width=0.225\textwidth]{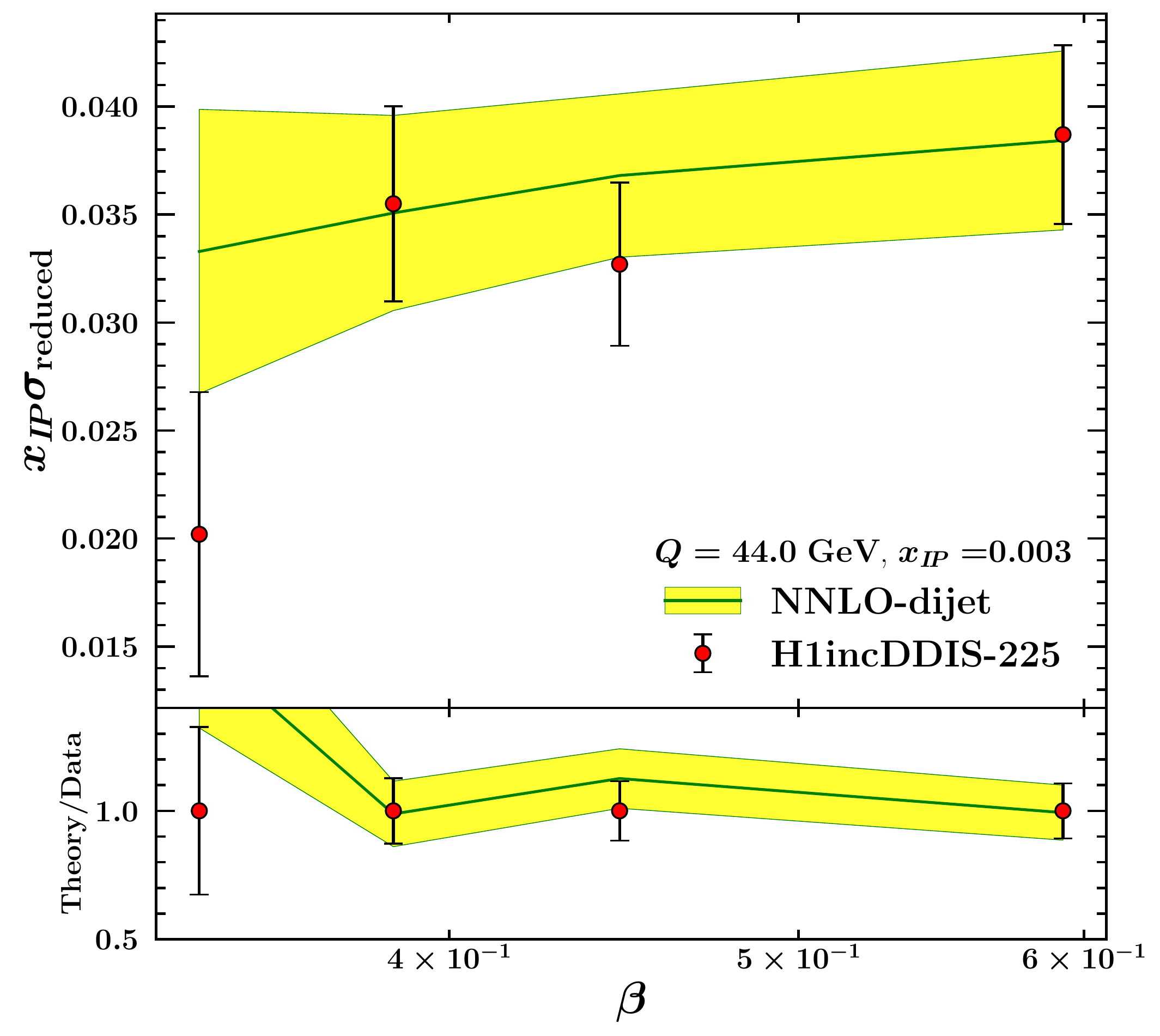}}\\
 \subfloat{\includegraphics[width=0.215\textwidth,height=0.155\textheight]{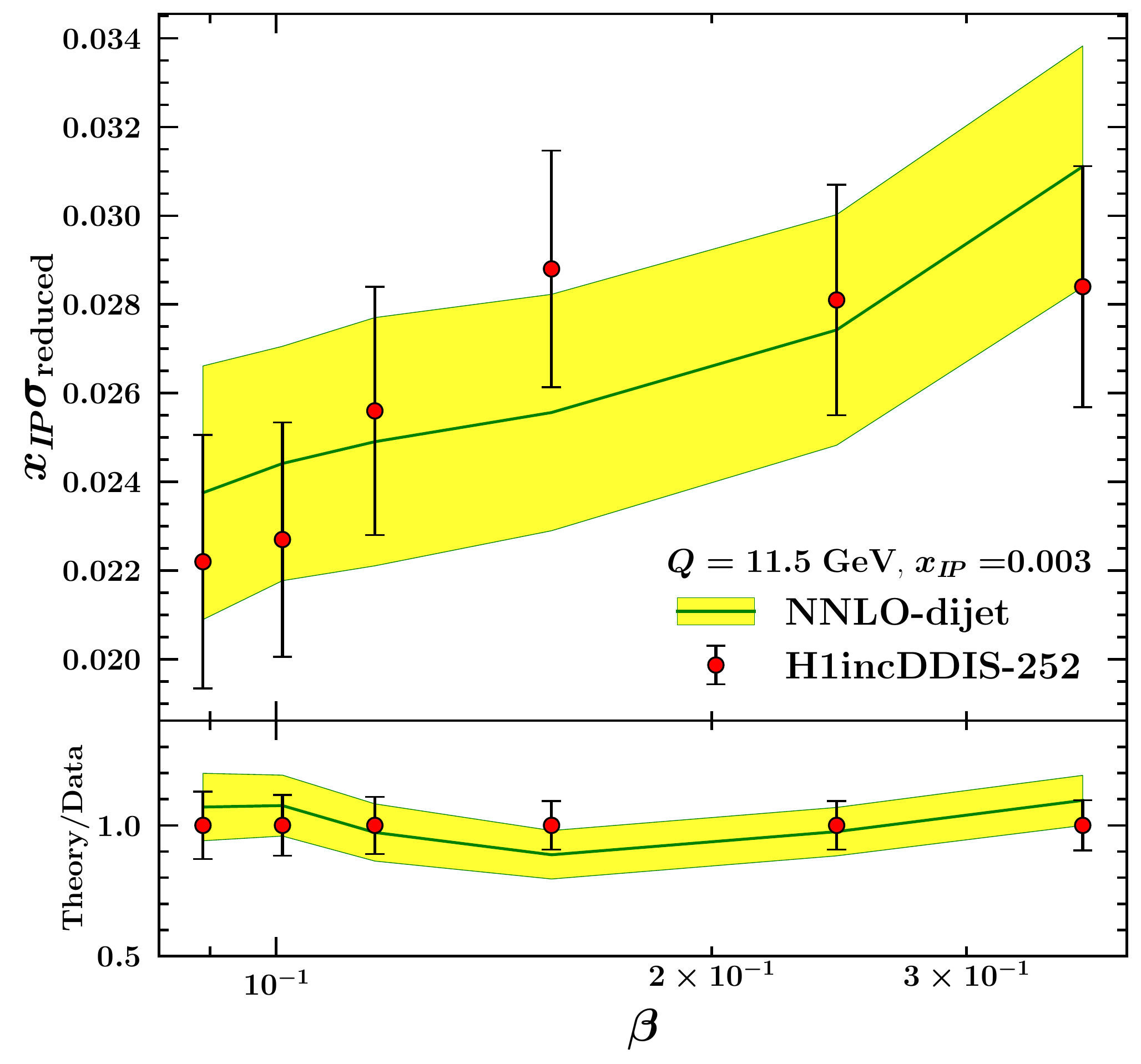}} 	
 \subfloat{\includegraphics[width=0.23\textwidth]{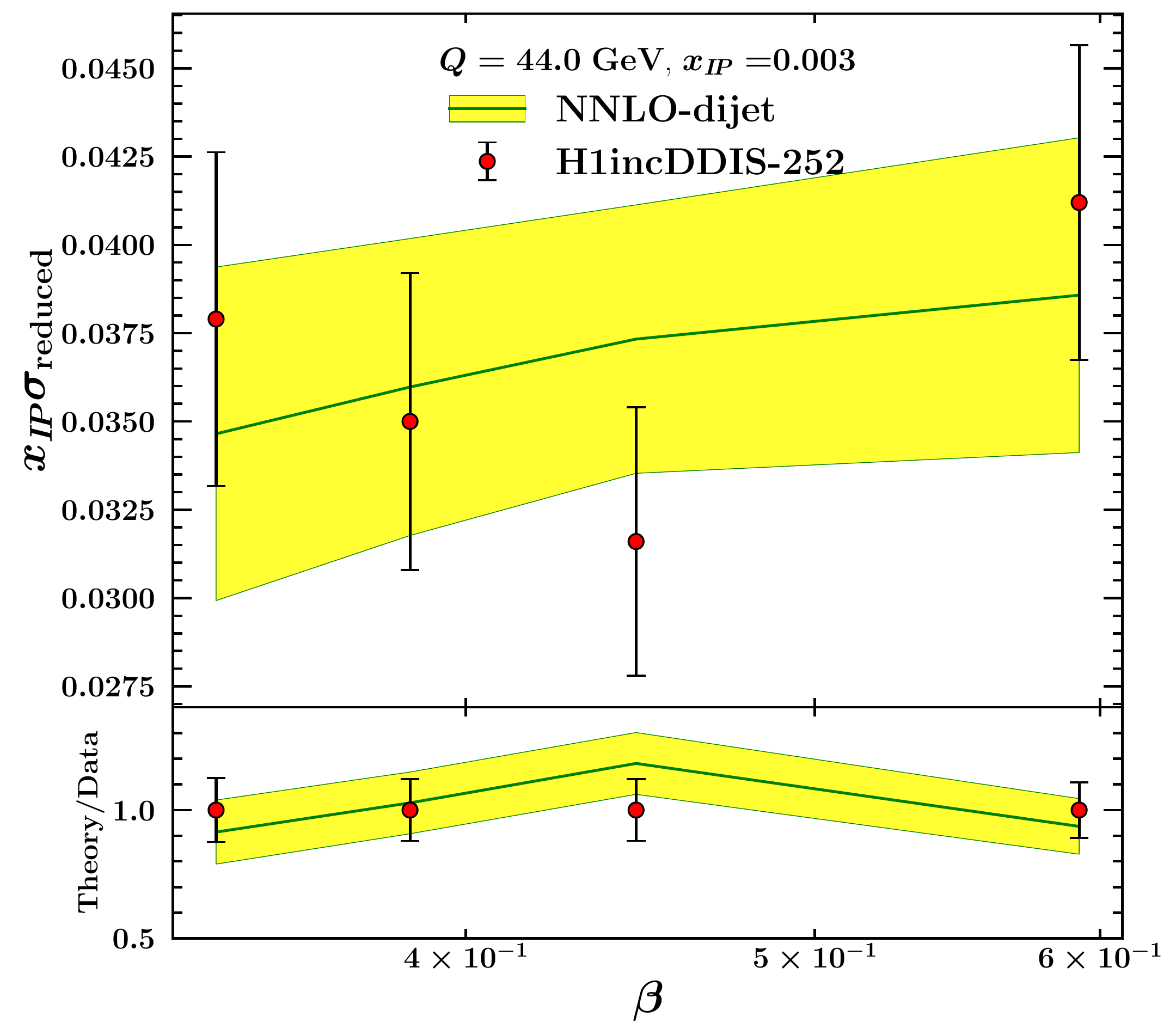}} 	
\begin{center}
\caption{ \small 
Comparison of the NNLO theory prediction for the inclusive diffractive 
cross section obtained using the {\tt SKMHS23-dijet}   
with the \text{H1-LRG-11} $\sqrt{s} = {225}$~GeV and \text{H1-LRG-11} $\sqrt{s} = {252}$~GeV 
inclusive diffractive DIS data sets. 
Both the absolute distributions (upper panel) and the data/theory ratios (lower panel) are shown. }
\label{fig:H1incDDIS-225-252}
\end{center}
\end{figure}

In Fig.~\ref{fig:H1incDDIS-225-252}, we compare the 
NNLO theory prediction for the inclusive cross section calculated 
using the {\tt SKMHS23-dijet} global QCD fit  
with the \text{H1-LRG-11} $\sqrt{s} = {225}$~GeV and \text{H1-LRG-11} $\sqrt{s} = {252}$~GeV 
inclusive diffractive DIS data sets.
The NNLO theory prediction are calculated and shown as a 
function of $\beta$ and for some selected values of $x_{\pom}$  and $Q$.
We show both the absolute distributions (upper panel) and the data/theory ratios (lower panel).
As one can see, the theoretical predictions and the data are in 
good agreement with the \text{H1-LRG-11} $\sqrt{s} = {225}$~GeV.
A small disagreement with the \text{H1-LRG-11} $\sqrt{s} = {252}$~GeV is found which reflects 
the origin of the large $\chi^2$ reported in Tab.~\ref{tab:chi2-all-SKMHS22dihet} for these data.

\begin{figure*}[htb]
\vspace{0.50cm}
\centering
\subfloat{\includegraphics[width=0.33\textwidth]{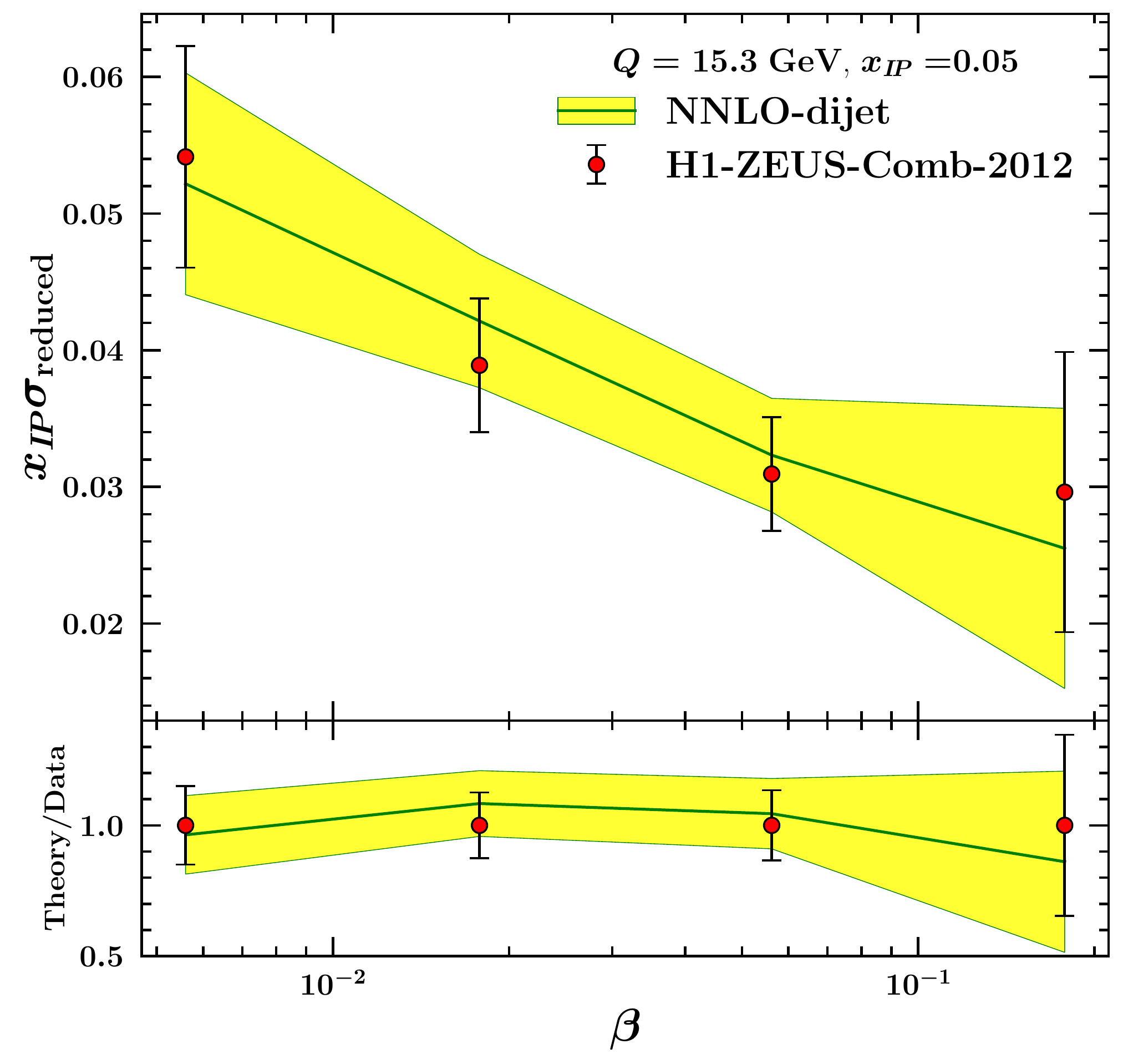}} 	
\subfloat{\includegraphics[width=0.33\textwidth,height=0.235\textheight]{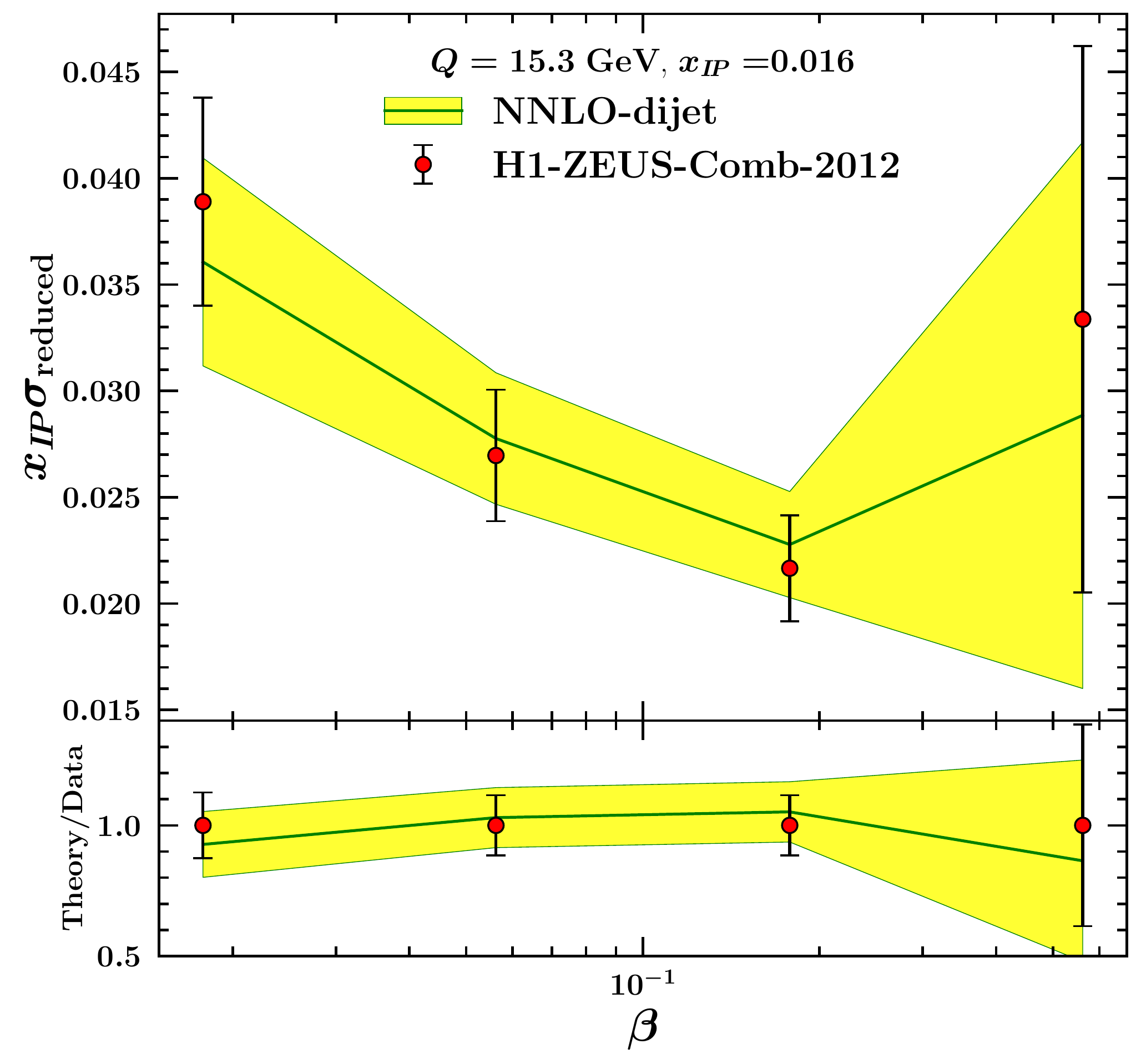}}
\subfloat{\includegraphics[width=0.33\textwidth]{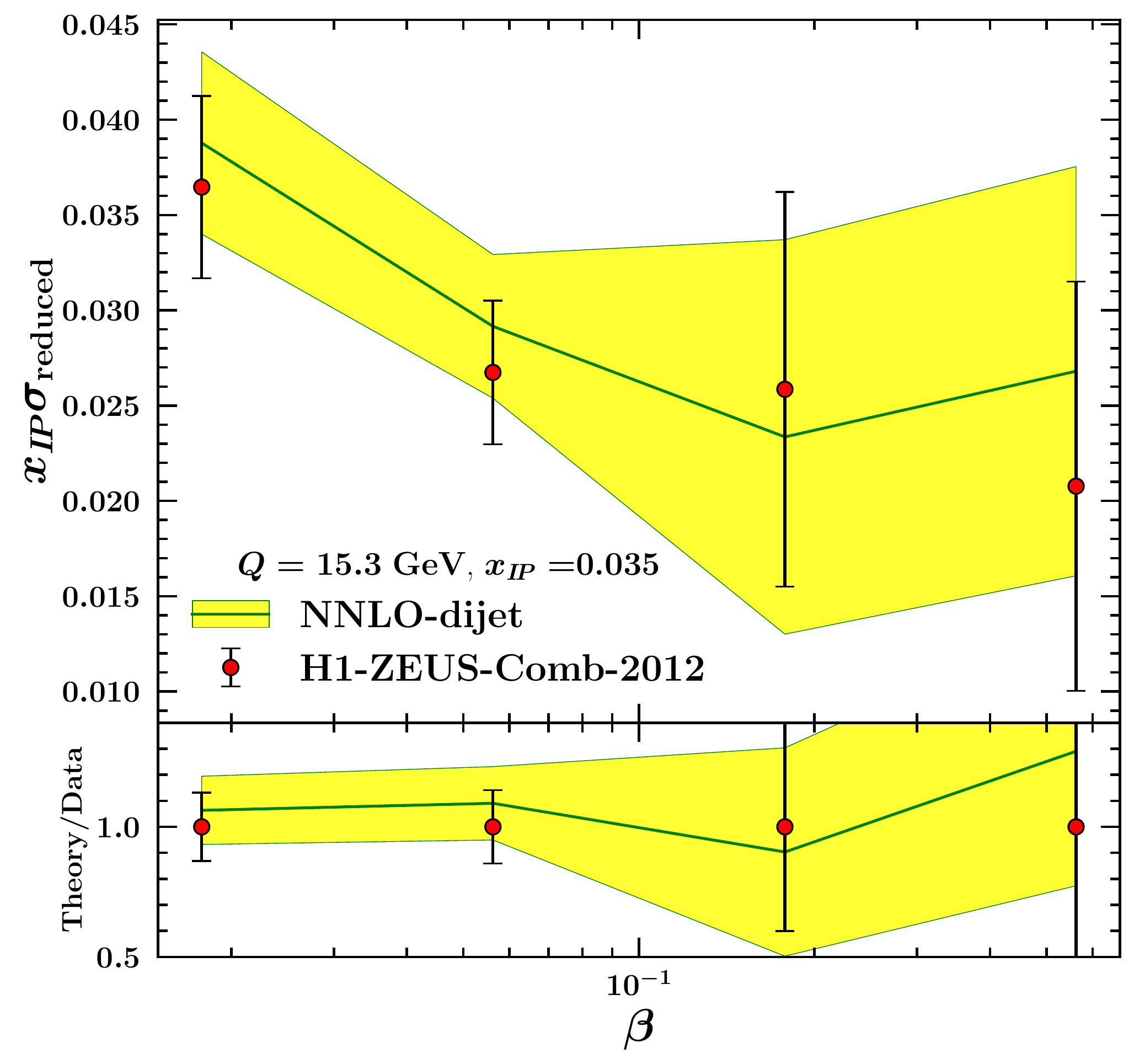}}\\	
\subfloat{\includegraphics[width=0.33\textwidth,height=0.235\textheight]{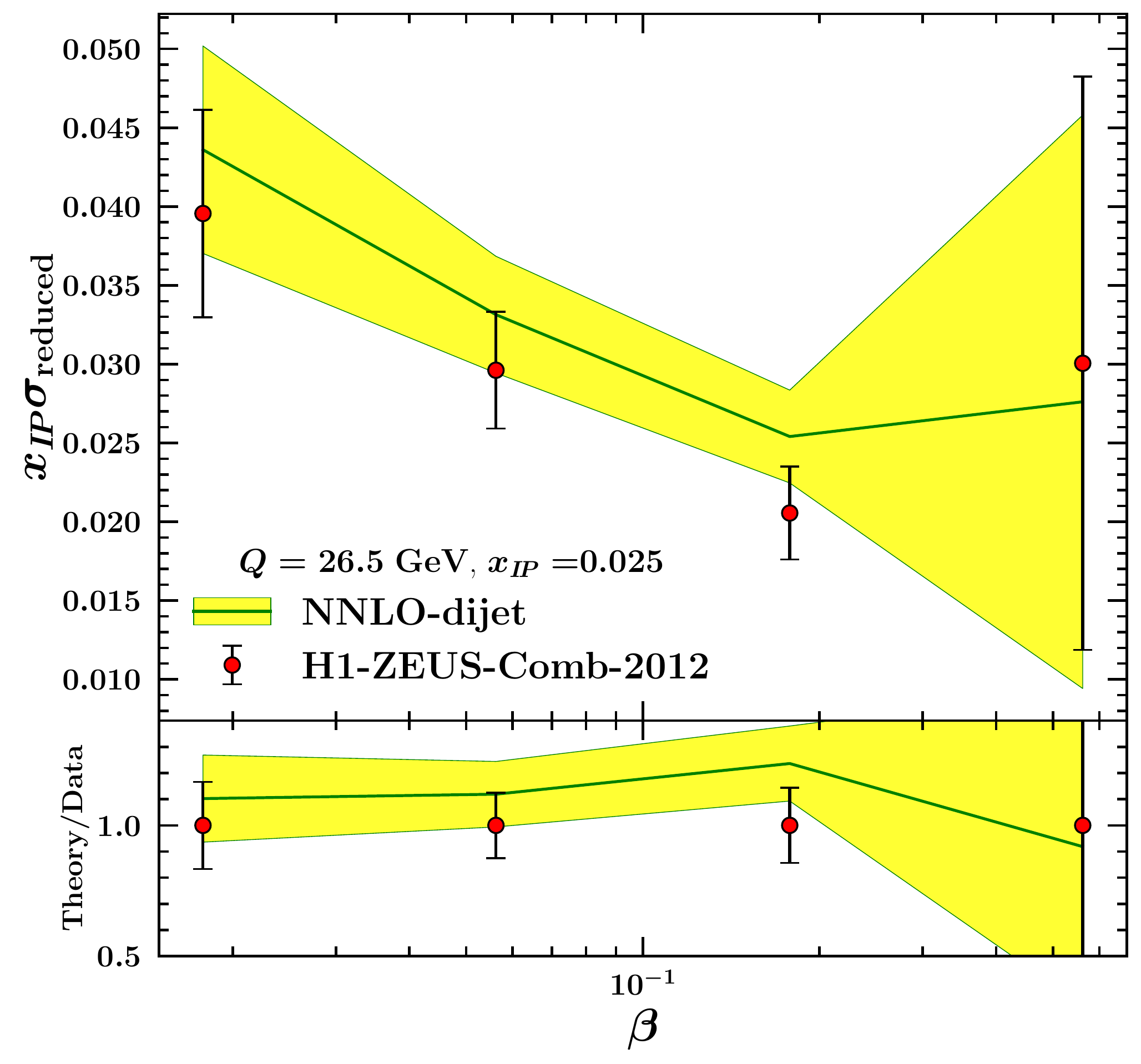}} 
\subfloat{\includegraphics[width=0.33\textwidth]{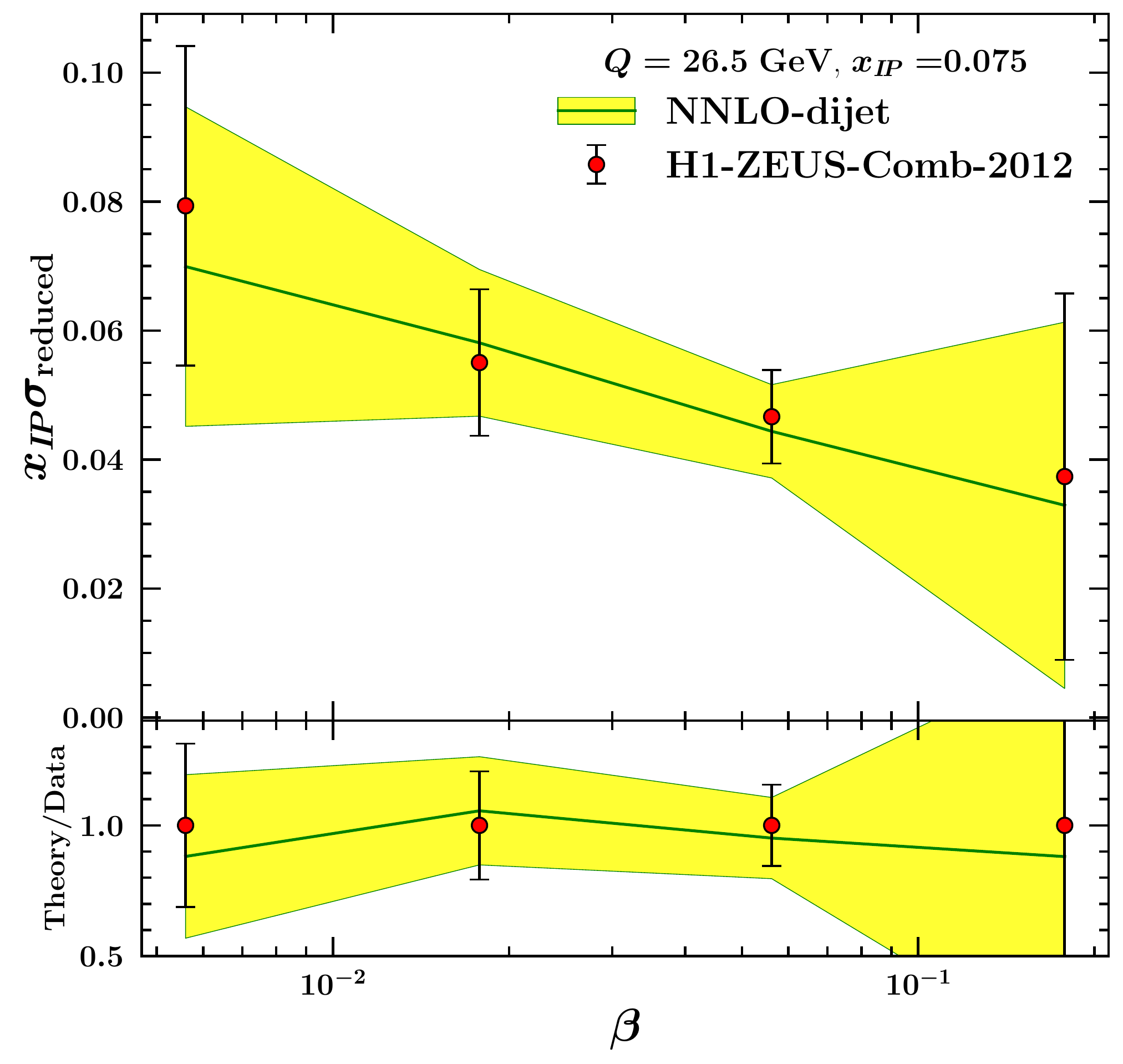}}	
\subfloat{\includegraphics[width=0.33\textwidth]{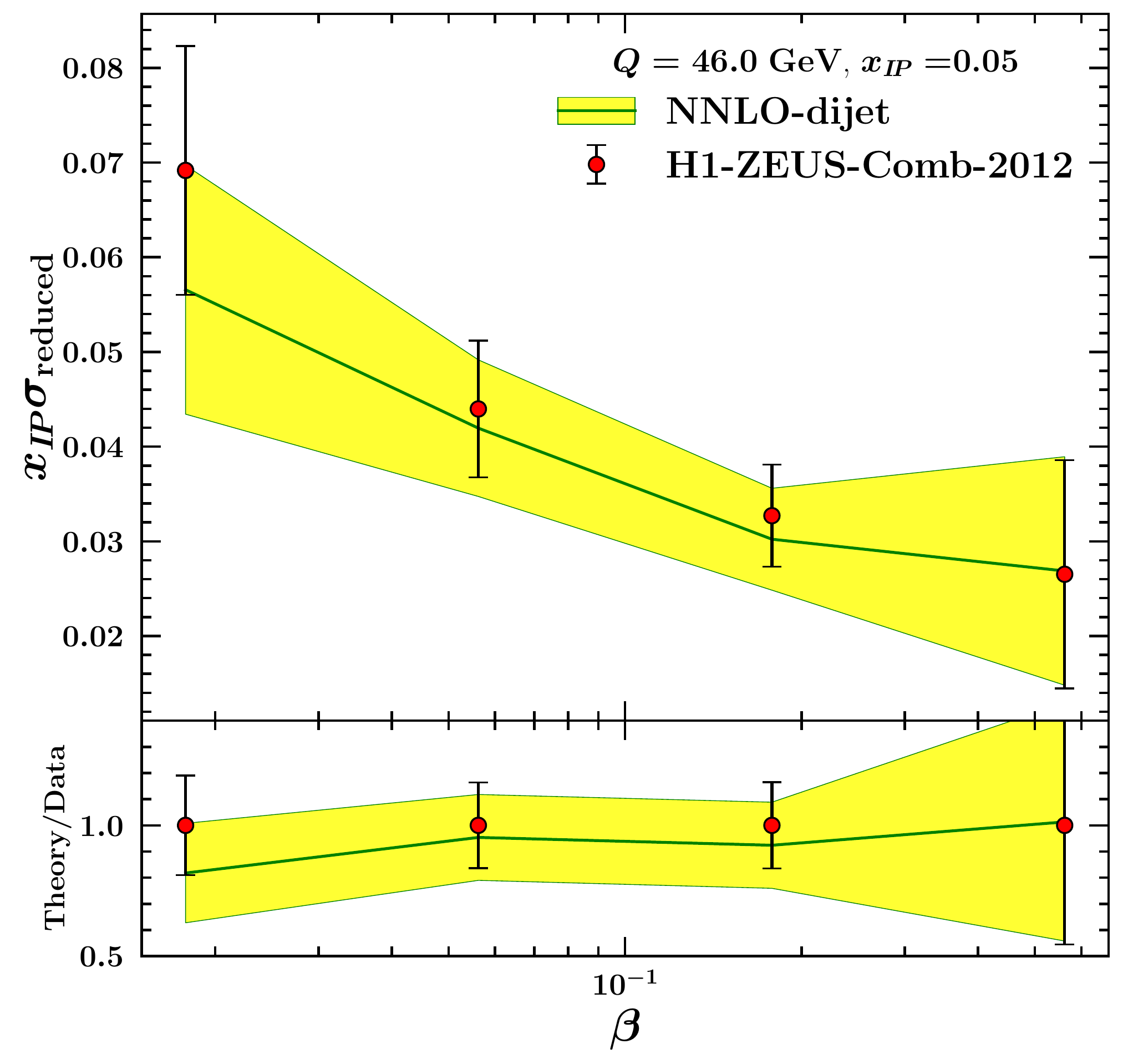}}	
\begin{center}
\caption{ \small 
Same as Fig.~\ref{fig:H1incDDIS-225-252} but this time in comparison with the H1/ZEUS combined data.}
\label{fig:H1incDDIS-Combined}
\end{center}
\end{figure*}

Finally, in Fig.~\ref{fig:H1incDDIS-Combined}, we present detailed 
comparisons between the {\tt SKMHS23-dijet}  NNLO theory 
predictions with the H1/ZEUS combined data.
The comparison are shown as a function of $\beta$ and for some 
selected values of $Q$ and  $x_{\pom}$.
The data/theory ratios also presented in the lower panel.
Again, an overall good agreement between the data
and the {\tt SKMHS23-dijet} theoretical predictions is 
achieved over the whole kinematical region.

\begin{figure*}[htb]
\vspace{0.50cm}
\centering
\subfloat{\includegraphics[width=0.33\textwidth]{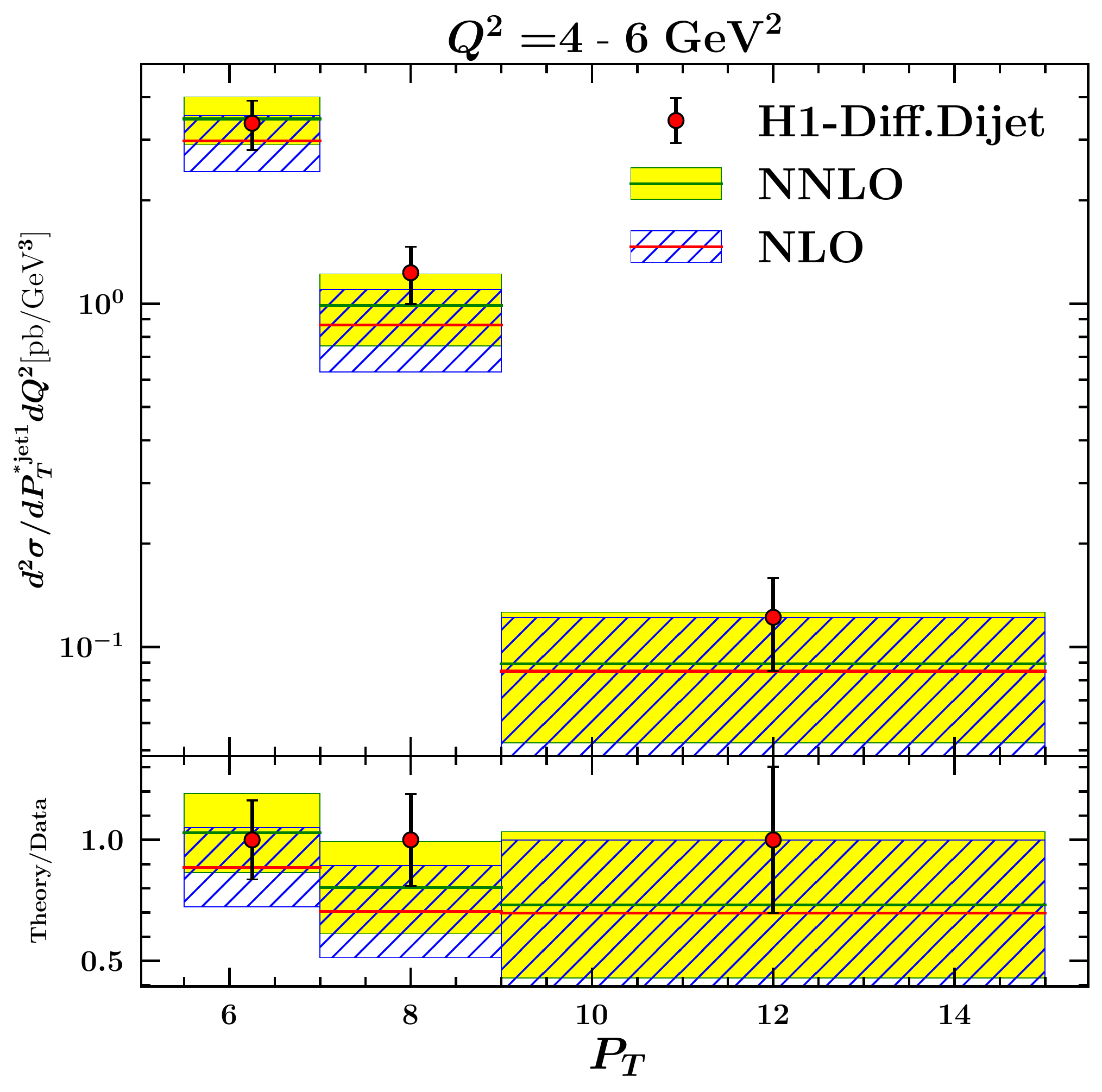}} 	
\subfloat{\includegraphics[width=0.33\textwidth,height=0.248\textheight]{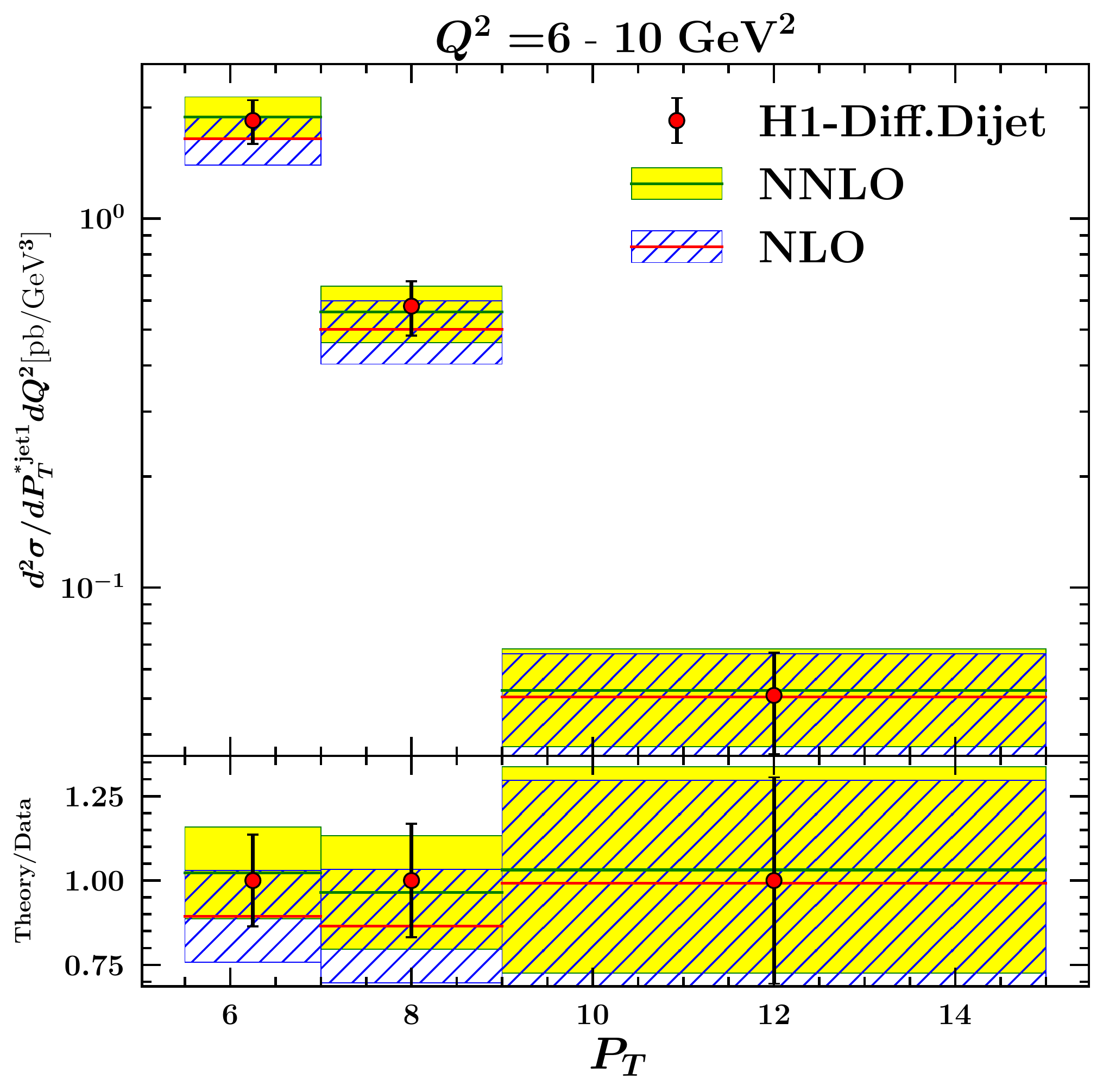}}
\subfloat{\includegraphics[width=0.33\textwidth]{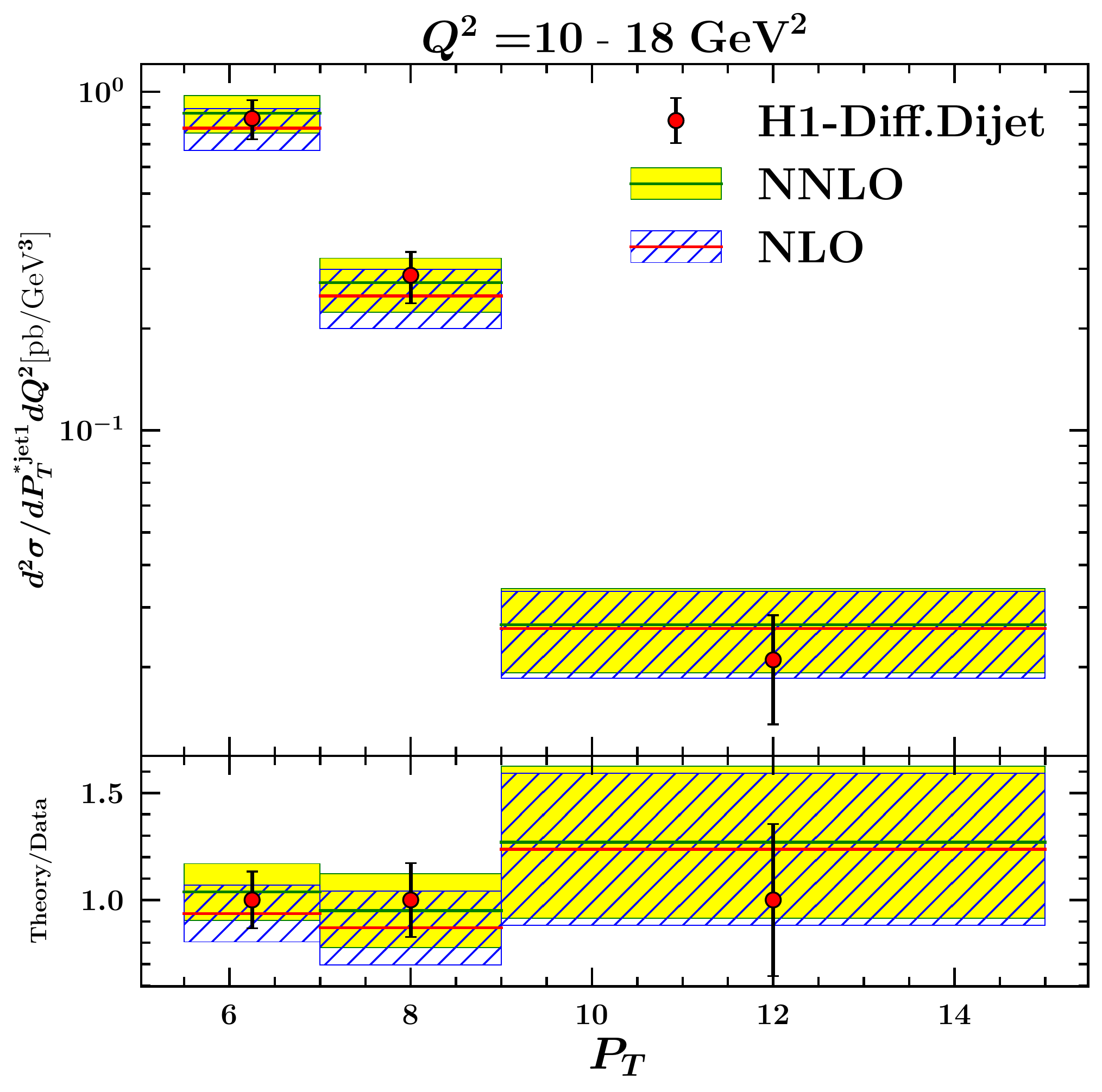}}\\	
\subfloat{\includegraphics[width=0.33\textwidth,height=0.248\textheight]{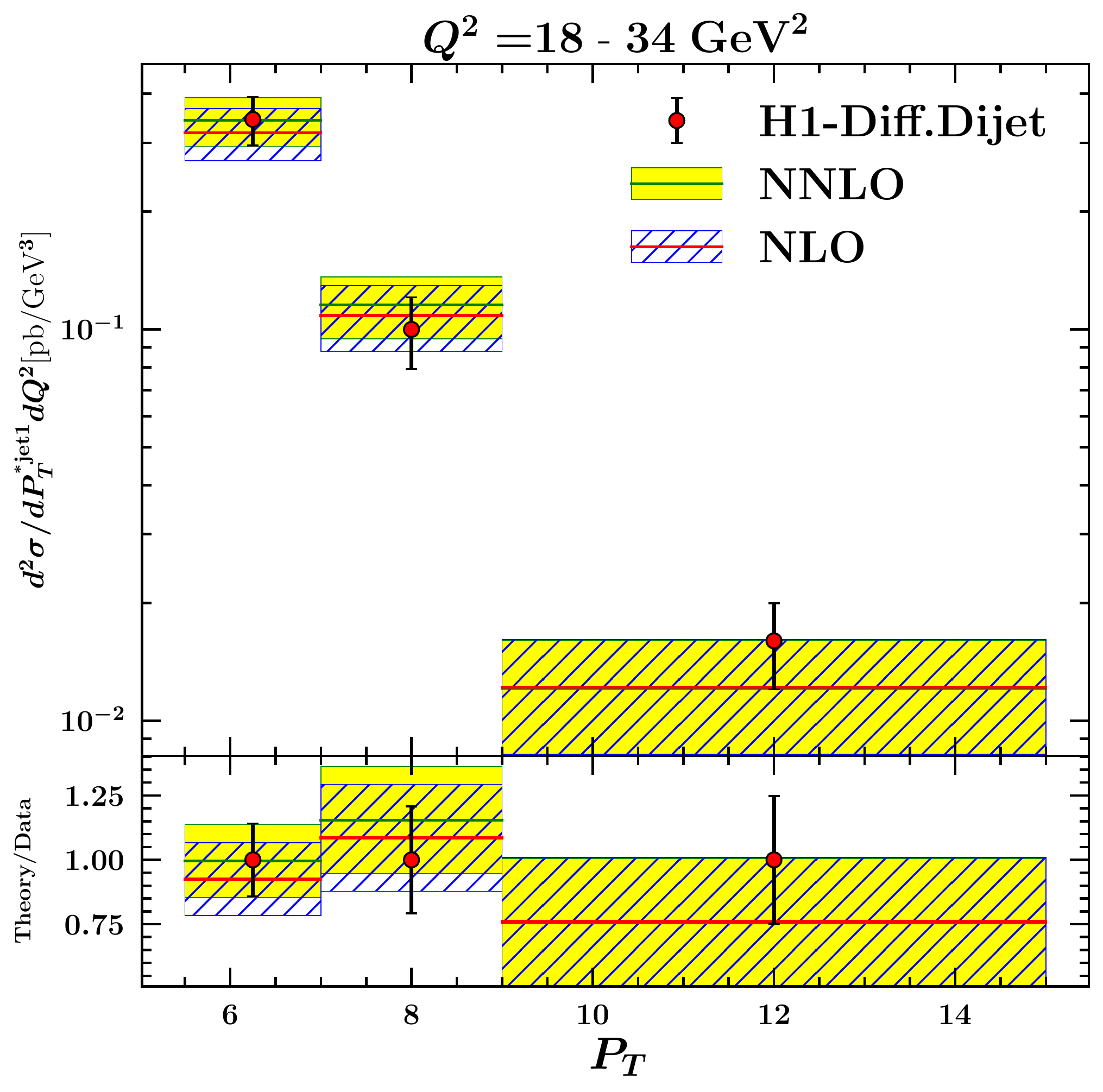}} 
\subfloat{\includegraphics[width=0.33\textwidth]{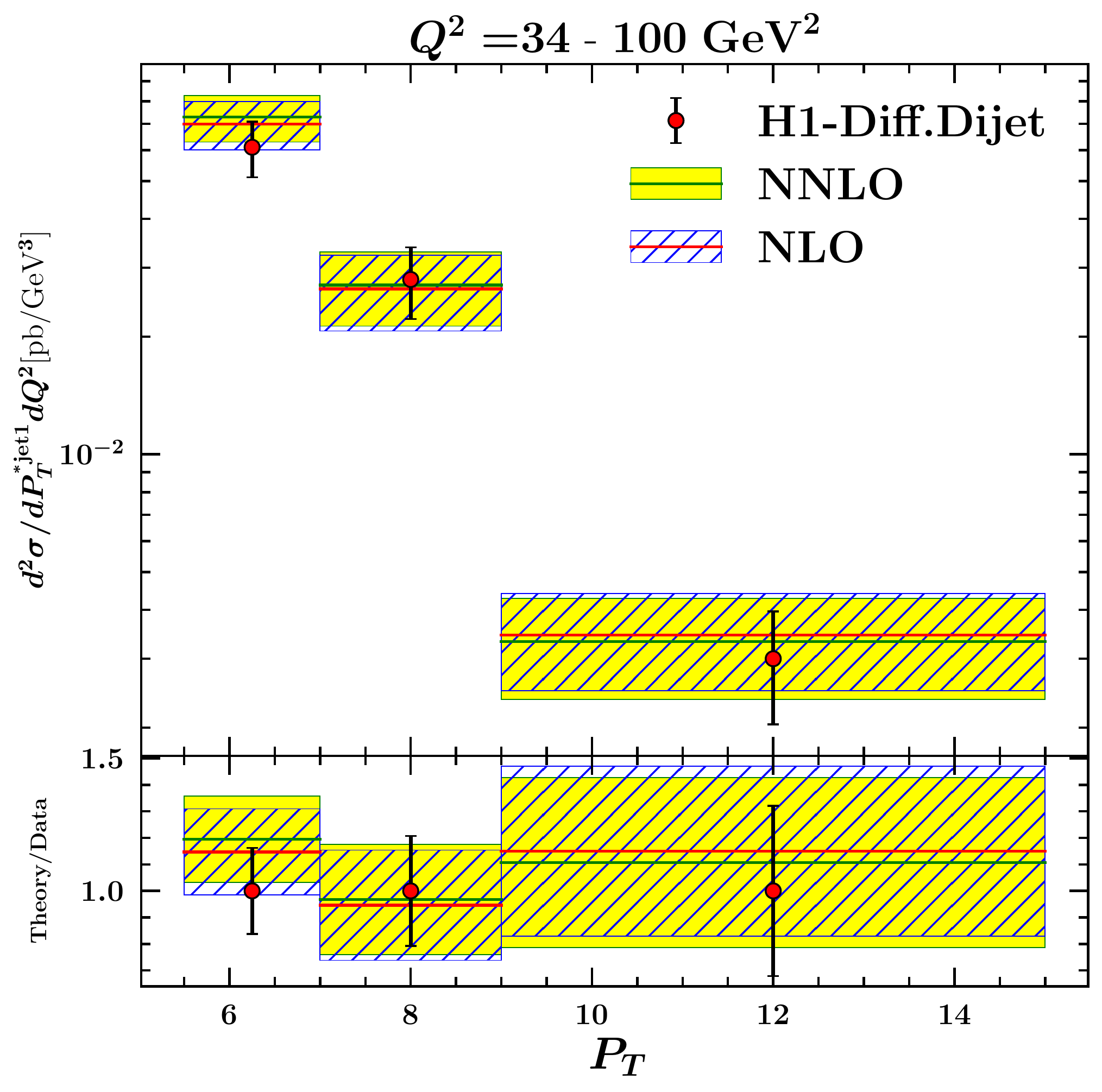}}	
\begin{center}
\caption{ \small 
Comparison of the NLO and NNLO theory prediction for diffractvie dijet production cross section 
calculated using the {\tt SKMHS23-dijet} diffractvie PDFs   
with the diffractive dijet production data published by H1 collaboration at HERA~\cite{H1:2012xlc}.
Both the absolute distributions (upper panel) and the data/theory ratios (lower panel) are shown as well. }
\label{fig:dijet}
\end{center}
\end{figure*}

Now we are in a position to turn our attention to a  detailed 
comparison with the newly added 
inclusive diffractive dijet production data published by the H1 
collaboration at HERA~\cite{H1:2012xlc}.
In Fig.~\ref{fig:dijet}, we compare the NLO and NNLO theory 
prediction for the diffractvie dijet production cross section 
calculated using the {\tt SKMHS23-dijet} diffractvie PDFs   
with the diffractive dijet production data. 
Both the absolute distributions (upper panel) and the data/theory 
ratios (lower panel) are shown. 
The comparisons are presented as a function of the transverse 
momentum $p_T$, and for different values of Q$^2$ from 4
to 100~GeV$^2$.
In general, a very good agreement between the data and the theoretical 
predictions is achieved for all values of Q$^2$.
As one can see, the NNLO predictions are very compatible with the data, consistent 
with the $\chi^2$ values per data points reported in Tab.~\ref{tab:chi2-all-SKMHS22dihet}.
For the case of the NLO fit, the $\chi^2/{\text {dof}} = 0.80$ is achieved, 
while for the NNLO fit, we obtained $\chi^2/{\text {dof}} = 0.66$.
The improvements upon inclusion of the NNLO accuracy is also reflected in the the data/theory 
comparison in Fig.~\ref{fig:dijet} and the smaller error bands in Fig.~\ref{fig:DPDF-compare}.

%
\section{Discussion and Conclusion}\label{Conclusion}

In this work, we have presented {\tt SKMHS23} and {\tt SKMHS23-dijet}, 
the first determination of diffractive PDFs  
up to the next-to-next-to-leading order accuracy in perturbative QCD 
taking into account the inclusive DIS and di-jet DIS data.
The data sets analyzed in this work include the  
combined H1 HERA-I and HERA-II LRG inclusive diffractive DIS data, 
H1 Low energy HERA-II LRG data, and more importantly the H1 HERA-II dijet LRG data.
We have discussed the quality of {\tt SKMHS23} and {\tt SKMHS23-dijet} 
QCD fits and shown that the inclusion
of QCD corrections up to the NNLO accuracy  improves the description
of the data. 
We have then examined the diffractive PDFs resulting from our QCD fits. 
We also highlighted their perturbative stability 
and observed a reduction of the diffractive PDFs uncertainties at NNLO
with respect to the NLO case.
Very good descriptions between the NLO and NNLO predictions based
on {\tt SKMHS23} and {\tt SKMHS23-dijet} and the data points are observed over 
a wide range of $x_{\pom}$ and $\beta$.
The extracted diffractive PDFs are also compared with 
the results available in the literature, where largely good agreement is found. 

In our {\tt SKMHS23} and {\tt SKMHS23-dijet} analysis we have introduced some methodological
improvements, and the theoretical framework applied in this work 
also features a number of further improvements.
As we discussed, a well-established fitting 
methodology is used to provide a faithful 
representation of the diffractive experimental
uncertainties, and to minimize any bias related to the
parametrization of the diffractive PDFs and to the 
minimization of the fitting procedure.

The theoretical calculations have been done at NLO and NNLO accuracy 
for both inclusive and jet production using the APFEL, NNLOJET and fastNLO schemes.
To consider the contribution from heavy quarks, we employed 
the FONLL-A and FONLL-C GM-VFNS approaches  which provide a proper theory input for 
such contributions at NLO and NNLO accuracy, respectively.

The H1 HERA-II dijet LRG data are also added to the data sample, to 
constrain the gluon component which is weekly 
constrained from the inclusive diffractive DIS data.
Hence, we expect that the determination of the gluon distribution is 
more reliable in our {\tt SKMHS23-dijet} QCD fit, since the dijet from
HERA-II are considered, which are directly sensitive to the gluon density.

The  {\tt SKMHS23} and {\tt SKMHS23-dijet} analyses presented in this work represents the first
step of a broader program. A number of updates and improvements are foreseen, 
and the {\tt SKMHS23} and {\tt SKMHS23-dijet} analyses presented in this article 
can be extended in several different directions.
The most important one is to repeat the analysis described here and present
a new combined QCD analysis of both recent data sets measured by
the H1 and ZEUS collaborations at HERA, and the expected observables 
from the future colliders considering the large 
hadron-electron collider (LHeC)~\cite{LHeC:2020van} on the top of the list, 
to examine the effect of such data on the extracted diffractive PDFs.

The {\tt SKMHS23} and {\tt SKMHS23-dijet} NLO and NNLO diffractive PDFs sets 
presented in this work 
are available in the standard {\tt LHAPDF} format~\cite{Buckley:2014ana} 
from the authors upon request.

%
\begin{acknowledgments}
%

Hamzeh Khanpour, Hadi Hashamipour and Maryam Soleymaninia thank the 
School of Particles and Accelerators, Institute 
for Research in Fundamental Sciences (IPM) for financial support of 
this project. Hamzeh Khanpour also is thankful to the Physics  
Department of University of Udine, and the University 
of Science and Technology of Mazandaran for the financial 
support provided for this research. 
Maryam Soleymaninia is thankful to the Iran Science Elites Federation
for the financial support.  This work was also supported in part
by the Deutsche Forschungsgemeinschaft (DFG, German Research
Foundation) through the funds provided to the Sino-German Collaborative
Research Center TRR110 ``Symmetries and the Emergence of Structure in QCD''
(DFG Project-ID 196253076 - TRR 110).  The work of UGM was 
supported in part by the Chinese
Academy of Sciences (CAS) President's International
Fellowship Initiative (PIFI) (Grant No. 2018DM0034) and 
by VolkswagenStiftung (Grant No. 93562).

\end{acknowledgments}
%

%
%

\clearpage


%

\end{document}